\pdfoutput=1
\documentclass[
reprint,
groupedaddress,
preprintnumbers,
nofootinbib,
amsmath,amssymb,
aps,
prd,
]{revtex4-2}

\usepackage[table,svgnames,dvipsnames]{xcolor}
\definecolor{Blue}{HTML}{1c1e94}
\usepackage[colorlinks=true,citecolor=Blue,linkcolor=Blue,urlcolor=Blue,pdfborder={0 0 0}]{hyperref}
\usepackage[normalem]{ulem}
\usepackage{aas_macros}
\usepackage{graphicx}
\usepackage{dcolumn}
\usepackage{bm}
\usepackage[mathlines]{lineno}
\usepackage{orcidlink}
\usepackage{cases}
\usepackage{booktabs}
\usepackage{comment}
\usepackage{multirow}
\usepackage{makecell}
\usepackage{siunitx}
\usepackage{booktabs}
\usepackage{dcolumn}
\graphicspath{{figs/}}

\usepackage{fontawesome5}
\makeatletter
\newcommand{\github}[1]{%
   \href{#1}{\faGithub}%
}
\makeatother

\newcommand\numberthis{\addtocounter{equation}{1}\tag{\theequation}}

\newcommand{\code}[1]{{\texttt{#1}}}
\renewcommand{\emph}[1]{\textit{#1}}

\usepackage[nameinlink,noabbrev]{cleveref}
\crefname{equation}{Eq.}{Eqs.}
\crefname{section}{Sec.}{Secs.}
\crefname{figure}{Fig.}{Figs.}
\crefname{table}{Tab.}{Tabs.}
\crefname{appendix}{App.}{Apps.}
\Crefname{figure}{Figure}{Figures}
\Crefname{equation}{Equation}{Equations}
\Crefname{section}{Section}{Sections}
\Crefname{table}{Table}{Tables}
\Crefname{appendix}{Appendix}{Appendices}

\newcommand{\<}{\left\langle}
\renewcommand{\>}{\right\rangle}
\newcommand{\diag}{\rm diag}

\newcommand{\fnl}{f_{\rm NL}}
\newcommand{\C}{\mathcal{C}}

\newcommand{\M}{\mathcal{M}}

\renewcommand{\P}{\mathcal{P}}
\newcommand{\R}{\mathcal{R}}
\newcommand{\nyq}{\mathrm{Nyq.}}
\newcommand{\kmin}{k_\mathrm{min}}
\newcommand{\kmax}{k_\mathrm{max}}
\newcommand{\knyq}{k_\nyq}
\newcommand{\knl}{k_\mathrm{NL}}

\newcommand{\eps}{\epsilon}

\DeclareMathOperator{\tr}{tr}
\DeclareMathOperator{\Cov}{\mathbf{Cov}}
\DeclareMathOperator{\F}{\mathbf{F}}

\def\lapl{\nabla^2}
\newcommand{\Om}{\Omega_m}

\newcommand{\rhob}{\bar\rho}
\newcommand{\nhat}{\hat{n}}
\newcommand{\khat}{\hat{k}}
\newcommand{\zhat}{\hat{z}}
\newcommand{\nbar}{\bar{n}}
\renewcommand{\ng}{n^{\rm g}}
\newcommand{\ngbar}{\nbar^{\rm g}}
\renewcommand{\d}{\delta}
\newcommand{\ddelta}{\d_{\mathrm{D}}}
\newcommand{\kdelta}{\d^{\mathrm{K}}}
\newcommand{\dm}{\d^{\mathrm{m}}}
\newcommand{\dg}{\d^{\rm g}}
\newcommand{\ds}{\d^{\rm s}}
\newcommand{\dtr}{\d^{\rm tr}}
\newcommand{\dtf}{\d^{\rm tf}}
\newcommand{\dtfm}{\left(\d^{\rm tf}\right)^m}
\newcommand{\D}{\Delta}

\newcommand{\Msunh}{\,h^{-1} M_\odot}

\newcommand{\hmpcinv}{\,h\,{\rm Mpc^{-1}}}

\newcommand{\vx}{\bm{x}}

\newcommand{\vk}{\bm{k}}

\newcommand{\vnhat}{\bm{\nhat}}

\newcommand{\vzhat}{\bm{\zhat}}

\definecolor{RoyalBlue}{rgb}{0.25,.41,.88}
\definecolor{WildStrawberry}{HTML}{EE2967}

\begin{document}

\title{Galaxy sizes as complementary (zero-)bias tracers of local primordial non-Gaussianity}

\author{Nhat-Minh Nguyen\orcidlink{0000-0002-2542-7233}}
\email{nhat.minh.nguyen@ipmu.jp}
\affiliation{Center for Data-Driven Discovery, Kavli IPMU (WPI), UTIAS, The University of Tokyo, Kashiwa, Chiba 277-8583, Japan}
\affiliation{Kavli IPMU (WPI), UTIAS, The University of Tokyo, 5-1-5 Kashiwanoha, Kashiwa, Chiba 277-8583, Japan}
\affiliation{Institute For Interdisciplinary Research in Science and Education, ICISE, Quy Nhon, 55121, Vietnam}
\author{Kazuyuki Akitsu\orcidlink{0000-0001-6473-3420}}
\affiliation{Theory Center, Institute of Particle and Nuclear Studies, High Energy Accelerator Research Organization (KEK), Tsukuba, Ibaraki 305-0801, Japan}
\author{Atsushi Taruya\orcidlink{0000-0002-4016-1955}}
\affiliation{Center for Gravitational Physics and Quantum Information, Yukawa Institute for Theoretical Physics, Kyoto University, Kyoto 606-8502, Japan}
\affiliation{Kavli IPMU (WPI), UTIAS, The University of Tokyo, 5-1-5 Kashiwanoha, Kashiwa, Chiba 277-8583, Japan}
\affiliation{Korea Institute for Advanced Study, 85 Hoegiro, Dongdaemun-gu, Seoul 02455, Republic of Korea}
\date{\today}

\begin{abstract}
The scale-dependent bias in halo and galaxy power spectra is a key signature of local primordial non-Gaussianity (local PNG), with PNG sensitivity scaling as $b_\phi/b_1$---the ratio of their responses to long-wavelength primordial potential $b_\phi$ and late-time density fluctuations $b_1$. For number density fluctuations, these responses are closely tied by the universality relation, limiting the achievable ratio. We show that size density fluctuations strongly violate this relation, thus evading the limit. For galaxy-mass halos, sizes have a vanishingly small density response but a sizable, negative local PNG response, implying an effective $b_\phi/b_1$ that is large in magnitude and opposite in sign to that of number counts. This makes galaxy sizes complementary probes of local PNG from the same galaxy sample, without any sample split. For a DESI-like survey, a multi-tracer analysis combining galaxy numbers and sizes improves the local-PNG detection significance by a factor of $\sim\!3.6$. Due to the sign flip, the number--size cross power spectrum further provides a handle on systematics in the event of a detection.
\end{abstract}

\preprint{IPMU26-0011,KEK-Cosmo-0414}

\maketitle

\section{Introduction}
\label{sec:intro}

Galaxies come in all shapes and sizes. Wide-field imaging surveys measure both properties but typically use only galaxy shapes for weak lensing measurements, discarding size information. Spectroscopic surveys are even more selective, using only galaxy numbers for galaxy clustering analyses while ignoring shape and size information altogether. Recent studies have begun exploring galaxy intrinsic shape alignment as large-scale structure tracers and cosmological signals \citep[e.g.,][]{Schmidt:2015xka,Akitsu:2020jvx,Kurita:2023qku,Kurita:2025hmp}.

In this work, we investigate the potential of galaxy intrinsic sizes as tracers of late-time large-scale structure and primordial non-Gaussianity (hereafter PNG). Using gravity-only simulations, we show that dark matter halo size perturbations respond to long-wavelength matter density perturbations at late times, as well as to long-wavelength gravitational potential perturbations in the early universe due to the presence of isotropic local PNG. Halo sizes therefore trace large-scale structure in close analogy to the halo density field: they admit the same symmetry-allowed bias expansion, but with generally different bias and stochastic coefficients, as first suggested by Ref.~\cite{Vlah:2019byq}. In particular, the PNG response of galaxy sizes need not follow the standard relation between $b_\phi$ and $b_1$ familiar from galaxy numbers, implying that the effective response direction---characterized by $b_\phi/b_1$---can differ strongly between sizes and numbers.

Correlations between galaxy sizes and large-scale structure have been reported in both \code{Horizon-AGN} hydrodynamical simulations \citep[][]{Johnston:2022nbv} and Hyper Suprime-Cam observations \cite{Ghosh:2024mlz}; Ref.~\cite{Behroozi:2021ovy} further showed that halos in simulations and galaxies in the Sloan Digital Sky Survey, when split by concentration (for halos) and a size proxy (for galaxies), cluster differently even on large scales. On the theoretical side, Refs.~\cite{Vlah:2020ovg,Ghosh:2020zfa} formulated these intrinsic correlations in terms of the power and angular power spectra, respectively.

Our main contribution is to show that galaxy size correlations provide a complementary handle on isotropic local PNG when combined with galaxy number correlations. The key insight is that the responses of the galaxy size correlation to local PNG can be very distinct relative to galaxy number correlations, enabling additional multi-tracer information when the galaxy size--size and size--number correlations are measured with sufficient signal-to-noise on large scales. To this end, we present a Fisher forecast for multi-tracer constraints on the amplitude of local PNG $\fnl$ that combines galaxy numbers and sizes, as well as their cross-correlations.

The remainder of this paper is organized as follows. In \cref{sec:formalism}, we first present the formalism of galaxy number and size bias for Gaussian and non-Gaussian initial conditions, then discuss two major observational effects that must be accounted for: projection and redshift-space distortions. We describe the simulations in \cref{sec:sims} and the measurements in \cref{sec:measurements}. In \cref{sec:forecast}, we forecast galaxy number--size multi-tracer constraints on local PNG. We discuss future developments necessary for applications on observational data in \cref{sec:discussion} before concluding in \cref{sec:conclusion}. The appendices provide further details of our measurements and forecasts, as well as supporting arguments. In particular, \cref{app:approx_Gaussian_bias_fits,app:approx_universality} present best-fit empirical relations between galaxy size and number bias coefficients.

\section{Formalism}
\label{sec:formalism}

While a galaxy is inherently a three-dimensional (hereafter 3D) object, its apparent size is two-dimensional (hereafter 2D) owing to projection along the observer's line of sight. In this section, we first introduce the estimator and bias formalisms for galaxy 3D sizes before turning to those for the observed galaxy 2D sizes in \cref{sec:formalism:observations:projection}.

Let us consider a galaxy comoving number density field within a narrow redshift range
\begin{equation}
    \ng(\vx)=\sum_g\ddelta(\vx-\vx_g)\label{eq:ng},
\end{equation}
where $\ddelta$ is the Dirac delta function and the sum $\sum_g$ runs over all galaxies in the sample.
The associated galaxy number density perturbation field $\dg$ is given by
\begin{equation}                     
    \dg(\vx)=\frac{\ng(\vx)}{\ngbar}-1\label{eq:dg_def}
\end{equation}
with the comoving mean number density $\ngbar=(1/V)\int d^3x\ng(\vx)$ where $V$ is the comoving volume of the galaxy sample.
\cref{eq:dg_def} is the usual observable in galaxy clustering analyses.

Considering a single galaxy as a set of point masses with the center of mass located at $\vx_g$, its three-dimensional size and shape can be characterized by the (mass-weighted) second-moment inertia tensor
\begin{equation}
    I_{ij}(\vx_g)\propto \sum_p w_p\left(x_{i,p}-x_{i,g}\right)\left(x_{j,p}-x_{j,g}\right),
    \label{eq:Iij_g}
\end{equation}
where $i,j\in\{x,y,z\}$ and the sum runs over all member particles (assumed to have the same mass). The radial weight $w_p$ is sometimes chosen to emphasize different regions of the mass distribution. In simulations, Eq.~\eqref{eq:Iij_g} can be evaluated directly. Two standard choices are the \emph{simple} tensor where $w_p=1$ and the \emph{reduced} tensor where $w_p\propto r_p^{-2}$ (often with $r_p$ taken as a normalized, iteratively updated ellipsoidal radius). For intrinsic alignment studies, the reduced form is more widely adopted to downweight the noisy halo outskirts \citep[e.g.,][]{Kurita:2020hap,Shi:2020cyz,Akitsu:2020fpg,Akitsu:2023eqa,vanHeukelum:2025njq}.
Here, we also adopt the trace of the reduced inertia tensor as the fiducial proxy for halo sizes. We discuss the relationship between this estimator and other halo properties, such as radius and concentration, in \cref{app:size_proxy}.

We decompose the 3D inertia tensor field into the trace $\dtr$ and trace-free $\dtf$ components as
\begin{align*}
    I_{ij}(\vx) &= \sum_g I_{ij}(\vx_g)\ddelta(\vx-\vx_g)\\
    &= \bar{T}\left[\frac{1}{3}\kdelta_{ij}\left(1+\dtr(\vx)\right)+\dtf_{ij}(\vx))\right]\numberthis\label{eq:Iij}
\end{align*}
where, similar to the number density in \cref{eq:ng} and the mean number density, we define the 3D trace field
\begin{equation}
    T(\vx)=\sum_g t(\vx_g)\ddelta(\vx-\vx_g)\label{eq:T3D},
\end{equation}
with $t\equiv\tr I_{ij}=I_{xx}+I_{yy}+I_{zz}$ for each individual galaxy, and the mean 3D trace
\begin{align*}
    \bar{T}&=(1/V)\int d^3xT(\vx)=\ngbar\<t\>\numberthis\label{eq:T3D_mean}.
\end{align*}
$\dtr$ and $\dtf_{ij}$ in \cref{eq:Iij} characterize the spin-0 galaxy size and spin-2 galaxy shape perturbations, respectively \citep{Vlah:2019byq,Vlah:2020ovg}.

\cref{eq:T3D} further implies that both $\dtr$ and $\dtf_{ij}$ are number-weighted.
In particular, the trace density perturbation field
\begin{equation}
    \dtr(\vx)=\frac{T(\vx)-\bar{T}}{\bar{T}}\label{eq:dtr_def}
\end{equation}
is dominated by perturbations in the galaxy number density $\ng$\footnote{This is because the average of the trace is not zero, different from e.g., the traceless shape and the peculiar velocity.}.
Therefore, we define the galaxy size density perturbation field $\ds$ as the difference between the galaxy trace $\dtr$ and number density perturbations $\dg$, i.e.
\begin{align*}
    \ds(\vx)=\dtr(\vx)-\dg(\vx)&=\frac{T(\vx)}{\bar{T}}-\frac{\ng(\vx)}{\ngbar}\\
    &=\frac{1}{\ngbar}\sum_g \left[\frac{t(\vx_g)}{\<t\>}-1\right]\ddelta(\vx-\vx_g)\numberthis\label{eq:ds_def}.
\end{align*}

As the shape density perturbations will become relevant when projection effect is considered, we also define it as the trace-free tensor field\footnote{Here, to keep the same normalization convention between sizes and shapes, we normalize the shape field by the spatial (or ensemble average) of the trace, rather than the per-object trace often adopted in weak lensing and intrinsic alignment literature.}
\begin{equation}
   \dtf_{ij}(\vx)=\frac{I_{ij}(\vx)-\frac{1}{3}\kdelta_{ij}T(\vx)}{\bar{T}}\label{eq:dtf_def}.
\end{equation}

\subsection{Gaussian bias}
\label{sec:formalism:3Dsize_Gaussian}

The observed galaxy numbers and sizes are modulated by long-wavelength matter perturbations. Their responses can be described by the effective-field-theory (EFT) bias expansions, which relate the corresponding observable field to the underlying matter field \citep[See][for a review]{Desjacques:2016bnm}.
In the EFT framework, perturbations in galaxy numbers $\dg$ and galaxy sizes $\ds$ are both scalar observables so their bias expansions assume the same form
\begin{equation}
    \d^{\rm g,s}(\vx,z)=\sum b^{\rm g,s}_OO(\vx,z)+\eps^{\rm g,s}(\vx,z),
    \label{eq:gaussian_bias_exp}
\end{equation}
where $O(\vx)$ are local gravitational observables allowed by Einstein's equivalence principle, each associated with an EFT bias coefficient $b_O$ while $\eps(\vx)$ encapsulates stochastic contributions from small-scale physics.
Below, we restrict to narrow redshift bins so that we can drop the redshift parameter $z$ in the equations.

In this paper, we restrict the bias expansions to leading order. For the spin-0 fields, that is the linear galaxy bias model
\begin{equation}
    \d^{\rm g,s}(\vx)=b^{\rm g,s}_1\dm(\vx)+\eps^{\rm g,s}(\vx)\label{eq:gaussian_linear_bias_spin0},
\end{equation}
where $b^{\rm g,s}_1$ are the Gaussian linear bias coefficients for galaxy numbers and sizes, respectively, and $\dm(\vx)\equiv[\rho(\vx)-\rhob]/\rhob$ is the perturbation in the matter density field $\rho(\vx)$.

For the spin-2 field, the leading-order bias expansion is the linear tidal alignment model\footnote{In past literature, $b^{\rm tf}_1$ is often denoted as $b_{\rm K}$ (or $b_{\rm TF[\Pi^{(1)}]}$ where $\mathrm{TF}[\Pi^{(1)}]=K^{(1)}_{ij}$).}
\begin{equation}
    \dtf_{ij}(\vx)=b^{\rm tf}_1K_{ij}(\vx)+\eps^{\rm tf}_{ij}(\vx)\label{eq:gaussian_linear_bias_spin2},
\end{equation}
where
\begin{equation}
    K_{ij}=\left(\frac{\partial_i\partial_j}{\nabla^2}-\frac{1}{3}\kdelta_{ij}\right)\d=\mathcal{D}_{ij}\d\label{eq:Kij}
\end{equation}
is the gravitational tidal tensor.

\subsection{Non-Gaussian bias}
\label{sec:formalism:3Dsize_PNG}

If primordial perturbations are not generated by the inflaton itself but rather by auxiliary fields as, e.g., in multifield inflation scenarios, then the primordial gravitational potential (Bardeen potential) can be parametrized as \citep{Gangui:1993tt,Komatsu:2001rj}
\begin{equation}
    \phi(\vx)=\phi_{\rm G}(\vx)+\fnl\left[\phi_{\rm G}^2(\vx)-\<\phi_{\rm G}^2(\vx)\>\right]\label{eq:phi_NG}.
\end{equation}
The second term on the right-hand side vanishes in the standard single-field inflation scenario. \cref{eq:phi_NG} defines PNG of the local type, or local PNG.

In the presence of local PNG, \cref{eq:gaussian_linear_bias_spin0} admits an additional contribution
\begin{equation}
    \d^{\rm g,s}(\vx)=b^{\rm g,s}_1\dm(\vx)+b^{\rm g,s}_\phi\fnl\phi_{\rm G}(\vx)+\eps^{\rm g,s}(\vx)\label{eq:PNG_linear_bias},
\end{equation}
where $b^{\rm g,s}_\phi$ are the non-Gaussian galaxy numbers and sizes linear bias.
The local PNG contribution modifies the large-scale galaxy number and size power spectra as
\begin{align*}
    P_{\rm gg,ss}(k)=&\left(b^{\rm g,s}_1\right)^2P_{\rm mm}(k)\\
    &+2b^{\rm g,s}b_\phi\fnl P_{\rm m\phi}(k) + \left(b^{\rm g,s}_\phi\right)^2\fnl^2 P_{\phi\phi}(k)\\
    &+P_{\eps^{\rm g,s}\eps^{\rm g,s}}(k)\numberthis\label{eq:PNG_linear_Pk_auto}
\end{align*}
where $k=|\vk|$ is the Fourier wavenumber.
The terms on the second line of \cref{eq:PNG_linear_Pk_auto} can be rewritten using the relation $\dm(\vk)=\M(k)\phi(\vk)$ as
\begin{align*}
    P_{\rm gg,ss}(k)=&\left[\left(b^{\rm g,s}_1\right)^2+\frac{2b^{\rm g,s}_1b^{\rm g,s}_\phi\fnl}{\M(k)}+\frac{\left(b^{\rm g,s}_\phi\right)^2\fnl^2}{\M(k)^2}\right]P_{\rm mm}(k)\\
    &+P_{\eps^{\rm g,s}\eps^{\rm g,s}}(k)\numberthis\label{eq:PNG_linear_Pk_auto_scale-dependent_bias}
\end{align*}
where $\M(k,z)=\left[2k^2T(k,z)\right]/\left(3\Om H_0^2\right)$ with $T$ being the matter transfer function and $\Om,\,H_0$ being the fractional matter density and the Hubble expansion rate today, respectively.
The $k$-dependence of $\M(k)$ on the right-hand side of \cref{eq:PNG_linear_Pk_auto_scale-dependent_bias} introduces the \emph{scale-dependent bias} \citep{Dalal:2007cu} in the galaxy number and size power spectra.

Below, we will assume the leading-order stochastic contributions to be isotropic and Poisson, i.e.
\begin{align}
    P_{\eps^{\rm g}\eps^{\rm g}}(k)&=P_{\eps^{\rm g}\eps^{\rm g}}=\frac{1}{\ngbar}\label{eq:Pgg_Poisson_shotnoise},\\
    P_{\eps^{\rm s}\eps^{\rm s}}(k)&=P_{\eps^{\rm s}\eps^{\rm s}}=\sigma_{\rm s}^2P_{\eps^{\rm g}\eps^{\rm g}}\label{eq:Pss_Poisson_shotnoise}
\end{align}
where $\sigma^2_{s}=\sigma^2_{\rm t}/\<t\>^2$ with $\sigma^2_{\rm t}$ being the sample variance.

Another observable is the number--size cross power spectrum $P_{\rm gs}$ given by
\begin{equation}
    P_{\rm gs}(k)=\left[b^{\rm g}_1b^{\rm s}_1+\frac{\left(b^{\rm g}_1b^{\rm s}_\phi+b^{\rm s}_1b^{\rm g}_\phi\right)\fnl}{\M(k)}+\frac{b^{\rm g}_\phi b^{\rm s}_\phi\fnl^2}{\M(k)^2}\right]P_{\rm mm}(k)\label{eq:PNG_linear_Pk_cross}
\end{equation}
where there is no stochastic contribution, i.e. $P_{\eps^{\rm g}\eps^{\rm s}}=0$ as, following the definition of $\ds$ in \cref{eq:ds_def}, we assume no correlation between $\eps^{\rm g}$ and $\eps^{\rm s}$ at the leading order.
In practice, since the size power spectrum $P_{\rm ss}$ is often very small (see below), the number--size cross power spectrum $P_{\rm gs}$ is the more relevant observable.

\subsection{Local PNG and zero-bias tracers}
\label{sec:formalism:zerobias_gaussian}

\cref{eq:PNG_linear_Pk_auto_scale-dependent_bias,eq:PNG_linear_Pk_auto_scale-dependent_bias} shows how local PNG modulates the power spectra of galaxy numbers and sizes as well as their cross power spectrum. Ref.~\cite{Castorina:2018zfk} argued that for a detection of the scale-dependent bias effect and local PNG ($\fnl\neq0$), the first terms on the right-hand sides act as noise. Therefore, the optimal strategy for detecting scale-dependent bias and local PNG in the cosmic variance limit\footnote{Even in the cosmic variance limit (i.e., neglecting the intrinsic $P_{\epsilon\epsilon}$), there appears a white-noise-like contribution on large scales from loop corrections \citep[e.g.,][]{Schmittfull:2018yuk,Cabass:2023nyo,Foreman:2024kzw,Kokron:2025yma,Akitsu:2025boy}.} should be to \emph{maximize the ratio $b_\phi/b_1$}. To this end, Ref.~\cite{Castorina:2018zfk} proposed splitting the galaxy samples and engineering galaxy weighting schemes to construct zero-bias galaxy samples. Recently, Ref.~\cite{Kokron:2025yma} also suggested exploiting a specific time window during reionization, where neutral hydrogen acts as a natural zero-bias tracer. In this work, we investigated galaxy size as another natural zero-bias tracer. One key advantage of galaxy sizes is that this information is already available for most (if not all) galaxy samples selected from wide-field imaging.

\subsection{Galaxy size as a zero-bias tracer for local PNG}
\label{sec:formalism:zerobias_SU}

To build some intuition, in this section, we adopt the peak-background split framework and the separate-universe approach, in which galaxy bias coefficients can be defined as responses to long-wavelength perturbations. Let us then consider some long-wavelength perturbations in the matter density $\d_L$ and the primordial gravitational potential $\fnl\phi_L$.

For galaxy number density perturbation field $\dg$, as defined by \cref{eq:dg_def}, we can define the Gaussian and non-Gaussian linear bias coefficients as the following responses
\begin{equation}
b^{\rm g}_{1,\rm L}(M,z)=\frac{\partial\ln \ngbar(M,z)}{\partial \d_L},\quad
b^{\rm g}_\phi(M,z)=\frac{\partial\ln \ngbar(M,z)}{\partial\fnl\phi_L},
\label{eq:bg_responses}
\end{equation}
where $b^{\rm g}_{1,\rm L}$ is the \emph{Lagrangian} linear bias. Here, $b^{\rm g}_{1,\rm L}$ quantifies the response of the galaxy number density to a long-wavelength density perturbation, which can be interpreted as a rescaling of the background density $\bar\rho_m\to \bar\rho_m(1+\d_L)$ \citep{Sirko:2005uz,Wagner:2014aka,Li:2015jsz,Baldauf:2015vio,Lazeyras:2015lgp,Desjacques:2016bnm}. Similarly, $b^{\rm g}_\phi$ quantifies the response of the galaxy number density to a long-wavelength primordial gravitational potential, which in the separate-universe picture corresponds to a rescaling of the primordial fluctuation amplitude $A_s\to A_s(1+4 \fnl\phi_L)$ \citep{Slosar:2008hx,Baldauf:2015vio,Desjacques:2016bnm}.

Considering the simplest version of the peak-background split picture, where halos and galaxies form from spherical collapses of density peaks above a scale-independent critical threshold density, we can write the galaxy numbers as $\ng(\nu>\nu_{\rm cr},z)$ where $\nu=\delta_{\rm cr}/\sigma(M,z)$ is the peak height. For a power-law matter power spectrum, $\ng(\nu>\nu_{\rm cr},z)$ exhibits universal behavior \citep[e.g.,][]{Press:1973iz,Bond:1990iw}. In that case, the Gaussian and non-Gaussian responses are both controlled by derivatives with respect to the same variable $\nu$, so that
\begin{align}
b^{\rm g}_{1,\rm L}
&=\frac{\partial\ln \ngbar(\nu)}{\partial\d_L}
=\frac{\partial\ln \ngbar}{\partial\nu}\frac{\partial\nu}{\partial\d_L},\\
b^{\rm g}_{\phi}
&=
2\frac{\partial\ln \ngbar(\nu)}{\partial\ln\sigma}
=
2\frac{\partial\ln \ngbar}{\partial\nu}\frac{\partial\nu}{\partial\ln\sigma}.
\label{eq:bg_chain_rule}
\end{align}
Using $\partial\nu/\partial\d_L=-\nu/\delta_{\rm cr}$ and $\partial\nu/\partial\ln\sigma=-\nu$, one obtains
\begin{equation}
b^{\rm g}_\phi(M,z)= 2\delta_{\rm cr}\,b^{\rm g}_{1,\rm L}(M,z).
\label{eq:universality_L}
\end{equation}
For a conserved tracer, the Eulerian and Lagrangian biases are related by $b^{\rm g}_1=1+b^{\rm g}_{1,\rm L}$, which yields the familiar ``universality relation'' for dark matter halos selected by mass \citep[][]{Dalal:2007cu}
\begin{equation}
b^{\rm g}_\phi(M,z)= 2\delta_{\rm cr}\left[b^{\rm g}_1(M,z)-1\right],
\label{eq:universality}
\end{equation}
where $\delta_{\rm cr}\simeq1.686$ is the threshold density for spherical collapse, linearly extrapolated to redshift $z=0$.

For galaxy samples, both galaxy formation history and sample selection lead to scatters around and deviations from the universality relation \citep{Barreira:2020kvh, Barreira:2023rxn}. For some tracer $A$, we can instead consider an extension of \cref{eq:universality}, e.g., of the form similar to Eq.~(18) of Ref.~\cite{Hadzhiyska:2025rez}:
\begin{equation}
b^A_\phi(M,z,\theta) = 2\delta_{\rm cr}\,c^{A}\left[b_1^{A}(M,z,\theta)-p^{A}\right],
\label{eq:ext_universality}
\end{equation}
where we have made explicit the \emph{assembly bias}, i.e. the fact that both the Gaussian and non-Gaussian biases depend on other intrinsic galaxy properties $\theta$ beyond mass \citep[e.g.,][]{Dalal:2008zd,Reid:2010vc,Lazeyras:2021dar,Lazeyras:2022koc}. Deviations of $c^{A}$ and $p^{A}$ from unity indicate that the response of tracer $A$ is not captured by a peak-height-only description, either because $A$ depends on additional halo properties beyond mass (hence beyond $\nu$), or because the selection does not correspond to a clean mass-selected sample. In fact, such deviations and secondary dependencies were suggested to serve as a guide on how to split galaxy samples in multi-tracer setups, where the constraints on $\fnl$ scale as $\sigma_{\fnl}\propto|b^B_1b^A_\phi-b^A_1b^B_\phi|$.

Here, we consider a complementary viewpoint in which secondary properties beyond mass directly serve as tracers. We therefore are interested in the clusterings and biases of the galaxy sizes themselves. For the galaxy size density perturbation field $\ds$, as defined by \cref{eq:ds_def}, the corresponding responses are
\begin{equation}
b^{\rm s}_1(M,z)=\frac{\partial\ln\langle t\rangle(M,z)}{\partial \d_L},\quad
b^{\rm s}_\phi(M,z)=\frac{\partial\ln\langle t\rangle(M,z)}{\partial\fnl\phi_L}.
\label{eq:bs_X}
\end{equation}
Unlike galaxy numbers, the size field defined in \cref{eq:ds_def} is a zero-mean internal property field rather than a conserved number density. Thus, there is no analogous Eulerian offset of $+1$.

If the mean halo trace depends on the peak height and additional internal variables $\theta_i$,
\begin{equation}
\langle t\rangle=\langle t\rangle(\nu,\{\theta_i\}),
\end{equation}
then
\begin{align}
b^{\rm s}_1
&=
\frac{\partial\ln\langle t\rangle}{\partial\nu}\frac{\partial\nu}{\partial\d_L}
+\sum_i \frac{\partial\ln\langle t\rangle}{\partial \theta_i}\frac{\partial \theta_i}{\partial\d_L},
\label{eq:bs_1_chain_X}\\
b^{\rm s}_\phi
&=
\frac{\partial\ln\langle t\rangle}{\partial\nu}\frac{\partial\nu}{\partial(\fnl\phi_L)}
+\sum_i \frac{\partial\ln\langle t\rangle}{\partial \theta_i}\frac{\partial \theta_i}{\partial(\fnl\phi_L)}.
\label{eq:bs_phi_chain_X}
\end{align}
If the peak height $\nu$ were the only relevant variable, one would recover the universality relation $b^{\rm s}_\phi\approx 2\delta_{\rm cr}\,b^{\rm s}_1$, so that the ratio between non-Gaussian and Gaussian responses of galaxy sizes would have comparable amplitude to that of galaxy numbers.
A useful way to understand why this expectation fails is to contrast the separate-universe interpretations of $\d_L$ and $\fnl\phi_L$.
For galaxy numbers, a long-wavelength density perturbation mainly acts as a shift of the local background density, or equivalently as an additive shift of the effective collapse threshold $\delta_{\rm cr}$.
This efficiently changes the abundance, but for an internal statistic at fixed final mass its leading abundance-like effect is removed once we condition on $(M,z)$.
This should be understood as a suppression mechanism rather than an exact cancellation: once the leading abundance-like contribution is removed, the remaining response is mediated by internal variables that respond only weakly to $\d_L$, so $b^{\rm s}_1$ is small but non-zero.

By contrast, local PNG induces a modulation of the small-scale fluctuation amplitude, which in the separate-universe picture can be viewed as a local rescaling of $A_s$ (or $\sigma_8$). In the excursion-set picture~\citep{Bond:1990iw, Zentner:2006vw}, this is not merely a shift of the effective collapse threshold $\delta_{\rm cr}$. Rather it rescales the small-scale variance along the entire random walk.
This changes formation histories and concentration-like variables more directly, even at fixed final mass, i.e. fixed $\delta_{\rm cr}$, so the different terms in \cref{eq:bs_phi_chain_X} can add coherently and yield a substantially larger $|b^{\rm s}_\phi|$.

This interpretation is built into the definition of the size field in \cref{eq:ds_def}: by construction, the leading number-density-like contribution is subtracted from the trace response, so that $\ds$ isolates the residual internal-property fluctuation rather than the dominant abundance-like piece itself.
This makes it natural that the size density field is much less sensitive to the leading threshold-crossing response than the number density field, while still allowing a nonzero response through $\nu$ and the additional variables $\theta_i$.
In \cref{app:size_concentration}, we illustrate this mechanism explicitly by taking halo concentration as a concrete example of an additional variable and showing that its mediated contribution is generally small for $b^{\rm s}_1$ but can be sizable for $b^{\rm s}_\phi$ --- a direct numerical signature of the asymmetric sensitivity of concentration-like assembly variables to $\d_L$ versus $\fnl\phi_L$ at fixed halo mass.

\begin{figure}[tb]
  \centering
  \includegraphics[width=0.8\linewidth]{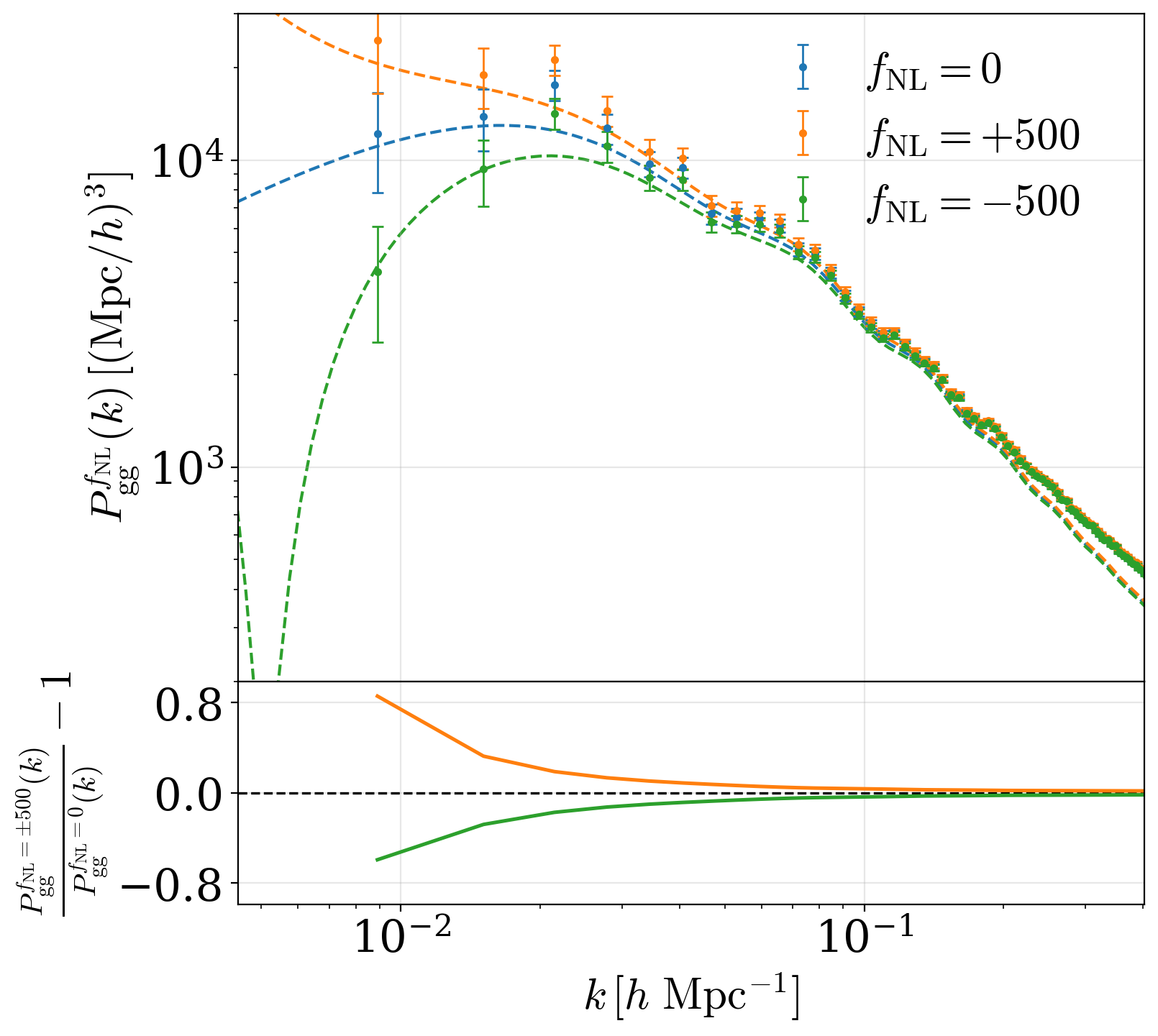}
  \vspace{0.15em}
  \includegraphics[width=0.8\linewidth]{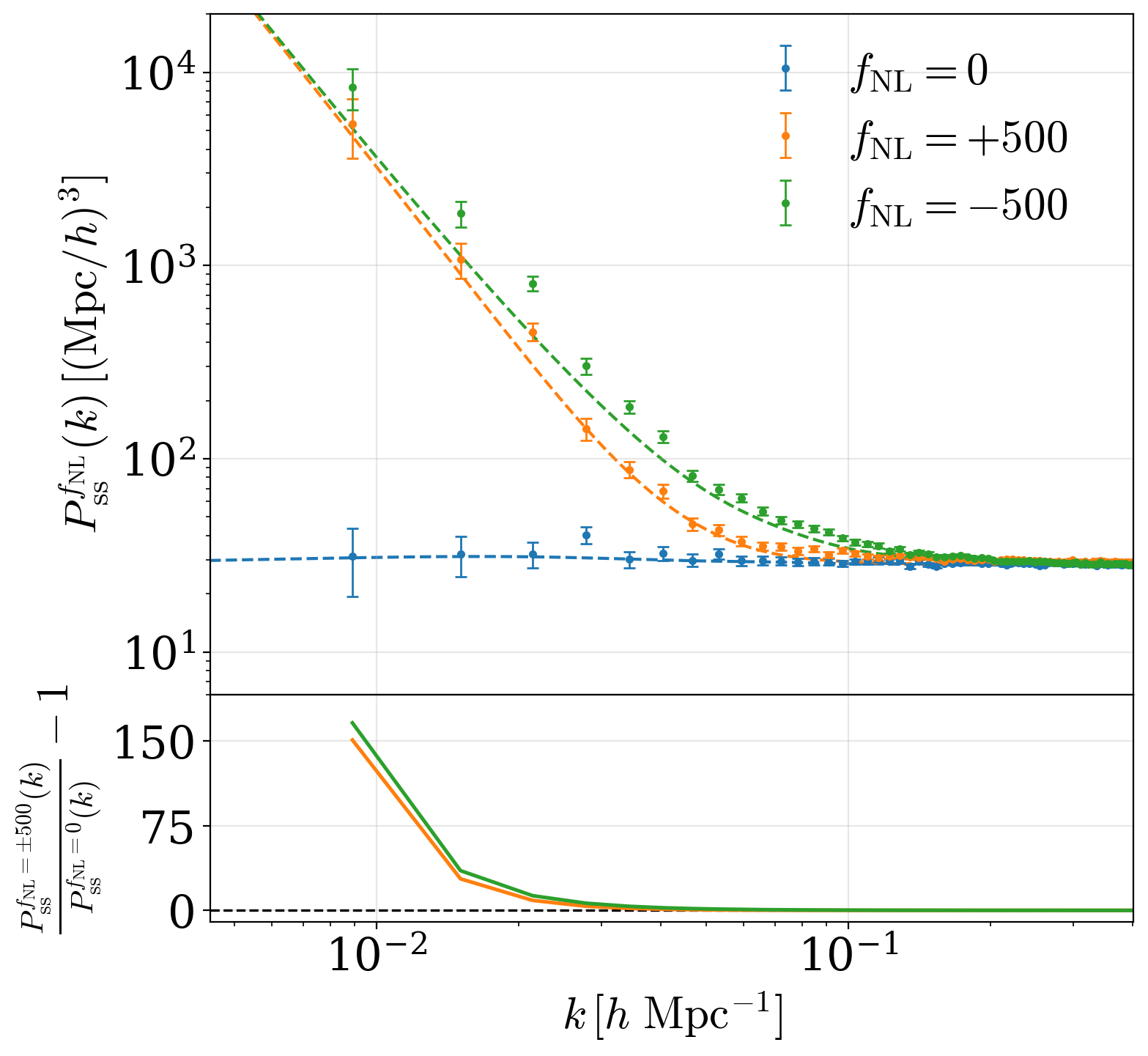}
  \caption{The auto power spectrum of galaxy numbers $P_{\rm gg}$ (upper panel) and sizes $P_{\rm ss}$ (lower panel) for $M_h\in[10^{11.5},\,10^{12.5})\,\Msunh$ at $z=0.9$. Error bars indicate the $1\sigma$ uncertainty estimated using Gaussian covariance. Dashed curves indicate the best-fit theoretical predictions.}
  \label{fig:Pk_auto_fnl}
\end{figure}

\begin{figure}[tb]
  \centering
  \includegraphics[width=0.8\linewidth]{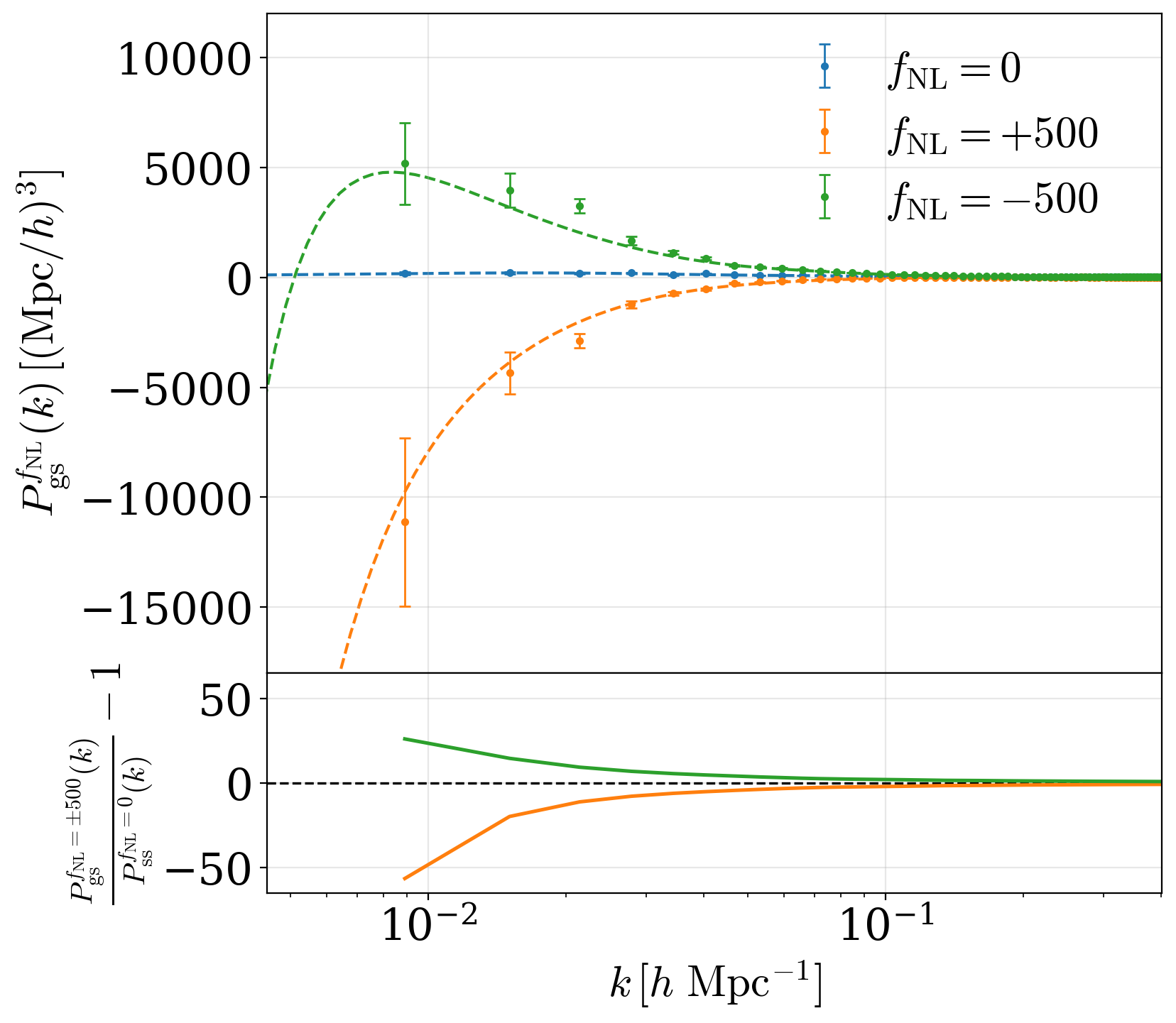}
  \caption{Similar to \cref{fig:Pk_auto_fnl} but for the cross power spectrum of galaxy numbers and sizes $P_{\rm gs}$.}
  \label{fig:Pk_cross_fnl}
\end{figure}

\cref{fig:Pk_auto_fnl,fig:Pk_cross_fnl} support the above intuition. 
In \cref{fig:Pk_auto_fnl}, we plot the galaxy number power spectrum $P_{\rm gg}$ and galaxy size power spectrum $P_{\rm ss}$, while in \cref{fig:Pk_cross_fnl}, we show the galaxy number--size power spectrum $P_{\rm gs}$, all measured in the galaxy rest frame of simulations with ($\fnl=\pm500$) and without ($\fnl=0$) local PNG for halos of mass $M_h\in[10^{11.5},\,10^{12.5})\,\Msunh$ at redshift $z=0.9$. The weak Gaussian response of galaxy sizes suppresses the background Gaussian contributions to the clustering signals in $P_{\rm ss}$ and $P_{\rm gs}$ on large scales, leaving clean local PNG imprints in both spectra, as can be seen from the signal-to-background ratio shown in the subpanels of \cref{fig:Pk_auto_fnl,fig:Pk_cross_fnl}. This novel observable naturally blends into the multi-tracer setup \citep{Seljak:2008xr}: the galaxy number and size information for the \emph{same} galaxy samples can be utilized for cosmic variance cancellation (see \cref{sec:forecast}). In addition, \cref{fig:Pk_cross_fnl} illustrates how the negative response of size to local PNG modulates $P_{\rm gs}$ in the opposite direction (relative to the local PNG modulation) of $P_{\rm gg}$ for the same $\fnl$ (see upper panel of \cref{fig:Pk_auto_fnl}), thereby suggesting that $P_{\rm gs}$ can serve as a cross validation for $\fnl$ detection in standard analyses using only $P_{\rm gg}$.

We additionally quantify the departure of both galaxy number and size responses from the universality relation in \cref{app:approx_universality}.

\subsection{Observational effects}
\label{sec:formalism:observations}

In this section, we consider the leading observational effects---namely, projection and redshift-space distortion along the observing direction---and how they modify the formalisms in \cref{sec:formalism:3Dsize_Gaussian,sec:formalism:3Dsize_PNG}. Notably, line-of-sight projection does not preserve a pure ``size'' observable: the projected trace mixes the spin-0 size fields ($\dtr$, $\ds$ in \cref{eq:dtr_def,eq:ds_def}) with the LOS component of the trace-free shape field ($\d^{\rm tf}_{zz}$ in \cref{eq:TrI^2D_ab}).
As a result, the projected size perturbations can inherit sensitivity not only to isotropic PNG but also to \emph{anisotropic} PNG sourced by higher-spin fields, which imprint characteristic angular dependence in the LOS multipoles.
Combining galaxy intrinsic sizes and shapes could therefore tighten constraints on anisotropic PNG \citep[e.g.,][]{Kurita:2023qku}.

\subsubsection{Projection}
\label{sec:formalism:observations:projection}

In observations, galaxy light profiles are measured only on the plane perpendicular to the line of sight (LOS). While galaxies possess intrinsic 3D sizes and shapes, the observed quantities are 2D projections of these 3D properties owing to this projection effect. At the same time, however, in spectroscopic surveys we can still map the spatial distribution of galaxies in 3D space; the density fields of these observables thus remain 3D fields, with coordinate vectors remaining 3D vectors.

We assume the distant observer approximation and fix the LOS to the direction along the $z$ axis, i.e. $\nhat=\zhat=(0,0,1)$. In this frame, the projector
$\P_{ij}(\vnhat)=\kdelta_{ij}-\nhat_i\nhat_j$ becomes globally diagonal, i.e. $\P_{ij}=\diag(1,1,0)$.

We write the trace $T^{\rm 2D}\equiv\tr I_{\rm ab}$, where $a,b\in[x,y]$, as
\begin{align*}
    T^{\rm 2D}(\vx)=\tr\P_{ij}I_{ij}(\vx)&=I_{xx}(\vx)+I_{yy}(\vx)\\
    &=\bar{T}\left[\frac{2}{3}\left(1+\dtr(\vx)\right)-\dtf_{zz}(\vx)\right]\numberthis\label{eq:TrI^2D_ab}.
\end{align*}
This implies that the projection effect mixes the information of galaxy size and shape perturbations along the LOS. Therefore, the projected galaxy sizes encode additional information about the LOS component of galaxy shapes, i.e. $\dtf_{zz}$ in our case.

Similar to \cref{eq:dtr_def,eq:ds_def}, we define the galaxy projected trace and size perturbations as
\begin{align}
\d^{\rm tr2D}(\vx)&=\frac{T^{\rm 2D}(\vx)-\overline{T^{\rm 2D}}}{\overline{T^{\rm 2D}}}=\dtr(\vx)-\frac{3}{2}\dtf_{zz}(\vx)\label{eq:dtr2D},\\
\d^{\rm s2D}(\vx)&=\frac{T^{\rm 2D}(\vx)}{\overline{T^{\rm 2D}}}-\frac{\ng(\vx)}{\ngbar}=\ds(\vx)-\frac{3}{2}\dtf_{zz}(\vx)\label{eq:ds2D},
\end{align}
where we have normalized both by
\begin{align*}
\overline{T^{\rm 2D}}&=(1/V)\int d^3xT^{\rm 2D}(\vx),\\
&=\ngbar\<t^{\rm 2D}_g\>\numberthis\label{eq:T2D_mean},
\end{align*}
which satisfies $\bar{T}/\overline{T^{\rm 2D}}=3/2$.
Because of the additional contribution from $\dtf_{zz}$ in \cref{eq:TrI^2D_ab}, we must further consider the perturbations in the galaxy projected shape field, which we define as
\begin{align}
\d^{\rm tf2D}_{\rm ab}(\vx)&=\frac{I_{\rm ab}(\vx)-\frac{1}{2}\kdelta_{\rm ab}T^{\rm 2D}(\vx)}{\overline{T^{\rm 2D}}}\label{eq:dtf2D_est},\\
&=\frac{3}{2}\left[\dtf_{\rm ab}(\vx)+\frac{1}{2}\kdelta_{\rm ab}\dtf_{zz}(\vx)\right]\label{eq:dtf2D_def}.
\end{align}

For simplicity, we first focus on the case of Gaussian initial conditions, i.e. $\fnl=0$.
At leading order in the bias expansion, \cref{eq:ds2D,eq:dtf2D_def} are related to the matter perturbations as
\begin{align}
\d^{\rm s2D}(\vk)&=\left[b^{\rm s}_1-L_{\ell=2}(\mu)b^{\rm tf}_1\right]\d_m(\vk)+\eps(\vk)\label{eq:gaussian_linear_bias_2Dprojected_spin0},\\
\d^{\rm tf2D}_{\rm ab}(\vk)&=\frac{3}{2}b^{\rm tf}_1\left[\khat_a\khat_b+\frac{1}{2}\kdelta_{\rm ab}\left(\mu^2-1\right)\right]\d_m(\vk)+\eps_{\rm ab}(\vk)\label{eq:gaussian_linear_bias_2Dprojected_spin2}
\end{align}
where we have defined $\mu=(\vk\cdot\vzhat)/k$ as the cosine angle between the Fourier unit wavevector and the LOS unit vector and $L_{\ell=2}$ as the Legendre polynomial of order two.
As the density fields and power spectra become LOS-dependent, we decompose the latter into multipoles:
\begin{equation}
    P(k,\mu)=\sum_{\ell=0}^{\ell_{\rm max}} P^\ell(k)L_\ell(\mu)\label{eq:multipole_decomp}
\end{equation}
where the multipoles are defined by the LOS integral
\begin{equation}
    P^\ell(k)=\frac{2\ell+1}{2}\int_{-1}^1d\mu P(k,\mu)L_\ell(\mu)\label{eq:multipole_def}.
\end{equation}

Below, we consider the monopole, quadrupole and hexadecapole ($\ell=0,2,4$) which, following \cref{eq:gaussian_linear_bias_2Dprojected_spin0,eq:gaussian_linear_bias_2Dprojected_spin2,eq:multipole_decomp,eq:multipole_def}, read
\begin{align}
    P^{\ell=0}_{\rm s2Ds2D}(k)&=\left[\left(b^{\rm s}_1\right)^2+\frac{1}{5}\left(b^{\rm tf}_1\right)^2\right]P_{\rm mm}(k) + P_{\eps^{\rm s2D}\eps^{\rm s2D}} \label{eq:Pss_ell0},\\
    P^{\ell=2}_{\rm s2Ds2D}(k)&=\left[-2b^{\rm s}_1b^{\rm tf}_1+\frac{2}{7}\left(b^{\rm tf}_1\right)^2\right]P_{\rm mm}(k)\label{eq:Pss_ell2},\\
    P^{\ell=4}_{\rm s2Ds2D}(k)&=\frac{18}{35}\left(b^{\rm tf}_1\right)^2P_{\rm mm}(k)\label{eq:Pss_ell4}
\end{align}
for the size--size power spectrum and
\begin{align}
    P^{\ell=0}_{\rm gs2D}(k)&=b^{\rm g}_1b^{\rm s}_1P_{\rm mm}(k)\label{eq:Pgs_ell0},\\
    P^{\ell=2}_{\rm gs2D}(k)&=-b^{\rm g}_1b^{\rm tf}_1P_{\rm mm}(k)\label{eq:Pgs_ell2},\\
    P^{\ell=4}_{\rm gs2D}(k)&=0\label{eq:Pgs_ell4}
\end{align}
for the number--size power spectrum.

For completeness, in \cref{app:shape_multipoles}, we also derive the number--shape $E$-mode cross power spectrum multipoles, i.e. $P^{\ell=0,2}_{\dg\gamma^{E}}$, and discuss their relation with the number--size quadrupole ($\ell=2$) in \cref{eq:Pgs_ell2}.

\subsubsection{Redshift-space distortion}
\label{sec:formalism:observations:RSD}
Galaxy surveys measure the radial distance to a given galaxy through the Doppler redshift, which includes a contribution from the peculiar motion of the galaxy. Therefore, the galaxy positions are distorted \cite{Kaiser:1987qv}. However, the galaxy sizes and shapes remain unchanged at the linear level. Linear redshift-space distortion (hereafter RSD) thus only modifies the galaxy number and size power spectra
\begin{align}
    P^S_{\rm gg}\left(k,\mu\right)&=\left[b^{\rm g}_1+f\mu^2\right]^2P_{\rm mm}(k)+P_{\eps^{\rm g}\eps^{\rm g}}\label{eq:PSgg},\\
    P^S_{\rm gs2D}\left(k,\mu\right)&=\left[b^{\rm g}_1+f\mu^2\right]\left[b^{\rm s}_1-b^{\rm tf}_1L_2(\mu)\right]P_{\rm mm}(k)\label{eq:PSgs}
\end{align}
where $f=d\ln D/d\ln a$ is the linear growth rate.

The galaxy number power spectrum multipoles are given by \cref{eq:multipole_def,eq:PSgg}
\begin{align}
    P^{S,\ell=0}_{\rm gg}(k)&=\left[\left(b^{\rm g}_1\right)^2+\frac{2}{3}b^{\rm g}_1f+\frac{1}{5}f^2\right]P_{\rm mm}(k) + P_{\eps^{\rm g}\eps^{\rm g}}\label{eq:PSgg_ell0},\\
    P^{S,\ell=2}_{\rm gg}(k)&=\left[\frac{4}{3}b^{\rm g}_1f+\frac{4}{7}f^2\right]P_{\rm mm}(k)\label{eq:PSgg_ell2},\\
    P^{S,\ell=4}_{\rm gg}(k)&=\frac{8}{35}f^2P_{\rm mm}(k)\label{eq:PSgg_ell4}
\end{align}
and the galaxy number--size power spectrum multipoles are given by \cref{eq:multipole_def,eq:PSgs}
\begin{align}
    P^{S,\ell=0}_{\rm gs2D}(k)&=\left[b^{\rm g}_1b^{\rm s}_1+\left(\frac{1}{3}b^{\rm s}_1-\frac{2}{15}b^{\rm tf}_1\right)f\right]P_{\rm mm}(k)\label{eq:PSgs_ell0},\\
    P^{S,\ell=2}_{\rm gs2D}(k)&=\left[-b^{\rm g}_1b^{\rm tf}_1+\left(\frac{2}{3}b^{\rm s}_1-\frac{11}{21}b^{\rm tf}_1\right)f\right]P_{\rm mm}(k)\label{eq:PSgs_ell2},\\
    P^{S,\ell=4}_{\rm gs2D}(k)&=-\frac{12}{35}b^{\rm tf}_1fP_{\rm mm}(k)\label{eq:PSgs_ell4}
\end{align}
respectively.

In the presence of local PNG, \cref{eq:PSgg_ell0,eq:PSgg_ell2,eq:PSgs_ell0,eq:PSgs_ell2} are simply modified such that $b^{\rm g,s}_1\to b^{\rm g,s}_1+b^{\rm g,s}_\phi\fnl/\M(k)$.
We further observe that the power spectrum hexadecapoles of size--size \cref{eq:Pss_ell4} and number--size \cref{eq:PSgs_ell4} are independent of $b^{\rm g,s}_1$, which explains their insensitivity to $\fnl$ in \cref{fig:PS_ell_PNG}.

\section{Simulations}
\label{sec:sims}

We run five triplets of \code{Gadget-2} simulations \citep{Springel:2005mi} in a volume of $V=1\,(h^{-1}\mathrm{Gpc})^3$ with $N_{\rm part}=4\times1290^3$ particles for $\fnl=[-500,0,+500]$, where each simulation triplet share the same initial phases to suppress sample variance.
The box size, mass resolution, and \(\fnl\) values are chosen to balance two requirements: (i) enough member particles even for low-mass halos to suppress random noise in the halo inertia-tensor estimators, and (ii) sufficient sensitivity to measure the local PNG modulation of the power spectra despite cosmic variance.

We assume the fiducial cosmology to follow the best-fit cosmology constrained by the combination of Planck CMB data (TTTEEE, low-$\ell$ EE and lensing) plus BAO and Pantheon supernova data sets\footnote{See \texttt{base\_plikHM\_TTTEEE\_lowl\_lowE\_lensing\_post\_BAO\_Pantheon} (Table.~2.20) in the Planck 2018 Results: Cosmological Parameter Tables at \url{https://wiki.cosmos.esa.int/planck-legacy-archive/images/b/be/Baseline_params_table_2018_68pc.pdf}} \citep{Planck:2018vyg}.

We generate the initial conditions for the simulations using \code{monofonIC}\footnote{\url{https://bitbucket.org/ohahn/monofonic/src/master/}} \citep{2020ascl.soft08024H}, which employs third-order Lagrangian perturbation theory (3LPT) to evolve the initial particle distribution to the starting redshift $z_{\rm start}=24.0$. This choice balances particle discreteness and displacement-truncation errors. We note that the implementation of PNG in the initial conditions also introduces aliasing, for which \code{monofonIC} applies a dealiasing correction following the Orszag 2/3 rule. We refer readers to Ref.~\cite{Adame:2025zhg} for further details which showed that the aliasing affects the late-time power spectrum only at the sub-percent level, and is thus negligible even without the dealiasing step.

We identify halos and estimate halo inertia tensors using \code{MPI-Rockstar}\footnote{\url{https://github.com/Tomoaki-Ishiyama/mpi-rockstar/}} \citep{Tokuue:2024ney}, a new, hybrid \code{MPI} and \code{openMP} implementation of the original phase-space friends-of-friend halo finder algorithm \code{Rockstar} \citep{Behroozi:2011ju}.
Following Ref.~\cite{Kurita:2020hap}, we adopt radial weights of the form $w_p=1/r^2$ in \cref{eq:Iij_g} and leave a systematic investigation of how this choice affects the signal-to-noise of the size perturbations for future work. Since the measured size bias coefficients depend on this choice, we also present results for the alternative weighting $w_p=1$ in \cref{app:trace_definition}.

We construct halo samples at four redshift snapshots, $z=[0.0,\,0.7,\,0.9,\,1.3]$, by dividing the main halos into three logarithmic mass bins,
$M_h\in[10^{11.5},10^{12.5}),\,[10^{12.5},10^{14.0}),\,[10^{14.0},10^{16.0})\,\Msunh$.
The first two bins span the halo mass range typically associated with emission-line galaxies (ELGs), bright galaxy sample (BGS), and luminous red galaxies (LRGs), while the last bin covers that of galaxy clusters.

\section{Measurements}
\label{sec:measurements}

We consider dark matter halos in gravity-only N-body simulations as proxies for galaxies and the halo 3D inertia tensor as a proxy for the galaxy 3D light distribution. In observations, there are complications that need to be further investigated: (1) how different measurement methods and proxies track galaxy sizes, (2) how different galaxy populations populate dark matter halos and (3) how gas physics and galaxy assembly history affect galaxy sizes. We leave these investigations for future study and make no distinction between halos and galaxies in the measurements in this section and the forecasts in \cref{sec:forecast}. We briefly discuss the complications and their implications in \cref{sec:discussion:proxies}.

\subsection{Three-dimensional galaxy size in the galaxy rest frame}
\label{sec:measurements:3Dsize}

In simulations, we have access to the halo comoving coordinates without RSD as well as their 3D sizes (and shapes) without projection. Therefore, we directly fit the theoretical predictions in \cref{sec:formalism:3Dsize_Gaussian,sec:formalism:3Dsize_PNG} to the corresponding power spectra measured in the galaxy rest frame. We estimate power spectra using \code{Pylians} \cite{Pylians} up to $\knyq=\pi N_{\rm grid}/L\simeq0.8\,\hmpcinv$ while computing the linear matter power spectrum and transfer function using \code{CAMB} \cite{Lewis:camb,Howlett:camb}.

\subsubsection{Gaussian bias}
\label{sec:measurements:3Dsize:Gaussian}

To measure the Gaussian linear bias coefficients $b^{\rm g,s}_1$, for each sample, we fit the measured galaxy number--matter and size--matter cross power spectra $P_{\rm gm,sm}(k)$ to the linear predictions:
\begin{equation}
    P_{\rm gm,sm}(k)=b^{\rm g,s}_1P_{\rm mm}(k)\label{eq:tree_Pgmsm}.
\end{equation}
Specifically, we define $q^{\rm g,s}(k)=P_{\rm gm,sm}(k)/P_{\rm mm}(k)$ and fit for $b^{\rm g,s}_1$ by minimizing
\begin{equation}
    \chi^2=\sum_{k_i<\kmax}\left[\frac{q^{\rm g,s}\left(k_i\right)-b^{\rm g,s}_1}{\sigma^2\left[q^{\rm g,s}(k_i)\right]}\right]\label{eq:chi2_fit}
\end{equation}
where $k_i$ is the central value of the wavenumbers $k$ in the bin $i$. We approximate $\sigma^2\left[q^{\rm g,s}(k_i)\right]$ by the variance of $q^{\rm g,s}(k_i)$ across five simulation realizations. We adopt $\kmax=0.06\hmpcinv$ to limit higher-order contributions; we have verified that varying $\kmax\in[0.04-0.08]\hmpcinv$ yields only sub-percent (and a few percent) variations in the best-fit bias coefficients for numbers (and sizes).

For a robust estimate of systematics due to higher-order contributions, we further include a higher-derivative term\footnote{See e.g. Ref.~\cite{Lazeyras:2015lgp} who included this term in their fiducial fits. We note that whether or not to include this term makes practically no difference for $b^{\rm g}_1$ but can shift $b^{\rm s}_1$ significantly in the high-mass bins.}
\begin{equation}
    P_{\rm gm,sm}(k)=\left[b^{\rm g,s}_1+b^{\rm g,s}_{\lapl\d}k^2\right]P_{\rm mm}(k)\label{eq:tree+higher-deriv_Pgmsm}
\end{equation}
where the coefficient $b^{\rm g,s}_{\lapl\d}$ associated with this counterterm is an additional free parameter. We report the shifts in the best-fit values (with and without $b^{\rm g,s}_{\lapl\d}$), i.e. $\D\hat{b}^{\rm g,s}_1=\hat{b}^{\rm g,s}_1\Bigr|_{b_{\lapl\d}\,\rm free}-\hat{b}^{\rm g,s}_1\Bigr|_{b_{\lapl\d}\,\rm=0}$ as systematics uncertainties and include those into the total uncertainties $\sigma_{\rm tot}=\sqrt{\sigma_{\rm stat}^2+\sigma_{\rm sys}^2}$ with the statistical uncertainties $\sigma_{\rm stat}$ estimated by the scatters between realizations.

\begin{figure}[tb]
  \centering
  \includegraphics[width=0.8\linewidth]{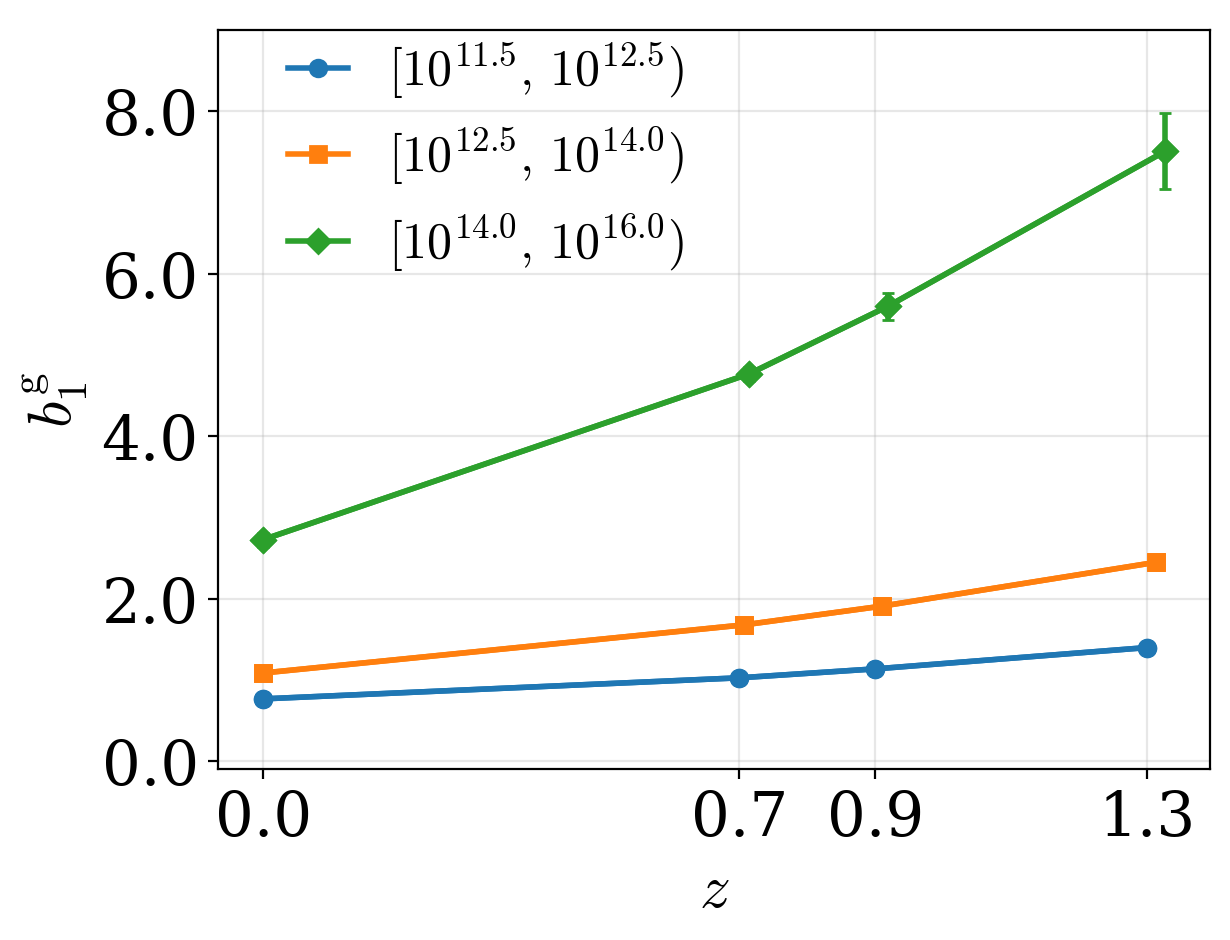}
    \vspace{0.15em}
  \includegraphics[width=0.8\linewidth]{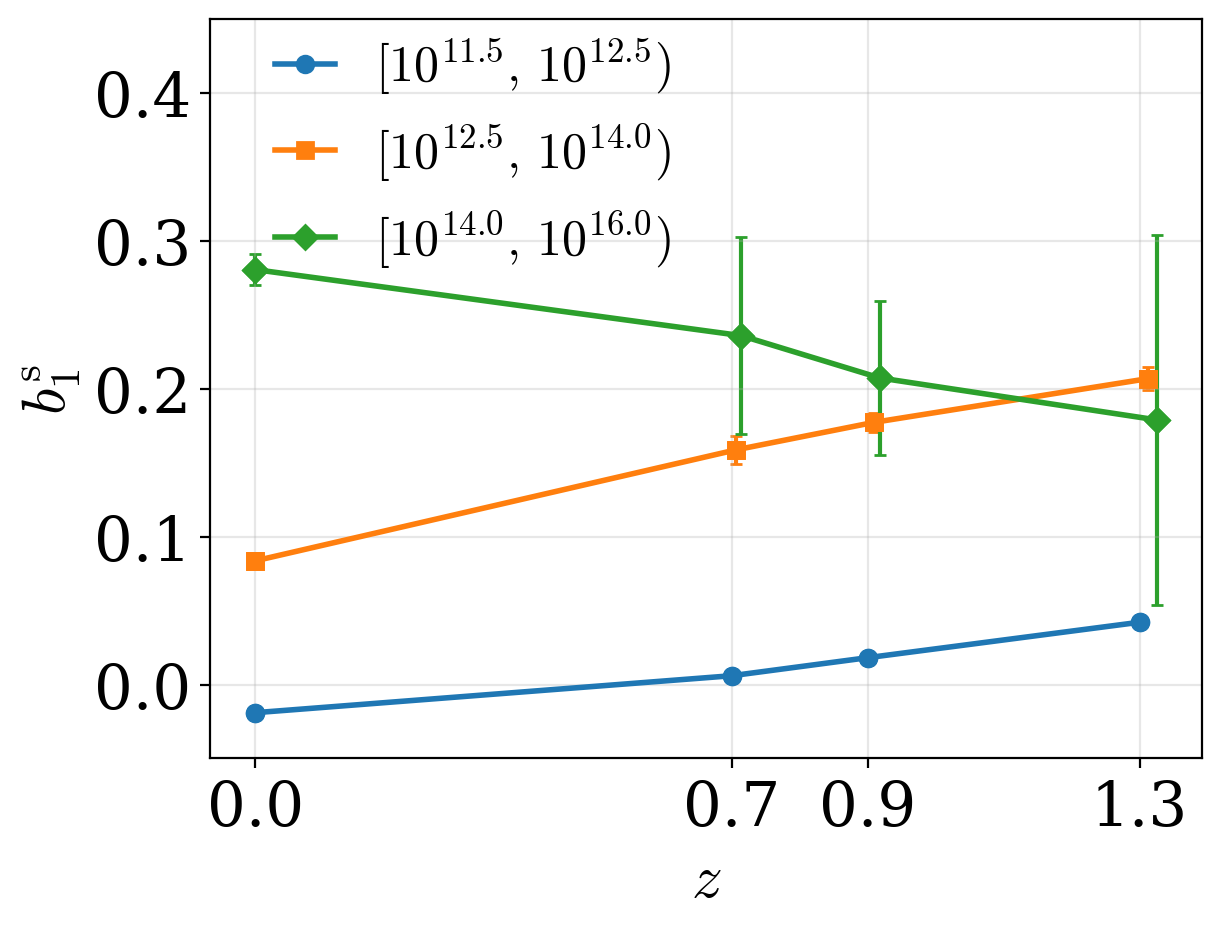}
  \caption{The Gaussian linear bias coefficients of galaxy number $b^{\rm g}_1$ (upper panel) and those of galaxy size $b^{\rm s}_1$ (lower panel) as functions of redshift for $M_h\in[10^{11.5},\,10^{12.5}),[10^{12.5},\,10^{14.0}),[10^{14.0},\,10^{16.0})\,\Msunh$. Error bars include statistical and systematic uncertainties (see text).
  }
  \label{fig:b1gs_z}
\end{figure}

In \cref{fig:b1gs_z}, we plot the best-fit values of the Gaussian linear bias coefficients for galaxy numbers and sizes as measured in the $\fnl=0$ simulations.
In particular, the measured Gaussian linear bias of galaxy size shown in the lower panel of \cref{fig:b1gs_z} confirms the argument regarding the amplitude of $b^{\rm s}_1$ in \cref{sec:formalism:zerobias_SU}: galaxy size responds weakly to the presence of long-wavelength perturbations for all mass bins considered.
 Our findings are further qualitatively consistent with those of Ref.~\cite{Johnston:2022nbv,Ghosh:2024mlz} who reported a positive correlation between galaxy sizes and (over) dense regions. In addition, we observe a reverse trend in the redshift evolution of the cluster-mass halo size bias.

\subsubsection{Non-Gaussian bias}
\label{sec:measurements:3Dsize:PNG}

For simulations with nonzero $\fnl$, we follow a similar approach to \cref{sec:measurements:3Dsize:Gaussian} while including the contributions from the non-Gaussianities of the Bardeen potential $\phi_{\rm G}$, as described in \cref{eq:PNG_linear_Pk_auto_scale-dependent_bias}.
Specifically, we fit the measured cross power spectra $P_{\rm gm}(k)$ and $P_{\rm sm}(k)$ to
\begin{align}
    P_{\rm gm}(k)&=\left[b^{\rm g}_1+\frac{b^{\rm g}_\phi\fnl}{\M(k)}\right]P_{\rm mm}(k)\label{eq:PNG_linear_Pk_gm},\\
    P_{\rm sm}(k)&=\left[b^{\rm s}_1+\frac{b^{\rm s}\phi\fnl}{\M(k)}\right]P_{\rm mm}(k)\label{eq:PNG_linear_Pk_tm}.
\end{align}
As before, systematics uncertainties are estimated by the shifts in best-fit values of $b^{\rm g,s}_1$ when adding $b^{\rm g,s}_{\lapl\d}k^2$ inside the square brackets of \cref{eq:PNG_linear_Pk_gm,eq:PNG_linear_Pk_tm}.

\begin{figure}[tb]
  \centering
  \includegraphics[width=0.8\linewidth]{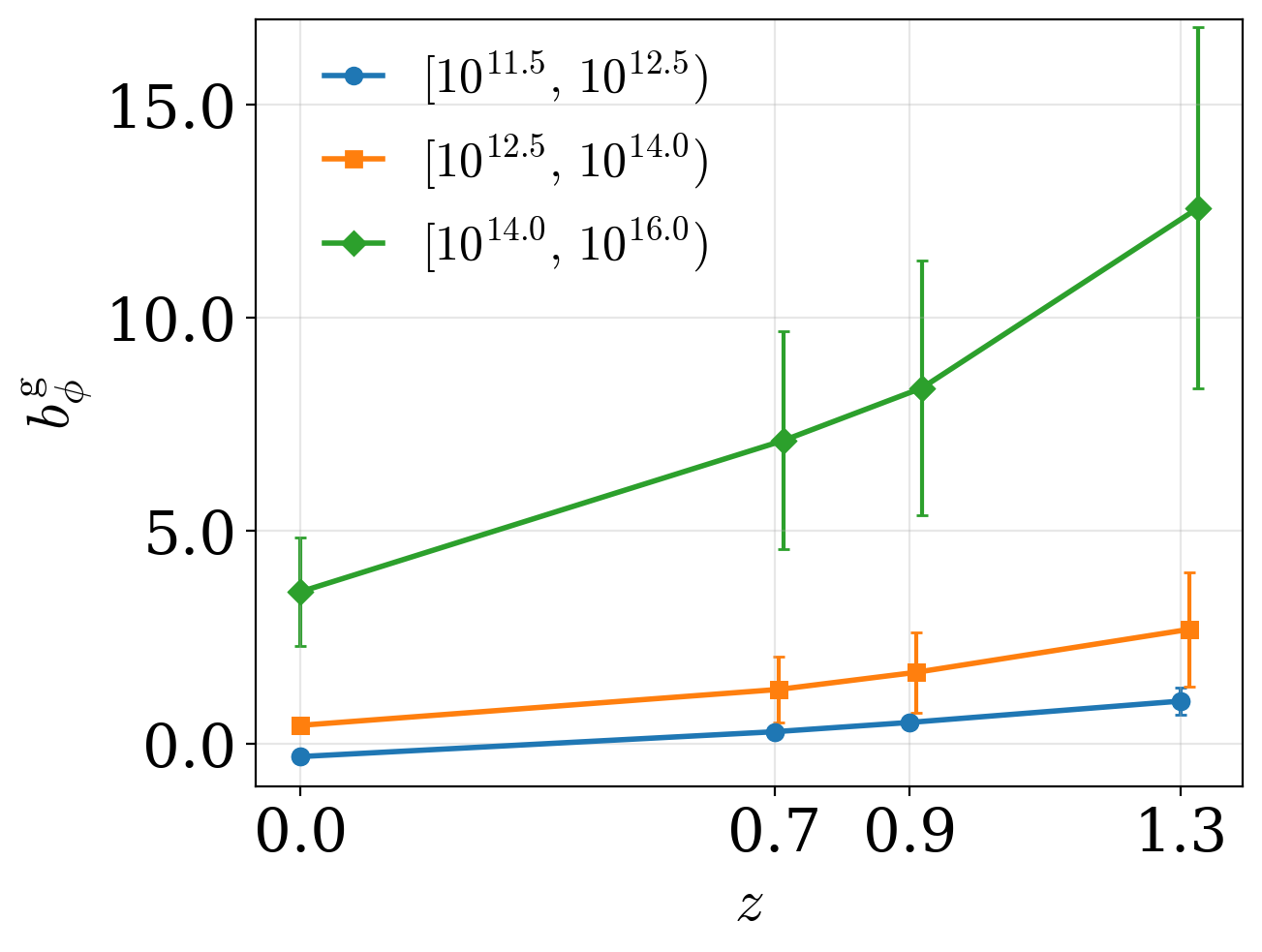}
    \vspace{0.15em}
  \includegraphics[width=0.8\linewidth]{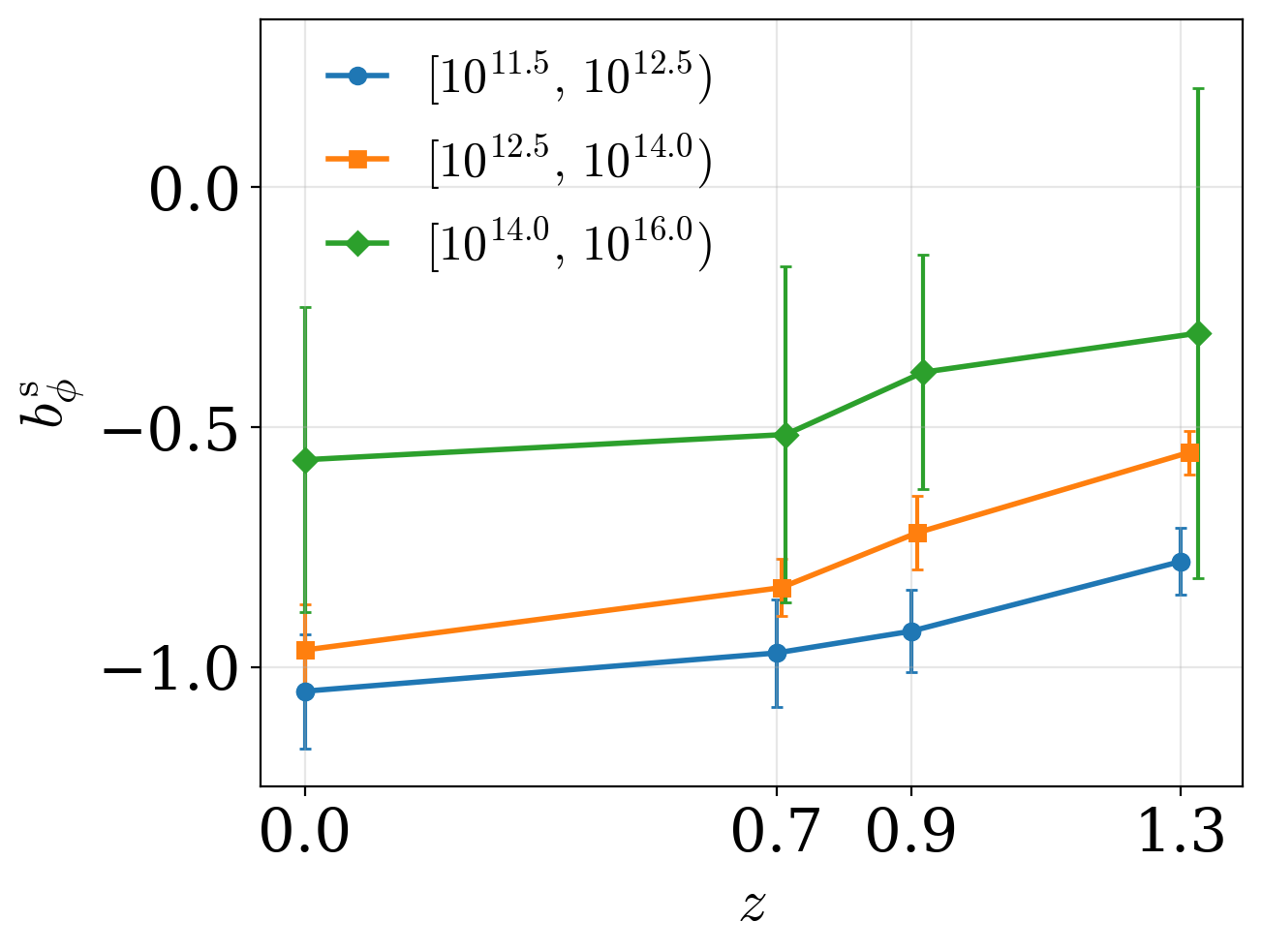}
  \caption{Similar to \cref{fig:b1gs_z} for non-Gaussian linear bias coefficients $b^{\rm g,s}_\phi$.
  }
  \label{fig:bphigs_z}
\end{figure}

We plot the best-fit non-Gaussian bias coefficients $b^{\rm g,s}_\phi$ as measured from the simulation triplets $\fnl=[-500,0,+500]$) in \cref{fig:bphigs_z}.
The size non-Gaussian bias is negative, as expected from the argument regarding the sign of $b^{\rm s}_\phi$ in \cref{sec:formalism:zerobias_SU}. Its amplitude, while remaining small for the high-mass bin, is relatively large for the intermediate- and low-mass bins which correspond to the LRG and ELG host halos, respectively. 
Our findings are further consistent with those of Ref.~\cite{Shiveshwarkar:2025nac} who investigated the universality of halos selected by concentration.

We summarize the best-fit values of the Gaussian and non-Gaussian bias coefficients $b^{\rm g,s}_1$ and $b^{\rm g,s}_\phi$ in \cref{tab:bias_values}. In \cref{app:approx_Gaussian_bias_fits}, we further provide the best-fit relations for $b^{\rm s}_1(b^{\rm g}_1)$ and $b^{\rm tf}_1(b^{\rm g}_1)$.
\begin{table}[t]
\centering
\caption{Gaussian and non-Gaussian linear bias coefficients $b^{\rm s}_{1,\phi}$ of halo size in different mass bins at different redshifts.}
\label{tab:bias_values}
\small
\setlength{\tabcolsep}{1pt}
\begin{tabular}{lcc|cc|cc|cc}
\toprule
Mass [$\Msunh$] & \multicolumn{2}{c}{$z=0.0$} & \multicolumn{2}{c}{$z=0.7$} & \multicolumn{2}{c}{$z=0.9$} & \multicolumn{2}{c}{$z=1.3$} \\
\cmidrule(lr){2-3}\cmidrule(lr){4-5}\cmidrule(lr){6-7}\cmidrule(lr){8-9}
$[M_{\rm min},M_{\rm max})$ & $b^{\rm s}_1$ & $b^{\rm s}_\phi$ & $b^{\rm s}_1$ & $b^{\rm s}_\phi$ & $b^{\rm s}_1$ & $b^{\rm s}_\phi$ & $b^{\rm s}_1$ & $b^{\rm s}_\phi$ \\
\midrule
$[10^{11.5}\!,\!10^{12.5})$ & $-0.02$ & $-1.05$ & $0.01$ & $-0.97$ & $0.02$ & $-0.93$ & $0.04$ & $-0.78$ \\
$[10^{12.5}\!,\!10^{14.0})$ & $0.08$ & $-0.96$ & $0.16$ & $-0.83$ & $0.18$ & $-0.72$ & $0.21$ & $-0.55$ \\
$[10^{14.0}\!,\!10^{16.0})$ & $0.28$ & $-0.57$ & $0.24$ & $-0.52$ & $0.21$ & $-0.39$ & $0.18$ & $-0.30$ \\
\bottomrule
\end{tabular}
\end{table}

\subsection{Two-dimensional galaxy size in redshift space}
\label{sec:measurements:2Dsize:PSell}

In this section, we compare the linear predictions for the size--size ($P^{S,\ell}_{\rm s2Ds2D}$, \cref{eq:Pss_ell0,eq:Pss_ell2,eq:Pss_ell4}) and number--size ($P^{S,\ell}_{\rm gs2D}$, \cref{eq:Pgs_ell0,eq:Pgs_ell2,eq:Pgs_ell4}) power spectrum multipoles to simulations with and without local PNG.
An additional ingredient in the models is the trace-free Gaussian linear bias coefficient $b^{\rm tf}_1$ for which we fit the three-dimensional power spectrum of shape-matter following the procedure described in \cref{app:shape-tidal_bias_measurement}.

\subsubsection{Gaussian initial conditions}
\label{sec:measurements:2Dsize:PSell:Gaussian}

\begin{figure}[tb]
  \centering
  \includegraphics[width=0.8\linewidth]{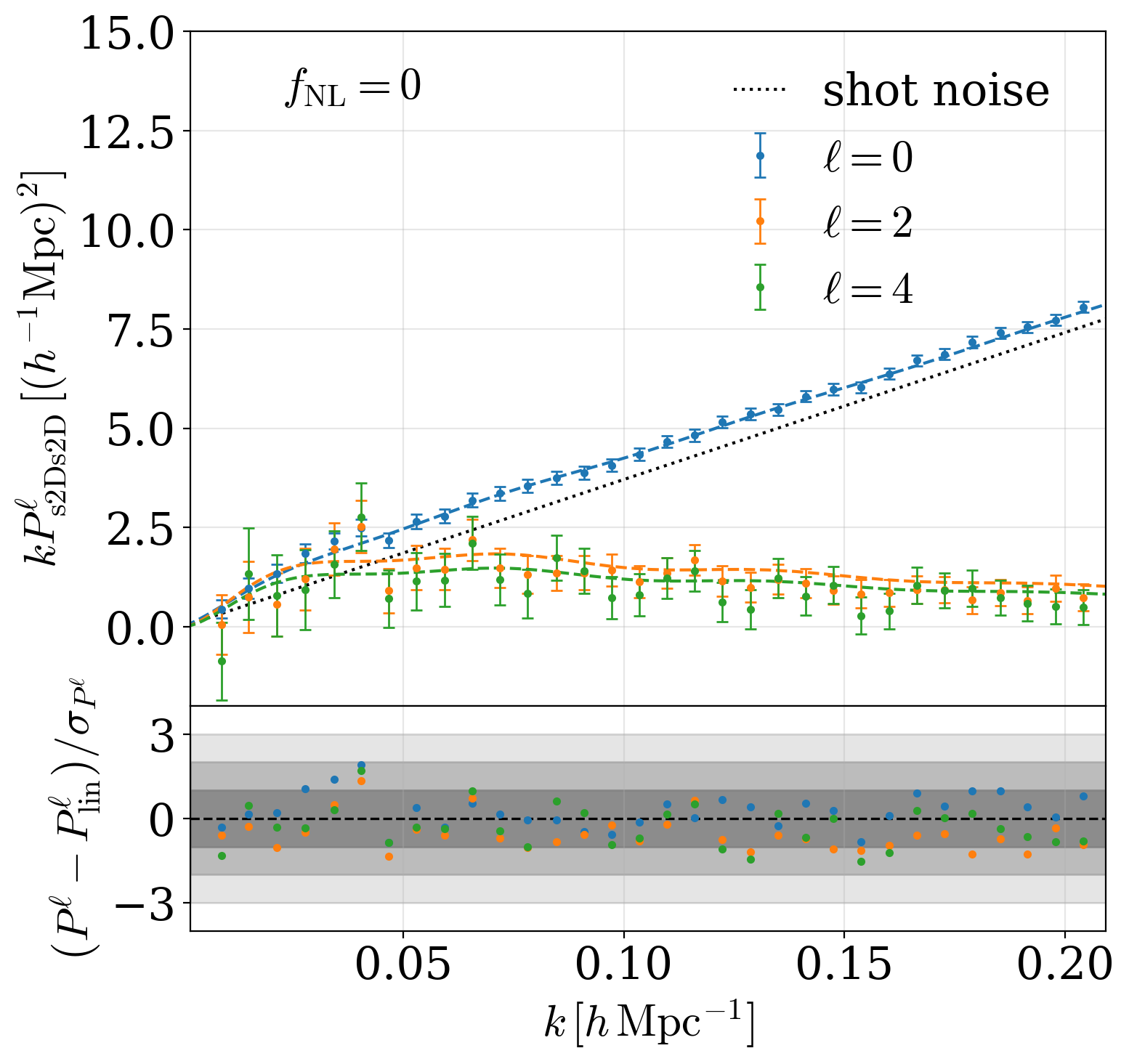}
    \vspace{0.15em}
  \includegraphics[width=0.8\linewidth]{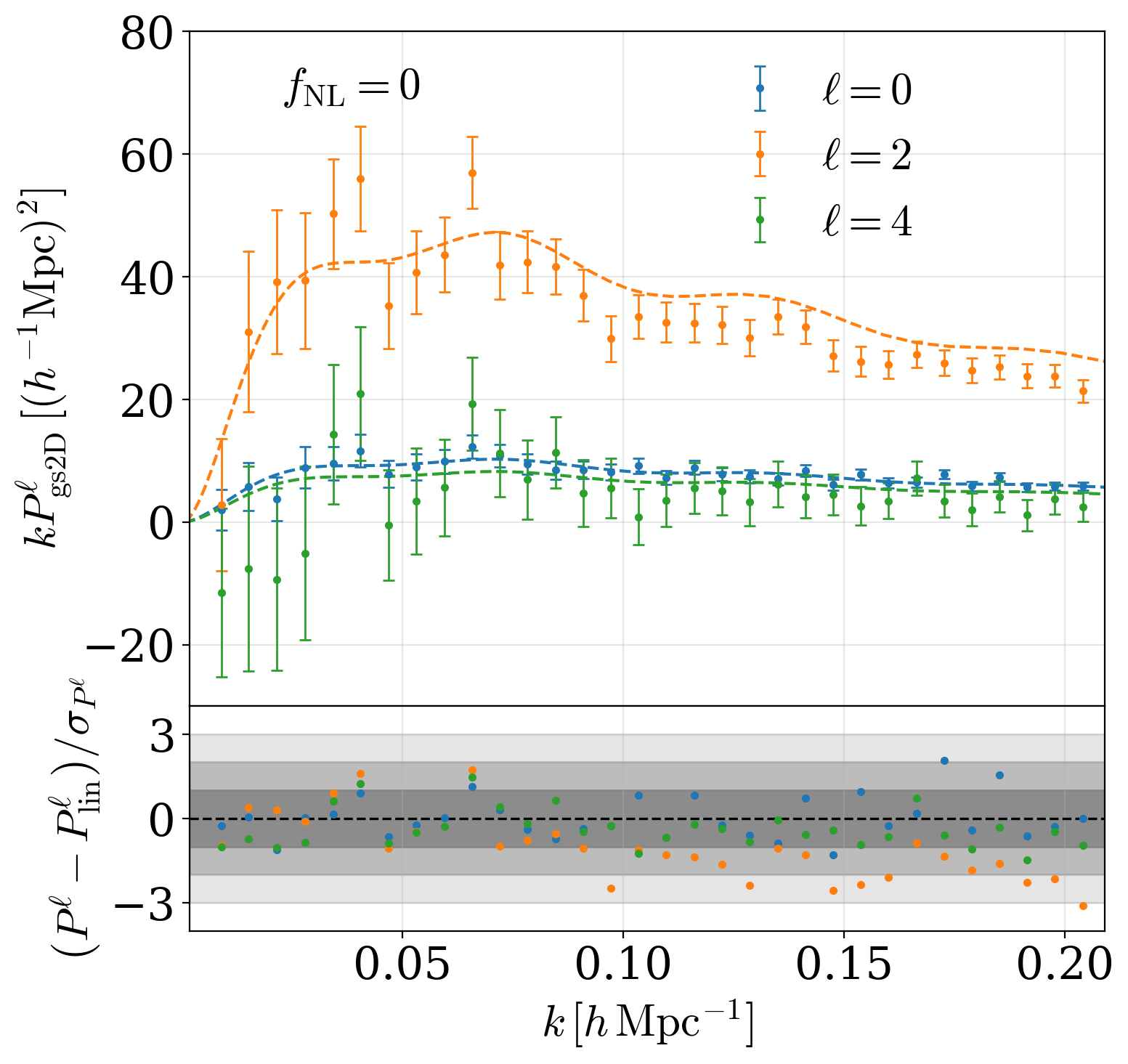}
  \caption{Size clustering in redshift space with Gaussian initial conditions for $M_h\in[10^{11.5},\,10^{12.5})\,\Msunh$ at $z=0.9$. Upper panel: $P^{S,\ell}_{\rm s2Ds2D}$ for $\ell=0,2,4$. Dashed curves indicate the best theoretical fits (including the Poisson stochastic contribution to the monopole). Points indicate measurements from a single simulation realization with $1\sigma$ Gaussian error bars (see \cref{app:covariance}). The dotted black line indicates the Poisson shot noise. The subpanel shows the data-theory misfits relative to the Gaussian $1\sigma$. Bands indicate the $1\sigma$-$2\sigma$-$3\sigma$ ranges. Lower panel:  $P^{S,\ell}_{\rm gs2D}$ in a similar notation.
  }
  \label{fig:PS_ell_Gaussian}
\end{figure}

In \cref{fig:PS_ell_Gaussian}, we plot the theoretical predictions against the corresponding measurements in a $\fnl=0$ realization for $M_h\in[10^{11.5},10^{12.5})[h^{-1}M_\odot]$ at $z=0.9$.
For the size power spectrum multipoles $P^{S,\ell}_{\rm s2Ds2D}$, the monopole (\cref{eq:Pss_ell0}) is dominated by the stochastic contribution $P_{\eps^{\rm s}\eps^{\rm s}}=(\sigma_{\rm s2D}/\<s2D\>)P_{\eps^{\rm g}\eps^{\rm g}}$.
For the number--size power spectrum multipoles $P^{S,\ell}_{\rm gs2D}$, there is no stochastic contribution, and we observe better the hierarchy and the structures of the multipoles. At $z=0.9$, all multipoles can be fitted reasonably well by linear theory up to the comoving wavenumber $k\simeq0.2\,\hmpcinv$; for reference, $\knl(z=0.9)=0.64\,\hmpcinv$ for the fiducial cosmology.

\subsubsection{Local PNG initial conditions}
\label{sec:measurements:2Dsize:PSell:PNG}

\begin{figure}[tb]
  \centering
  \includegraphics[width=0.8\linewidth]{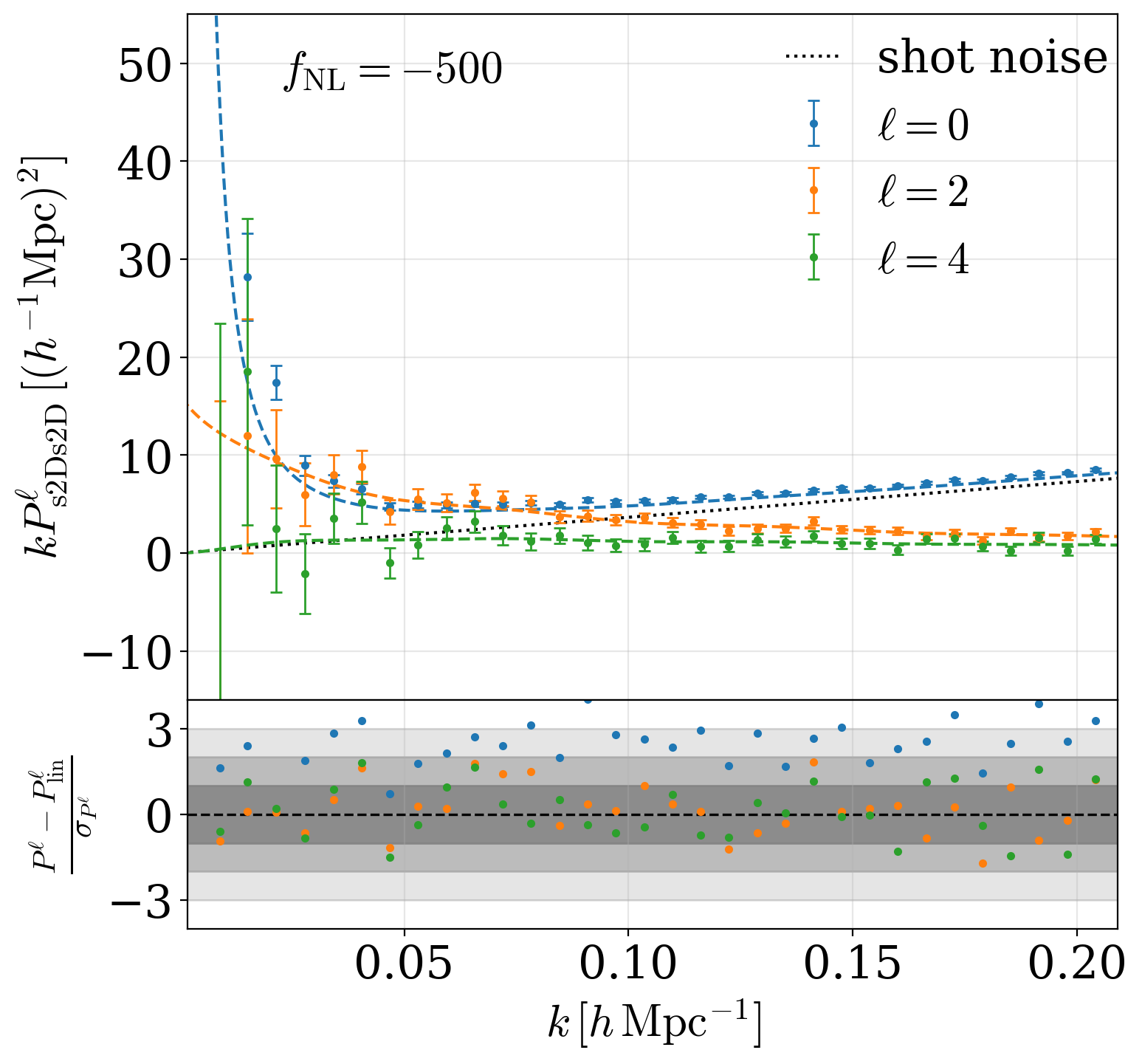}
    \vspace{0.15em}
  \includegraphics[width=0.8\linewidth]{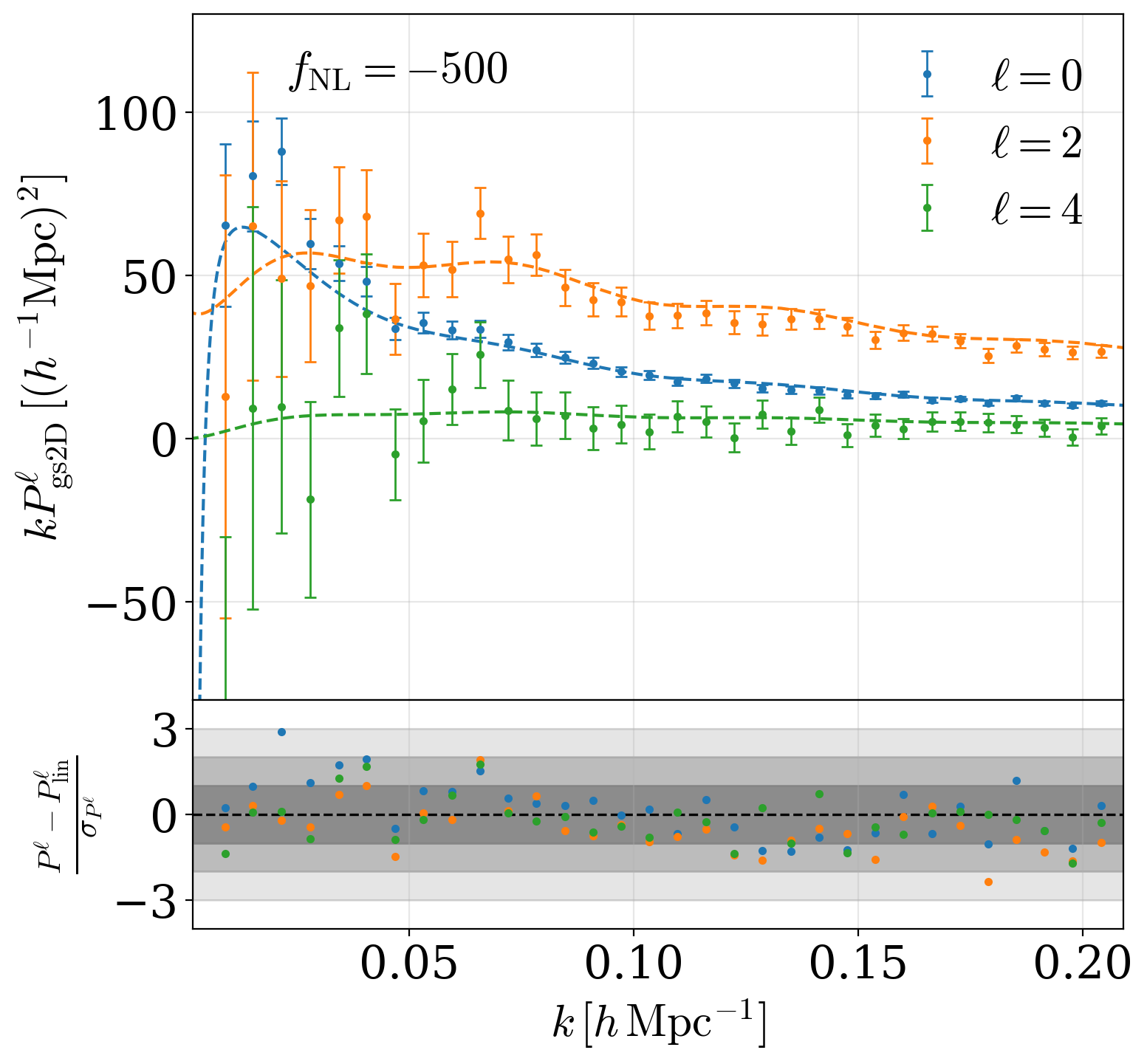}
  \caption{Similar to \cref{fig:PS_ell_Gaussian} with non-Gaussian initial conditions.
  }
  \label{fig:PS_ell_PNG}
\end{figure}

\cref{fig:PS_ell_PNG} shows a comparison similar to that in \cref{fig:PS_ell_Gaussian}, for the same mass bin at the same redshift, but in a $\fnl=-500$ realization. The low-$k$ modulations of local PNG are most visible in the monopoles (\cref{eq:Pss_ell0,eq:Pgs_ell0}), though also noticeable in the quadrupoles (\cref{eq:Pss_ell2,eq:Pgs_ell2}); they do not appear in the hexadecapole, as expected from \cref{eq:Pss_ell4,eq:Pgs_ell4}.

\section{Forecasts}
\label{sec:forecast}
Combining the galaxy numbers and sizes from the \emph{same} galaxy sample opens the door for a novel multi-tracer setup where there is no sample split involved.
In this section, we analyze and compare the detection significance (hereafter DS) of local PNG between the \emph{single-tracer (ST)} case---using only galaxy numbers, i.e. $(A,)=(\dg,)$---and the \emph{multi-tracer (MT)} case---using both galaxy numbers and sizes as tracers, i.e. $(A,B)=(\dg,\ds)$. Due to the perfect degeneracy between $b_\phi$ and $\fnl$, we set up our Fisher forecasts for constraints on the parameter combination $b^{\rm g}_\phi\fnl$ in the single-tracer case and on the combinations $\bm{\theta}_{\rm NL}^{\!\top}=(b^{\rm g}_\phi\fnl,b^{\rm s}_\phi\fnl)$ in the multi-tracer case.

A helpful way to visualize multi-tracer configurations that would maximally improve the DS is to note that, in the large-scale and cosmic-variance-limited limit, the constraining power is controlled by how \emph{linearly independent} the two tracers are in their responses to $\d_L$ and $\phi$.
Specifically, for two tracers $A$ and $B$, Ref.~\cite{Barreira:2023rxn} showed that the Fisher information on $b_\phi\fnl$ scales with $\bigl|\,b^{A}_{1}\,b^{B}_{\phi}-b^{B}_{1}\,b^{A}_{\phi}\,\bigr|$ (instead of $\bigl|b^{A}_{1}-b^{B}_{1}\bigr|$, which would follow if the two tracers follow the universality relation).
For our multi-tracer setup, $(A,B)=(\dg,\ds)$ and the measured combination of a suppressed Gaussian size bias $b^{\rm s}_1$ together with a sizable---and negative---$b^{\rm s}_\phi$ naturally yields a large $\bigl|b^{\rm g}_{1}b^{\rm s}_{\phi}-b^{\rm s}_{1}b^{\rm g}_{\phi}\bigr|$ (see also \cref{app:approx_universality}). We thus expect that the number--size multi-tracer case to significantly outperform the single-tracer case.

Following Ref.~\cite{Barreira:2022sey,Barreira:2023rxn}, we evaluate the DS as
\begin{equation}
    \mathrm{DS^{ST}}=\langle b^{\rm g}_\phi\fnl\rangle/\sigma_{b^{\rm g}_\phi\fnl}\label{eq:DS_st}
\end{equation}
for the single-tracer case and where $\<\>$ and $\sigma$ denote the fiducial mean value and the $1\sigma$ uncertainty on $b^{\rm g}_\phi\fnl$
\begin{equation}
\mathrm{DS^{MT}}=
\sqrt{
\left(\begin{smallmatrix}
\langle b^{\rm g}_\phi \fnl\rangle\\
\langle b^{\rm s}_\phi \fnl\rangle
\end{smallmatrix}\right)^{\!\top}
\cdot\F_{\theta_{\rm NL}}
\cdot\left(\begin{smallmatrix}
\langle b^{\rm g}_\phi \fnl\rangle\\
\langle b^{\rm s}_\phi \fnl\rangle
\end{smallmatrix}\right)
}
\label{eq:DS_mt}
\end{equation}
for the multi-tracer case where $\F$ is the Fisher matrix which we define as
\begin{equation}
\F_{\bm{\theta}_{\rm NL}}=\sum_{i,j}\frac{\partial \mathbf{D}(k_i)}{\partial\bm{\theta}_{\rm NL}^{\!\top}}\cdot\Cov^{-1}[\mathbf{D}(k_i),\mathbf{D}(k_j)]\cdot\frac{\partial \mathbf{D}^{\!\top}(k_j)}{\partial\bm{\theta}_{\rm NL}}\label{eq:Fisher_matrix},
\end{equation}
for the multi-tracer data vector
\begin{equation}
\mathbf{D}^{\!\top}(k)
=\Bigl[
\widehat P^{(\ell)}_{\rm AA}(k),\,
\widehat P^{(\ell)}_{\rm AB}(k),\,
\widehat P^{(\ell)}_{\rm BB}(k)
\Bigr]_{\ell\in\{0,2,4\}}
\label{eq:multi-tracer_data_vec}
\end{equation}
whose covariance matrix is given by \cref{eq:cov_full_block}.

We restrict the model of galaxy bias and RSD to the tree-level and linear regimes. The observables we consider are the monopoles, quadrupoles and hexadecapoles of the galaxy number, size and number--size power spectra. We derive an analytic covariance for the observables by considering only the leading Gaussian contribution.

We consider a simplified setup of the complete Dark Energy Spectroscopic Instrument (hereafter DESI) \citep[][]{DESI:2016fyo}, covering a sky area of 14,000 square degrees, with LRGs in two redshift bins $z_1\in[0.4-0.6]$, $z_2\in[0.6-0.8]$ and ELGs in two redshift bins $z_3\in[0.8-1.1]$, $z_4\in[1.1-1.6]$. For simplicity, we ignore the LRGs in the $z_3$ bin where they overlap with the ELGs. We summarize the survey specifications assumed for our analysis in \cref{tab:survey_specs}. For the fiducial cosmology, we assume the same cosmology as in our simulations in \cref{sec:sims} with a local PNG amplitude $\fnl=+1$. Note that the fiducial value of $\fnl$ does not affect the ratio between the DS in \cref{eq:DS_st,eq:DS_mt}, hence the improvement factors we report below.

The fiducial bias values are taken from \cref{tab:bias_values}, except for the $z_1$ bin, where we linearly interpolate between the $z=0.0$ and $z=0.7$ values. We emphasize again that the bias coefficients do not depend on the actual value of $\fnl$.
\begin{table}
\centering
\begin{tabular}{|l|cccc|}
\hline
Tracer & $10^4\ngbar [h^3\mathrm{Mpc}^{-3}]$ & $z$ range & $z_{\rm eff}$ & Area [deg$^2$] \\\hline
LRG1  & 5.0 & $[0.4,0.6)$ & 0.5 & 14,000\\
LRG2  & 5.0 & $[0.6,0.8)$ & 0.7 & 14,000\\
ELG1  & 6.0 & $[0.8,1.1)$ & 0.9 & 14,000\\
ELG2  & 4.0  & $[1.1,1.6]$ & 1.3 & 14,000\\
\hline
\end{tabular}
\caption{Survey specifications assumed for the forecast and analysis of information content in \cref{sec:forecast}.
\label{tab:survey_specs}}
\end{table}
For the analysis cutoff scales, we choose $\kmin=0.003\hmpcinv$ and $\kmax=0.08\hmpcinv$ following the fiducial values adopted by Ref.~\cite{Chaussidon:2024qni} in their local PNG analysis using the DESI DR1 LRG and quasar samples.
We assume the power spectra of the stochasticity to follow Poisson shot noise, namely \cref{eq:Pgg_Poisson_shotnoise} for $P_{\rm gg}$ and \cref{eq:Pss_Poisson_shotnoise} for $P_{\rm s2Ds2D}$.

\begin{figure}[tb]
  \centering
  \includegraphics[width=0.8\linewidth]{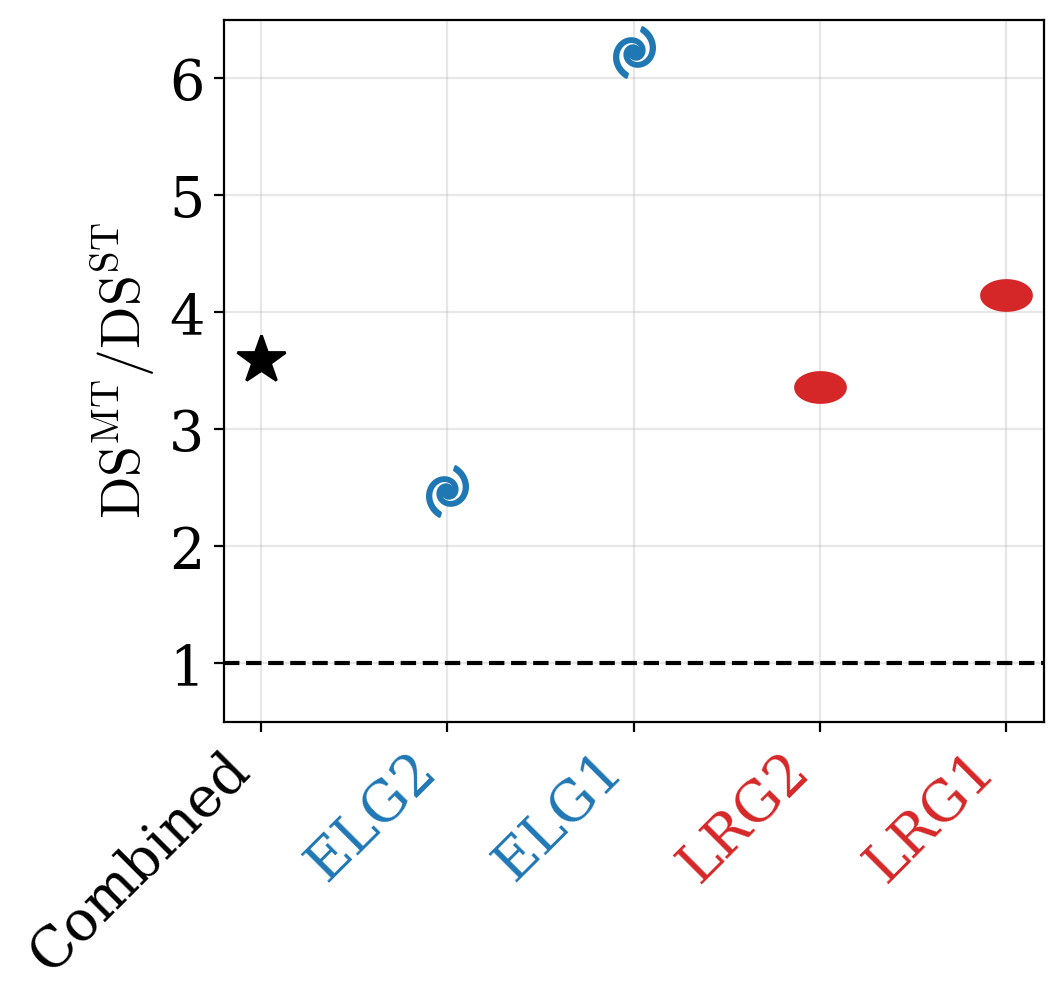}
  \caption{Forecast of the improvement in local PNG detection significance (DS) with a galaxy number--size multi-tracer setups for DESI-like galaxy samples. The improvement factor---defined by the ratio of the DS in the multi-tracer case over that in the single-tracer case, $\mathrm{DS^{MT}}/\mathrm{DS^{ST}}$---is shown for the LRG bins (red ellipticals), ELG bins (blue spirals) and for all galaxy bins combined (black star). The horizontal dashed line indicates no improvement over the galaxy number single-tracer setup.}
  \label{fig:DS_forecast}
\end{figure}
In \cref{fig:DS_forecast}, we plot the ratio between detection significances in the multi-tracer and single-tracer setup, $\mathrm{DS^{MT}}/\mathrm{DS^{ST}}$. These improvement factors indicate the additional clustering information galaxy sizes add to the detection of local PNG through their own auto correlations and their cross-correlations with galaxy numbers. The improvement factors appear sensitive to galaxy number density hence more stable for the LRG bins (3.4-4.2) while less so for the ELG bins (2.5-6.2). With all LRG and ELG samples combined (assuming zero overlap hence no correlation between any galaxy bin), the number--size multi-tracer setup achieves an improvement of 3.6.
We emphasize that the improvement factors do not depend on the fiducial value of $\fnl$ assumed in the forecast.

\section{Discussion}
\label{sec:discussion}

\subsection{From halo size proxies to observed galaxy sizes}
\label{sec:discussion:proxies}

Our analysis uses distinct dark matter halos as galaxy proxies and traces of the reduced inertia tensor as fiducial halo-size proxies.
This choice isolates the symmetry-allowed bias structure of the size field and emphasizes the inner mass distribution that is most directly associated with the luminous galaxy.
However, baryonic physics can modify the internal structure of halos and, crucially, the mapping between halo-based size proxies and observed light-profile-based estimates of galaxy sizes \citep{Chua:2018sbi,Karmakar:2023dtt}.
Establishing observational robustness therefore requires calibrating this halo--galaxy size mapping and validating the measured size biases in hydrodynamical simulations, ideally including separate-universe variants \citep[e.g.][]{Barreira:2020kvh}, while quantifying the dependence on the size definition and sample selection.

In observations, the size of an individual galaxy is inferred from the projected surface-brightness field convolved with the point-spread function (PSF).
Two classes of methods are commonly used \citep{Massey:2006ha,Singh:2015sva}: moment-based estimators that measure PSF-corrected weighted quadrupole moments \citep[e.g.,][]{Hirata:2003cv}, and profile-fitting estimators that forward-model the PSF-convolved image with parametric light profiles and define size through fitted parameters such as the effective radius \citep[e.g.,][]{Meert:2014pta}.
Both classes effectively down-weight the low-surface-brightness outskirts, either explicitly through the window function or implicitly through the finite signal-to-noise ratio and model choice.
Therefore, different estimators weight different parts of the galaxy light profile and can change both the stochasticity of $\ds$ and its large-scale bias responses.

This flexibility is a feature rather than a nuisance.
For a given scientific goal, the size estimator can be tailored to optimize performance while controlling the systematics.
In the context of local-PNG analyses, the optimal choice is the definition that maximizes the multi-tracer response misalignment $|b^{\rm g}_1 b^{\rm s}_\phi - b^{\rm s}_1 b^{\rm g}_\phi|$ which controls cosmic-variance cancellation \citep{Barreira:2023rxn}. In single-tracer analyses, this is closely related to enhancing the PNG-induced scale dependence (relative to the Gaussian clustering amplitude).
Motivated by the fact that common observational estimators suppress contributions from faint outskirts, we adopted the (trace of the) \emph{reduced} inertia tensor as our fiducial proxy in this work.
In \cref{app:trace_definition}, we additionally report measurements using the simple inertia tensor, and in \cref{app:size_proxy}, we show the correlation between our fiducial proxy and other commonly used halo size measures (virial radius, scale radius, and half-mass radius).
We will present an analogous comparison of galaxies in hydrodynamical simulations in a follow up.

\subsection{Robustness to realistic galaxy samples}
\label{sec:discussion:selection}

Observed galaxy samples are defined by luminosity, color, and completeness cuts; they therefore occupy a broad, selection-dependent distribution of halo masses. In such cases, the effective biases of the observed sample are weighted averages over halo mass. For galaxy numbers, that is
\begin{equation}
    b^{\rm g}_{\rm eff}(z)=\frac{\int_{M_{\min}}^{M_{\max}} dM \, \frac{dn(M,z)}{dM}\, b^{\rm g}(M,z)\, S(M)}
{\int_{M_{\min}}^{M_{\max}} dM \, \frac{dn(M,z)}{dM}\, S(M)},
\label{eq:sample_bgeff}
\end{equation}
where $S(M)$ is the sample selection function. For galaxy sizes, however, the effective bias is more subtle:
\begin{align}
b^{\rm s}_{\rm eff}(z)
=&\frac{\int dM\, \frac{dn}{dM}\, S(M)\,\langle t\rangle
\left[b^{\rm g}(M,z)+b^{\rm s}(M,z)\right]}
{\int dM\, \frac{dn}{dM}\, S(M)\,\langle t\rangle}
\nonumber\\
&-\frac{\int dM\, \frac{dn}{dM}\, S(M)\, b^{\rm g}(M,z)}
{\int dM\, \frac{dn}{dM}\, S(M)}.
\label{eq:sample_bseff}
\end{align}
Unlike \cref{eq:sample_bgeff}, \cref{eq:sample_bseff} is not a simple weighted average of the narrow-bin size biases. Rather, it is the difference between a trace-weighted average of $b^{\rm g}+b^{\rm s}$ and a number-weighted average of $b^{\rm g}$. As a result, the size bias measured in a broad mass bin need not lie between the values measured in its narrow sub-bins. In particular, because the first term is weighted by $\langle t\rangle$, the sample-wise normalization in \cref{eq:dtr_def,eq:ds_def} can enhance the contribution of the larger, more massive halos within a broad interval.

In this work, $S(M)$ is simply a top-hat function in mass. One might therefore worry that the broad-bin effective bias could dilute the distinctive response pattern we identify, or that a small bin-averaged $b^{\rm s}_1$ could arise from accidental cancellation among sub-bin contributions rather than from a genuinely weak response. We address both concerns in \cref{app:mass_binning} by remeasuring $(b^{\rm s}_1, b^{\rm s}_\phi)$ in five narrow sub-bins and in a single broad interval spanning the full mass range. The hierarchy $b^{\rm s}_1\lesssim 0.2$ and $b^{\rm s}_\phi\sim -1$ persists across all bins and redshifts, confirming that the suppressed Gaussian size response is genuine rather than a cancellation artifact. The appendix also illustrates concretely how the nontrivial weighting structure of \cref{eq:sample_bseff} shapes the broad-bin effective responses, and how these depend on the normalization convention used to define the size field. We leave a more detailed investigation of alternative normalization schemes and realistic selection functions $S(M)$ to future work.

Taken together, these points indicate that the qualitative signature identified here should survive in realistic luminosity- and color-selected samples whose halo-mass support lies predominantly within the mass ranges we probe.
We discuss the sensitivity of the models and measurements to these assumptions in \cref{sec:discussion:limitations} below.

\subsection{Theoretical limitations and extensions}
\label{sec:discussion:limitations}

A few modeling assumptions and approximations we adopt in this work should be relaxed as the sensitivity improves with future data.

First, we restrict the galaxy bias expansion to linear order. For the number density field, higher-order Gaussian contributions (from operators such as $\d^2$ and the tidal field $K^2$) have been well studied \citep[e.g.,][]{McDonald:2006mx,McDonald:2009dh,Desjacques:2016bnm}; their counterparts for the size field remain unmeasured but are allowed by the same symmetries and will contribute at the same perturbative order. Whether the suppression of the linear Gaussian response $b^{\rm s}_1$ extends to a similarly suppressed second-order response remains an open empirical question beyond the scope of this paper. Our systematic checks by varying the fitting $k$ range and allowing for an additional higher-derivative counterterm, however, do not find evidence for any correction of the order of $b^{\rm s}_1$. Nevertheless, including these terms, in principle, will extend the $\kmax$ in the analyses. Such an extended analysis range might not significantly help with the local PNG constraint but potentially will improve other cosmological constraints.

Second, we assume the stochastic contributions to be Poisson 
(\cref{eq:Pgg_Poisson_shotnoise,eq:Pss_Poisson_shotnoise}). For the halo size samples we consider in this work, the sample-mean normalization in \cref{eq:dtr_def,eq:ds_def} shifts the effective contribution toward the more massive halos in the sample, which are known to exhibit sub-Poisson stochasticity driven by halo exclusion 
\citep{Smith:2006ne,Hamaus:2010im,Baldauf:2013hka}.
On the other hand, the galaxy stochasticity can differ substantially from halo stochasticity, behaving either super- or sub-Poisson depending on the samples. Furthermore, the galaxy--halo size mapping necessarily introduces additional scatter. Characterizing the stochastic contributions of the galaxy size field in hydrodynamical simulations is therefore a near-term priority.

\subsection{Beyond size: intrinsic-property tracers}
\label{sec:discussion:generalization}

The size field $\ds$ studied here is one instance of a broader construction.
Given any intrinsic galaxy property $q$ measured per object---size, concentration, color, morphology, or a star-formation indicator---one can form a number-weighted intrinsic-property density field and characterize its large-scale clustering through Gaussian and PNG response coefficients.
Wide-field surveys already provide multiple such observables for the \emph{same} galaxies, enabling multi-tracer analyses by reusing the same objects with different weights rather than engineering many disjoint subsamples.

The galaxy number--size combination is particularly powerful for local PNG because the two fields share the same volume and objects while exhibiting markedly different response directions in the $(b_1, b_\phi)$ plane: sizes supply an almost zero Gaussian response paired with a large, negative PNG response, whereas number counts supply a positive Gaussian response paired with a positive PNG response whose amplitude roughly tracks $b_1$ via the universality relation.
The galaxy number--size cross-spectrum additionally provides a sign-specific internal consistency check: for a given sign of $\fnl$, genuine local PNG predicts a correlated and sign-definite modulation across $P_{\rm gg}$, $P_{\rm ss}$, and $P_{\rm gs}$ which is difficult to mimic with observational or modeling systematics.

More generally, the optimal strategy for searches of local PNG need not be to maximize bias or split samples aggressively, but rather to identify sets of intrinsic-property tracers whose response vectors span a large area in $(b_1, b_\phi)$ space while maintaining low stochasticity on large scales.
Exploiting this information for new physics discoveries requires no new observations, only a change in how we combine and analyze existing data.

\section{Conclusion}
\label{sec:conclusion}

Wide-field imaging surveys routinely measure galaxy sizes, yet this information is typically discarded in large-scale structure analyses. We set out to ask whether it can instead sharpen constraints on local primordial non-Gaussianity (local PNG), and in particular whether galaxy sizes can act as a genuinely distinct tracer when measured for the same objects used in number-count analyses.

Using gravity-only simulations with dark matter halos as galaxy proxies, we have presented the first measurements of the linear Gaussian and local-PNG bias coefficients of the galaxy size field $\delta_s$. For galaxy-mass halos, sizes respond only weakly to long-wavelength matter density perturbations ($0.0 \lesssim b^{\rm s}_1\lesssim 0.2$) while exhibiting sizable and negative responses to primordial potential fluctuations ($b^{\rm s}_\phi \sim -1$; \cref{fig:b1gs_z,fig:bphigs_z} and \cref{tab:bias_values}). This pattern persists across the full mass and redshift range we probe and is robust to mass binning (see \cref{app:mass_binning}). In contrast, galaxy numbers remain close to the universality relation, so the effective PNG response direction of sizes is markedly different from that of counts.

This response misalignment gives galaxy sizes two concrete roles in local-PNG searches: it increases information, and it provides a cross-check. On the information side, sizes supply additional constraining power when combined with galaxy numbers in multi-tracer setups---without requiring any tracer split. For a combination of DESI-like LRG and ELG samples, we forecast an improvement in local-PNG detection significance by a factor of $\sim\!3.6$ (\cref{sec:forecast}). On the cross-check side, because the PNG response of sizes is opposite in sign to that of number counts, the number--size cross-spectrum $P_{\rm gs}$ provides a sign-specific internal consistency test: for a given sign of $\fnl$, genuine local PNG predicts a correlated modulation across $P_{\rm gg}$, $P_{\rm ss}$, and $P_{\rm gs}$ whose directional pattern is difficult to mimic with typical observational systematics.

Galaxy sizes thus provide genuinely independent information on local PNG beyond what number counts alone can access, precisely because their bias responses violate the universality relation that limits the number-count signal. This makes them a promising additional probe for ongoing and upcoming surveys targeting PNG as one of their key science goals \citep[e.g.,][]{DESI:2016fyo,Euclid:2024yrr,Dore:2014,Wang:2021oec}.

Unlocking the full observational potential of this idea will require calibrating the halo-to-galaxy size mapping---accounting for the galaxy--halo connection, line-of-sight projection, and observational size estimation---and characterizing stochastic contributions and higher-order corrections in hydrodynamical simulations. These are tractable steps, and the payoff is significant: the formalism developed here treats number counts, sizes, and shapes on equal footing within a single analysis framework, opening a unified multi-tracer route to both isotropic PNG through counts and sizes and anisotropic PNG through shapes---all from the same galaxy catalog. As a step toward this goal, we provide in \cref{app:approx_Gaussian_bias_fits,app:approx_universality} empirical relations between size, shape, and number-count bias coefficients as directly applicable inputs for future forecasts.

\medskip

\begin{acknowledgments}
The authors thank Fabian Schmidt and Jingjing Shi for their comments on the draft.
NMN and KA thank Masahiro Takada; over lunch in his office, they discovered that they had independently arrived at the idea of galaxy sizes as biased tracers---and that field-level inference was not their only shared pursuit \citep{Nguyen:2024yth,Akitsu:2025boy}.
NMN thanks Peter Behroozi, Neal Dalal, Jingjing Shi, Masahiro Takada and Zvonimir Vlah for helpful discussions, also Misao Sasaki for his inspiring lecture series on inflation.
NMN dedicates this paper to Alex Barreira and Titouan Lazeyras.
This work was supported by World Premier International Research Center Initiative (WPI Initiative), MEXT, Japan.
NMN acknowledges support from the Japan Foundation for Promotion of Astronomy Research Grant and the Japan Society for the Promotion of Science (JSPS) KAKENHI Grant Number 25K23373.
KA is supported by Fostering Joint International Research (B) under Contract No.~21KK0050 and the JSPS KAKENHI Grant No.~JP24K17056.
AT acknowledges support from JSPS KAKENHI Grant Numbers JP23K20844 and JP23K25868, and ISHIZUE 2025 of Kyoto University. 
This work was performed on the \code{Raven} cluster at the Max Planck Computing and Data Facility (MPCDF). NMN thanks the MPCDF staff for their support.
This study utilized the following open-source libraries: \href{https://www.h5py.org/}{\code{h5py}} \cite{andrew_collette_2022_6575970}, \href{https://matplotlib.org/}{\code{Matplotlib}}, \href{https://numpy.org/}{\code{NumPy}}, \href{https://pandas.pydata.org/}{\code{pandas}}, \href{https://papermill.readthedocs.io/en/latest/index.html}{\code{papermill}} and \href{https://scipy.org/}{\code{SciPy}}.
\end{acknowledgments}

\clearpage
\onecolumngrid
\appendix
\crefalias{section}{appendix}
\crefalias{subsection}{appendix}

\section{Gaussian covariance for power spectrum multipoles}
\label{app:covariance}

To estimate the uncertainties on the measurements of the power spectrum multipoles and especially to perform the Fisher forecast in \cref{sec:forecast}, we have used the analytic Gaussian covariance of the power spectrum multipoles.
In this appendix, we provide the derivation of the covariance for both single-tracer and multi-tracer setups.

For a survey redshift bin of comoving volume $V_{\rm survey}$, the number of independent Fourier modes in a thin spherical shell $[k-\D k/2,k+\D k/2]$ is
\begin{equation}
N_k \;\equiv\; \frac{1}{2}\frac{V_{\rm survey}}{(2\pi)^3}\int_{k-\D k/2}^{k+\D k/2} d^3q
\;\simeq\frac{V_{\rm survey}\,k^2\D k}{4\pi^2}\,.
\label{eq:Nk_thinshell}
\end{equation}
The extra factor of $1/2$ accounts for the $\delta_A^{\ast}(\vk)=\delta_A(-\vk)$ constraint for a real field $A$.

\subsection{General covariance}
\label{app:general_cov}

We neglect survey window effects and non-Gaussian contributions so that the covariance is diagonal in $k$-bins. The covariance
between two multipole estimators $\widehat P_{\rm AB}^{\ell}(k_i)$ and $P_{CD}^{\ell'}(k_j)$ is diagonal in $k$ and, according to the Wick's theorem, given by
\begin{align*}
\Cov\!\left[\widehat P_{\rm AB}^{\ell}(k_i),\widehat P_{CD}^{\ell'}(k_j)\right]
&=
\frac{\kdelta_{ij}}{N_{k_i}}\frac{(2\ell+1)(2\ell'+1)}{2}
\int_{-1}^{1}d\mu
L_\ell(\mu)L_{\ell'}(\mu)\,
\Bigl[P_{AC}(k_i,\mu)\, P_{BD}(k_i,\mu)
+ P_{AD}(k_i,\mu)\, P_{BC}(k_i,\mu)
\Bigr],\\
&=\frac{\kdelta_{ij}}{N_{k_i}}\Bigl[\C^{(\ell,\ell')}[P_{AC},P_{BD}]+\C^{(\ell,\ell')}[P_{AD},P_{BC}]\Bigr]\numberthis\label{eq:cov_general},
\end{align*}
where we have introduced the bilinear form
\begin{align*}
    \C^{(\ell,\ell')}[X,Y]&=\frac{(2\ell+1)(2\ell'+1)}{2}\\
&\times\int_{-1}^{1}d\mu L_\ell(\mu)L_{\ell'}(\mu)\,
X(k,\mu)Y(k,\mu)
\numberthis\label{eq:bilinear_form}
\end{align*}
for any $[X,Y]$ pair of power spectra.
We evaluate the $\mu$-integrands in \cref{eq:bilinear_form} analytically by multipole expanding $X$ and $Y$ up to $\ell_{\rm max}=4$ (see \cref{eq:multipole_decomp}). The six independent components are as follows:
\begin{align}
\C^{(0,0)}[X,Y] &= X^{(0)} Y^{(0)} + \frac{1}{5}X^{(2)}Y^{(2)} + \frac{1}{9}X^{(4)}Y^{(4)}\label{eq:C00},\\
\C^{(0,2)}[X,Y] &= X^{(0)} Y^{(2)} + X^{(2)} Y^{(0)}
+ \frac{2}{7}X^{(2)} Y^{(2)}
+ \frac{2}{7}\!\left[X^{(2)} Y^{(4)} + X^{(4)} Y^{(2)}\right]
+ \frac{100}{693}X^{(4)} Y^{(4)}\label{eq:C02},\\
\C^{(0,4)}[X,Y] &= X^{(0)} Y^{(4)} + X^{(4)} Y^{(0)}
+ \frac{18}{35}X^{(2)} Y^{(2)}
+ \frac{20}{77}\!\left[X^{(2)} Y^{(4)} + X^{(4)} Y^{(2)}\right]
+ \frac{162}{1001}X^{(4)} Y^{(4)}\label{eq:C04},\\
\C^{(2,2)}[X,Y] &= 5X^{(0)} Y^{(0)}
+ \frac{10}{7}\!\left[X^{(0)} Y^{(2)} + X^{(2)} Y^{(0)}\right]
+ \frac{10}{7}\!\left[X^{(0)} Y^{(4)} + X^{(4)} Y^{(0)}\right]\nonumber\\
&+ \frac{15}{7}X^{(2)} Y^{(2)}
+ \frac{60}{77}\!\left[X^{(2)} Y^{(4)} + X^{(4)} Y^{(2)}\right]
+ \frac{8945}{9009}X^{(4)} Y^{(4)}\label{eq:C22},\\
\C^{(2,4)}[X,Y] &= \frac{18}{7}\!\left[X^{(0)} Y^{(2)} + X^{(2)} Y^{(0)}\right]
+ \frac{100}{77}\!\left[X^{(0)} Y^{(4)} + X^{(4)} Y^{(0)}\right]\nonumber\\
&+ \frac{108}{77}X^{(2)} Y^{(2)}
+ \frac{1789}{1001}\!\left[X^{(2)} Y^{(4)} + X^{(4)} Y^{(2)}\right]
+ \frac{900}{1001}X^{(4)} Y^{(4)}\label{eq:C24},\\
\C^{(4,4)}[X,Y] &= 9X^{(0)} Y^{(0)}
+ \frac{180}{77}\!\left[X^{(0)} Y^{(2)} + X^{(2)} Y^{(0)}\right]
+ \frac{1458}{1001}\!\left[X^{(0)} Y^{(4)} + X^{(4)} Y^{(0)}\right]\nonumber\\
&+ \frac{16101}{5005}X^{(2)} Y^{(2)}
+ \frac{1620}{1001}\!\left[X^{(2)} Y^{(4)} + X^{(4)} Y^{(2)}\right]
+ \frac{42849}{17017}X^{(4)} Y^{(4)}\label{eq:C44}.
\end{align}

Setting $A=B=C=D$ in \cref{eq:cov_general} yields the
Gaussian covariance for the auto power spectrum $\widehat P^{\ell}_{\rm AA}$:
\begin{equation}
\Cov\!\left[\widehat P^{\ell}_{\rm AA}(k_i),\widehat P^{\ell'}_{\rm AA}(k_j)\right]
=
\frac{2\kdelta_{ij}}{N_{k_i}}\,
\C^{(\ell,\ell')}\!\left[P_{\rm AA},P_{\rm AA}\right].
\label{eq:cov_auto_compact}
\end{equation}

Setting
$C=B$ and $D=A$ in \cref{eq:cov_general} yields the
Gaussian covariance for the cross power spectrum $\widehat P^{\ell}_{\rm AB}$,
\begin{align*}
\Cov\!\left[\widehat P^{\ell}_{\rm AB}(k_i),\widehat P^{\ell'}_{\rm AB}(k_j)\right]
=
\frac{\kdelta_{ij}}{N_{k_i}}\Bigl[&\C^{(\ell,\ell')}\!\left[P_{\rm AA},P_{\rm BB}\right]\\
+&\C^{(\ell,\ell')}\!\left[P_{\rm AB},P_{\rm AB}\right]\Bigr]\numberthis\label{eq:cov_cross_compact}.
\end{align*}

\subsection{Multi-tracer covariance}
\label{app:cov_full}

Consider now the full data vector for the multi-tracer case in \cref{eq:multi-tracer_data_vec}, using \cref{eq:cov_general,eq:bilinear_form}, the full covariance $\Cov[\mathbf{D}(k),\mathbf{D}(k)]$ can be expressed
in a $3\times3$-block form
\begin{equation}
\Cov[\mathbf{D}(k_i),\mathbf{D}(k_j)]
=
\frac{\kdelta_{ij}}{N_{k_i}}
\begin{pmatrix}
2\,\mathbf{C}[P_{\rm AA},P_{\rm AA}]
&
2\,\mathbf{C}[P_{\rm AA},P_{\rm AB}]
&
2\,\mathbf{C}[P_{\rm AB},P_{\rm AB}]
\\[4pt]
2\,\mathbf{C}[P_{\rm AB},P_{\rm AA}]
&
\mathbf{C}[P_{\rm AB},P_{\rm AB}]
+
\mathbf{C}[P_{\rm AA},P_{\rm BB}]
&
2\,\mathbf{C}[P_{\rm AB},P_{\rm BB}]
\\[4pt]
2\,\mathbf{C}[P_{\rm AB},P_{\rm AB}]
&
2\,\mathbf{C}[P_{\rm BB},P_{\rm AB}]
&
2\,\mathbf{C}[P_{\rm BB},P_{\rm BB}]
\end{pmatrix}\label{eq:cov_full_block},
\end{equation}
where each block $\mathbf{C}[X,Y]$ is a $3\times3$ symmetric matrix whose entries are given by \cref{eq:C00,eq:C02,eq:C04,eq:C22,eq:C24,eq:C44}.

\section{Mass binning and number weighting}
\label{app:mass_binning}

As discussed in \cref{sec:discussion:selection}, broad galaxy samples probe a range of halo masses, so their effective biases are selection-weighted averages over the halo mass function; see \cref{eq:sample_bgeff,eq:sample_bseff}. To address the two concerns raised there, we remeasure the responses in five narrow mass bins, $M_h\in[10^{11.5},10^{12.0}),\,[10^{12.0},10^{12.5}),\,[10^{12.5},10^{13.0}),\,[10^{13.0},10^{13.5}),\,[10^{13.5},10^{14.0})\,\Msunh$, as well as in a single broad bin, $M_h\in[10^{11.5},10^{14.0})\,\Msunh$, excluding cluster-mass halos with $M_h\geq10^{14.0}\,\Msunh$.

\begin{figure}[tb]
  \centering
  \includegraphics[width=0.49\linewidth]{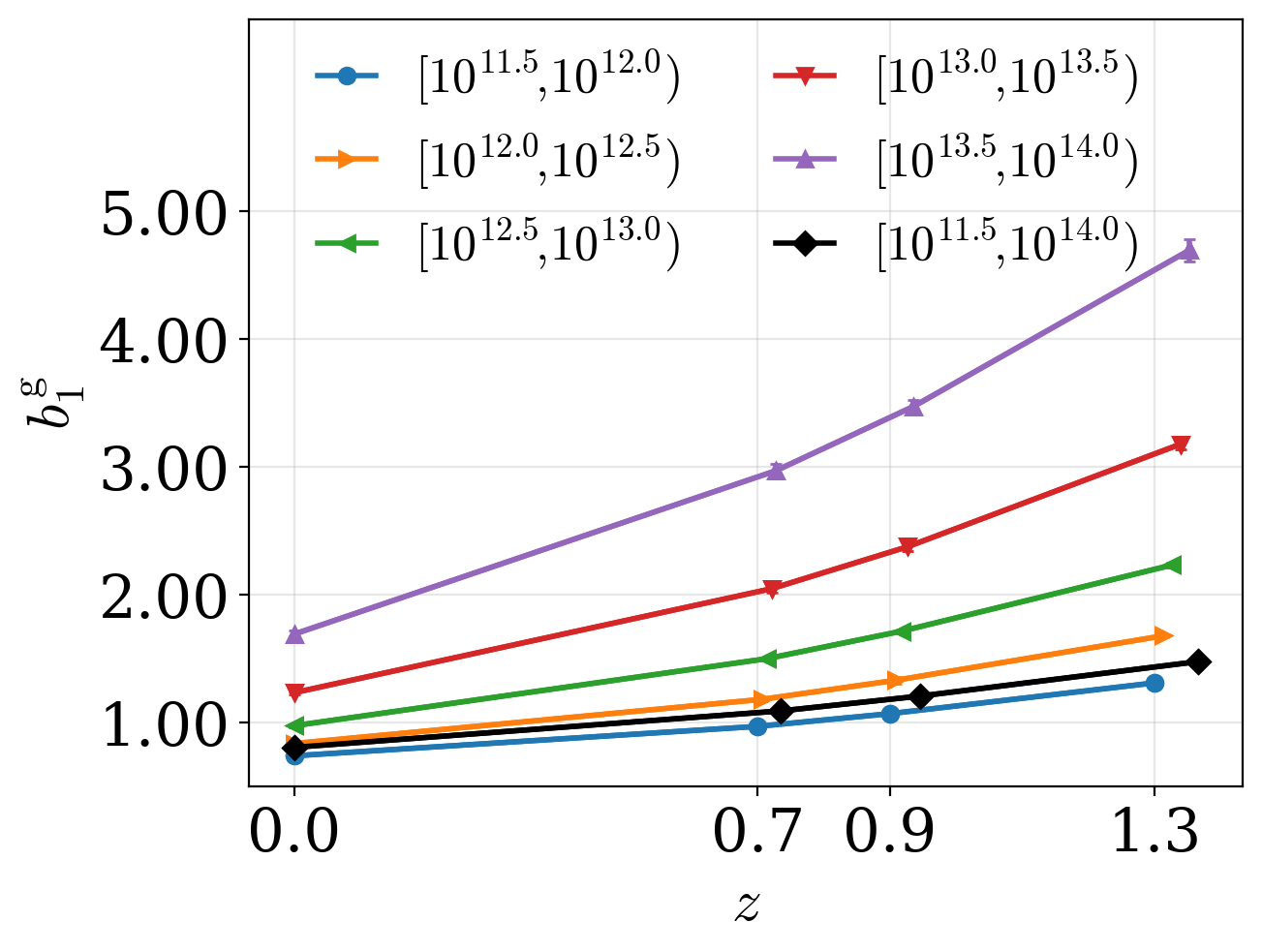}\hfill
  \includegraphics[width=0.49\linewidth]{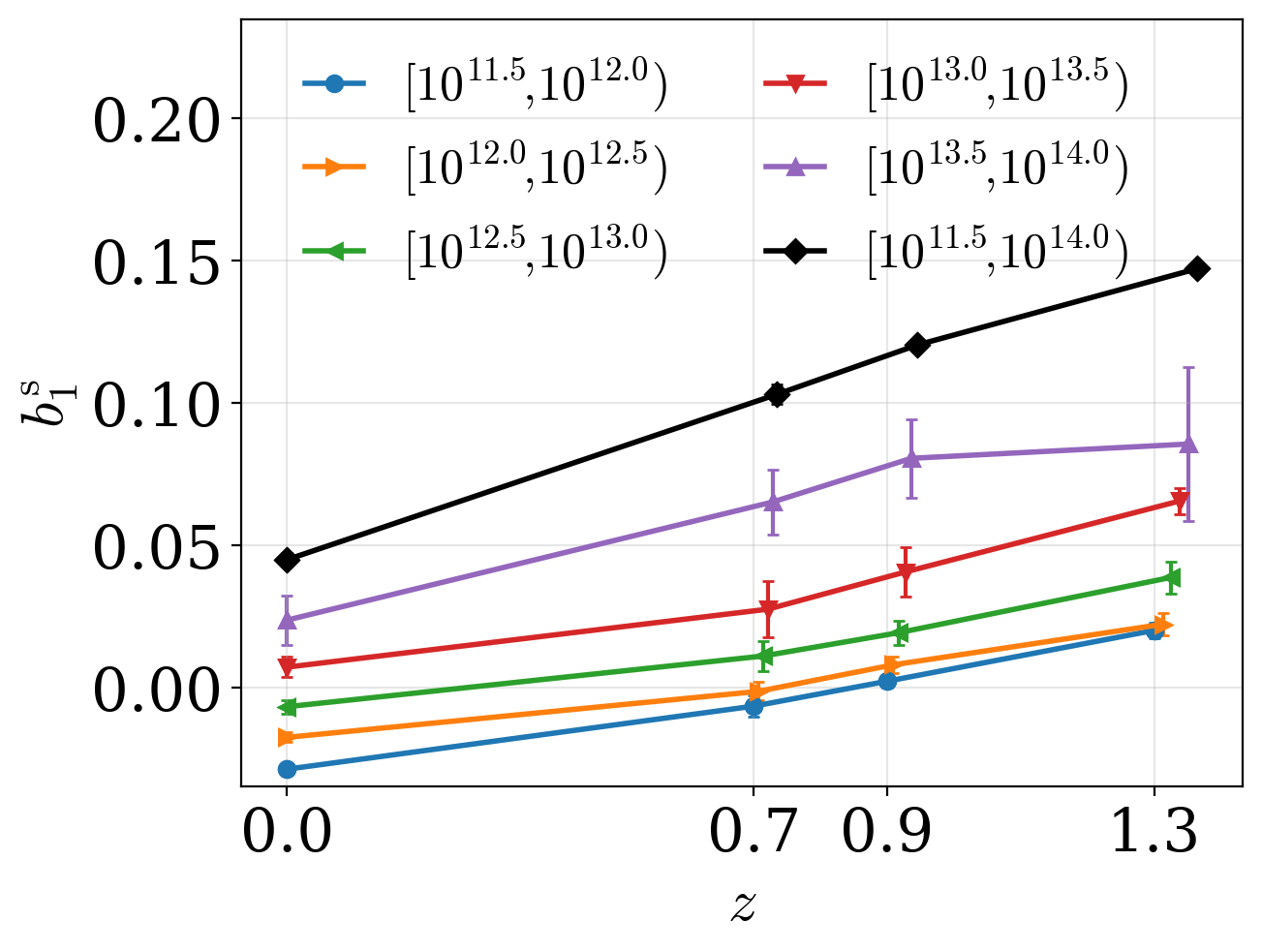}
  \vspace{0.3em}
  \includegraphics[width=0.49\linewidth]{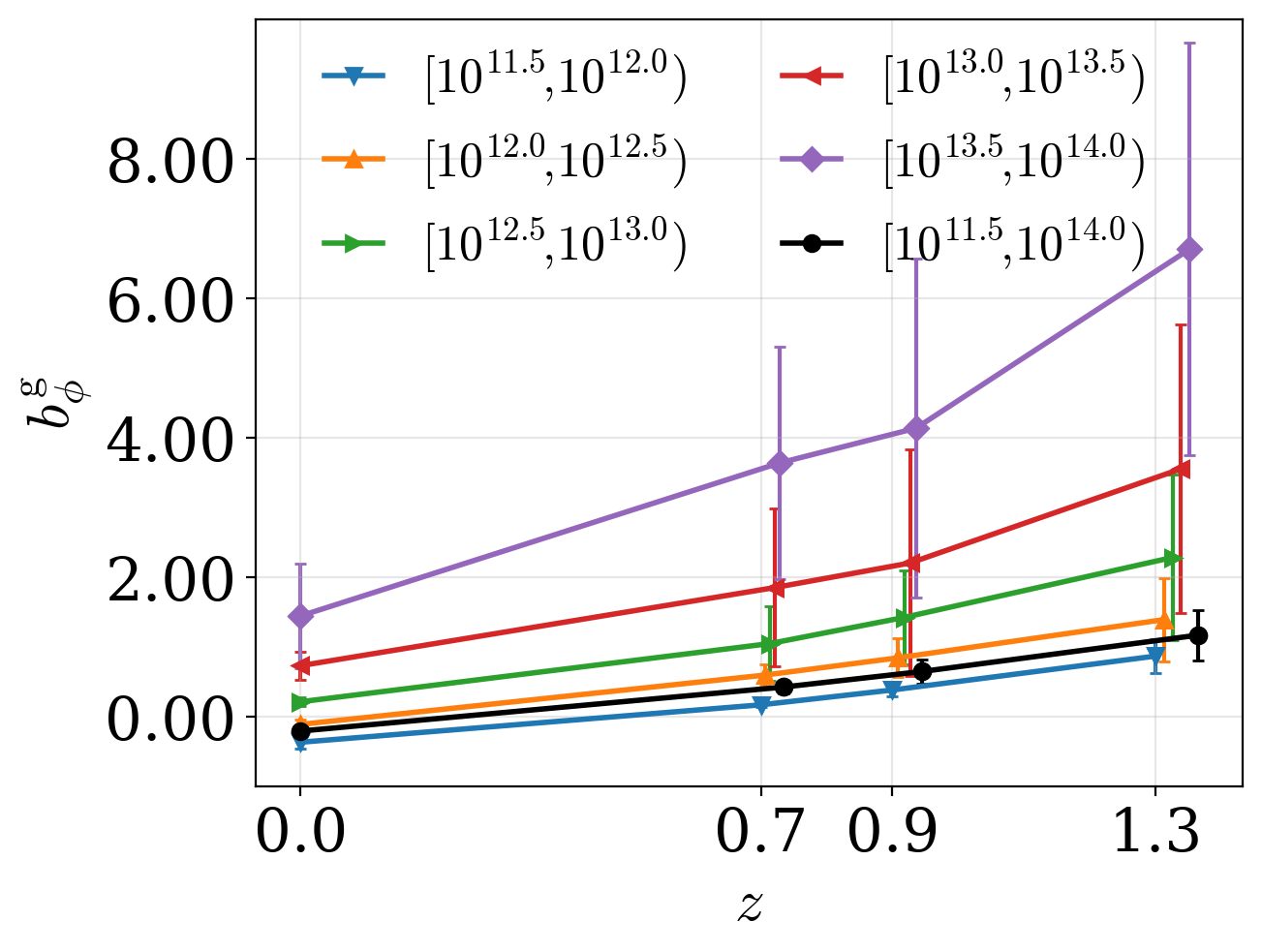}\hfill
  \includegraphics[width=0.49\linewidth]{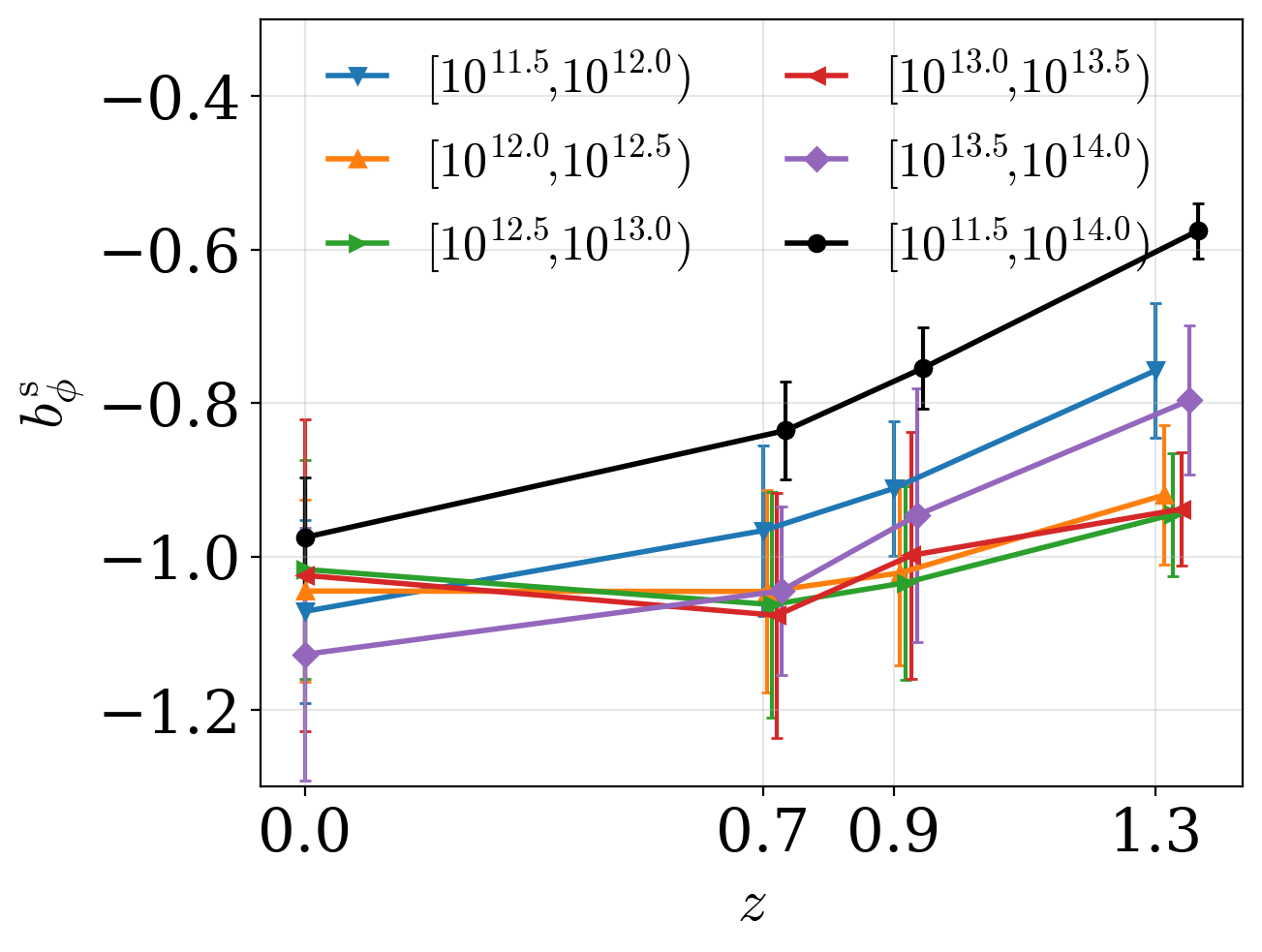}
  \caption{Upper panels: The Gaussian linear bias coefficients $b^{\rm g}_1$ (left) and $b^{\rm s}_1$ (right) as a function of mass bins and redshifts, similar to \cref{fig:b1gs_z}, but with fine (color) and broad mass bins (black). Points are slightly offset horizontally for visibility. Lower panels: Similar to the upper panels but for the non-Gaussian linear bias coefficients $b^{\rm g}_\phi$ (left) and $b^{\rm s}_\phi$ (right).}
  \label{fig:b1phigs_z_narrow_and_broad_Mbins}
\end{figure}

We plot the Gaussian bias coefficients $b^{\rm g}_1$ and $b^{\rm s}_1$ in the upper-left and upper-right panels of \cref{fig:b1phigs_z_narrow_and_broad_Mbins}, respectively. Similarly, we plot the non-Gaussian responses $b^{\rm g}_\phi$ and $b^{\rm s}_\phi$ in the lower-left and lower-right panels.
\cref{fig:b1phigs_z_narrow_and_broad_Mbins} answers both questions. First, the suppressed Gaussian size response and sizable negative PNG size response persist across all narrow bins and the broad bin alike, confirming that the small value of $b^{\rm s}_1$ reflects a genuinely weak density response rather than accidental cancellation. Notably, the broad-bin $b^{\rm s}_1$ lies above all narrow-bin values at every redshift, and the broad-bin $b^{\rm s}_\phi$ is less negative than all narrow-bin values---both direct manifestations of the nontrivial weighting structure of \cref{eq:sample_bseff}. These measurements further consolidate the conclusion that galaxy sizes are promising complementary zero-bias tracers of local PNG.

Second, the contrast between number and size responses follows directly from the different weighting structures in \cref{eq:sample_bgeff,eq:sample_bseff}. For galaxy numbers, the effective response is dominated by the number weighting $dn/dM$, and therefore by the more abundant low-mass halos. For galaxy sizes, by contrast, the additional factor $\langle t\rangle$ in \cref{eq:sample_bseff} shifts the effective weight toward the larger, rarer, and more massive halos. This difference arises because the trace and size perturbation fields are normalized by the sample mean $\langle t\rangle$ (see \cref{eq:dtr_def,eq:ds_def});
In this sense, the broad-bin values in \cref{fig:b1phigs_z_narrow_and_broad_Mbins} lying outside the narrow-bin envelope is not a failure of robustness but the expected signature of $\langle t\rangle$-weighted averaging. The figure therefore illustrates concretely that the inferred size response depends not only on the sample selection $S(M)$, but also on the normalization convention used to define the size field.

\section{Approximate relations between Gaussian bias coefficients}
\label{app:approx_Gaussian_bias_fits}

While the main text focuses primarily on the non-Gaussian responses of galaxy sizes and on how they can be leveraged to search for local PNG, it is also useful in a more general context to summarize the empirical relations between the Gaussian bias coefficients of galaxy sizes or shapes and those of galaxy numbers

In this appendix, we provide approximate fitting functions for the size--number bias relation, $b^{\rm s}_1(b^{\rm g}_1)$, and the shape--number bias relation, $b^{\rm tf}_1(b^{\rm g}_1)$. To better resolve these relations, we divide halos into finer mass bins than in our fiducial analysis in the main text. Specifically, we use the narrow mass bins defined in \cref{app:mass_binning} and fit the rational function
\begin{equation}
y(x)=\frac{a+b\,x}{1+c\,x}
\label{eq:rational_form}
\end{equation}
to the measured pairs $(b^{\rm s}_1,b^{\rm g}_1)$ and $(b^{\rm tf}_1,b^{\rm g}_1)$, accounting for the measurement uncertainties on both variables in each case.

\begin{figure}[tb]
  \centering
  \includegraphics[width=0.49\linewidth]{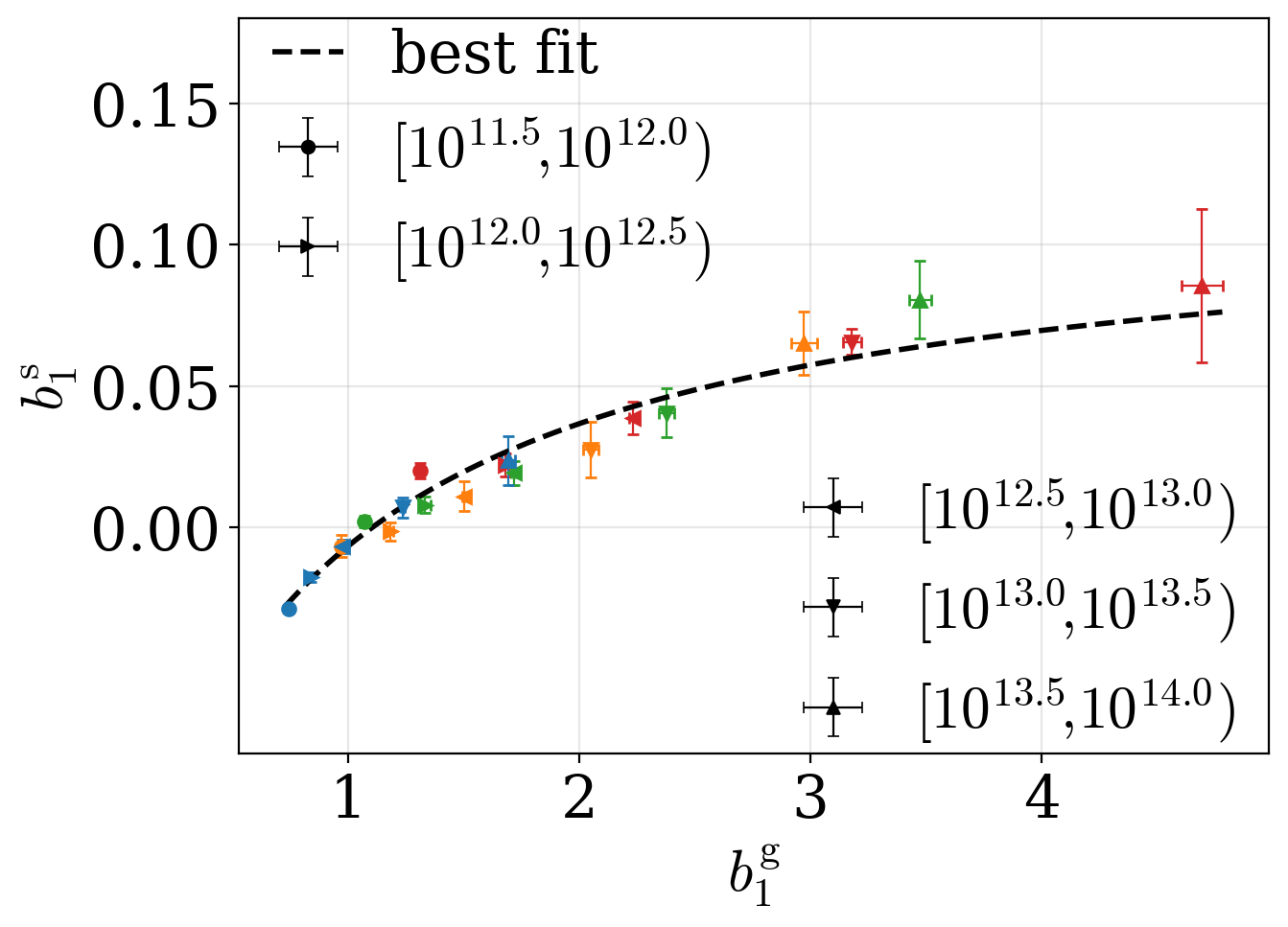}
  \includegraphics[width=0.49\linewidth]{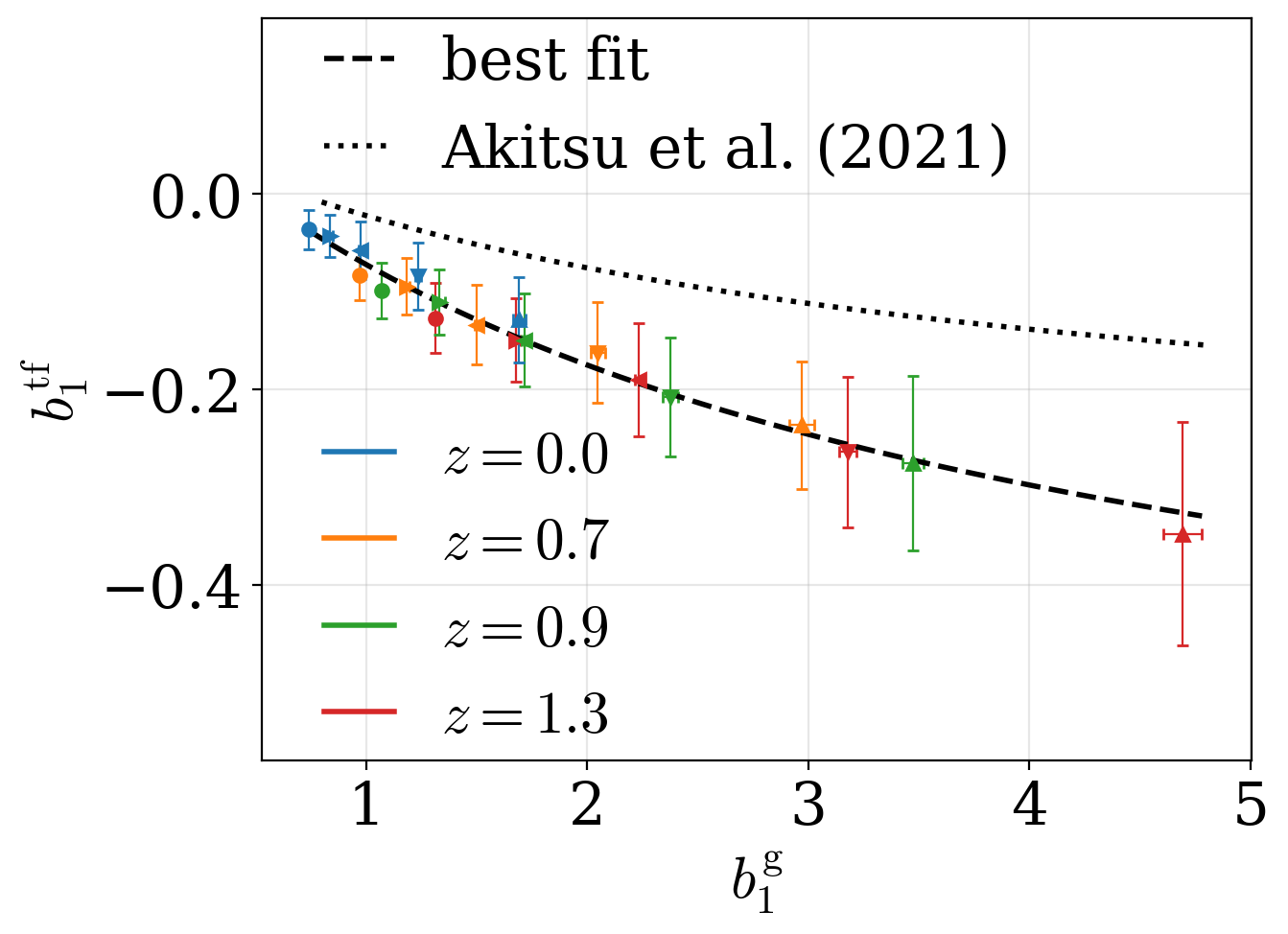}
  \caption{Left panel: The measured relation between $b^{\rm s}_1$ and $b^{\rm g}_1$. Right panel: The measured relation between $b^{\rm tf}_1$ and $b^{\rm g}_1$. In both panels, points indicate bias measurements at different redshifts, as indicated by the color coding, while markers indicate the narrow halo mass bins defined in \cref{app:mass_binning}; the dashed black curves indicate the best-fitting rational function in \cref{eq:rational_form}. In the right panel, the dotted curve indicates the corresponding fit reported in Ref.~\cite{Akitsu:2020fpg} for comparison.
  }
  \label{fig:b1s_fit_b1tf_fit}
\end{figure}

The best fits are shown in \cref{fig:b1s_fit_b1tf_fit}.
We find
\begin{equation}
b^{\rm s}_1=\frac{-0.15293+0.13831b^{\rm g}_1}{1+1.18647b^{\rm g}_1}
\label{eq:b1s_b1g_rational}
\end{equation}
and
\begin{equation}
b^{\rm tf}_1=\frac{0.07432-0.15014b^{\rm g}_1}{1+0.29092b^{\rm g}_1}.
\label{eq:b1tf_b1g_rational}
\end{equation}

To our best knowledge, there is no previous result to compare \cref{eq:b1s_b1g_rational} with. As shown in the left panel of \cref{fig:b1s_fit_b1tf_fit}, the measurements from different redshifts approximately collapse onto a single one-parameter curve when plotted against $b^{\rm g}_1$, suggesting that $b^{\rm g}_1$ captures most of the dependence of $b^{\rm s}_1$ over the $[10^{11.5}-10^{14.0})\,\Msunh$ range. The reduced chi square $\chi^2=2.05$ however indicates that either there is significant scatter or the error bars on bias parameters are underestimated still.

For \cref{eq:b1tf_b1g_rational}, we obtain a very good fit with the reduced chi square $\chi^2=0.11$. We can further compare our measurements and best fit with those presented in Ref.~\cite{Akitsu:2020fpg}. For ease of comparison, we show the rational best fit from Ref.~\cite{Akitsu:2020fpg} as the dotted curve in the right panel of \cref{fig:b1s_fit_b1tf_fit}. We find that our measurements, and hence our best-fit relation, are systematically lower (i.e. more negative) than theirs over the mass range considered. The offset increases with $b^{\rm g}_1$, although it remains statistically consistent within $2\sigma$.

While we measure the bias by fitting the power spectra, Ref.~\cite{Akitsu:2020fpg} ran separate-universe simulations and measured the tidal alignment bias as responses to the long-wavelength tidal modes.
This allowed them to use smaller boxes with twice the mass resolution over this mass range.
Furthermore, they used a modified spherical-overdensity halo finder (\code{AHF} \citep[][]{Knollmann:2009pb}) while we use a phase-space friends-of-friend halo finder (\code{MPI-Rockstar} \citep[][]{Tokuue:2024ney}). We leave a detailed investigation of this discrepancy for future work.

In summary, we consider \cref{eq:b1s_b1g_rational} and \cref{eq:b1tf_b1g_rational} as convenient empirical interpolations of our measurements, rather than as fundamental relations. They should therefore be useful to approximate and translate between Gaussian bias parameters within the specified halo mass range, and should not be extrapolated beyond that.

\section{Approximate universality relations for number count and size bias}
\label{app:approx_universality}

To better understand the gain from combining galaxy number and size clustering information in a multi-tracer setup, it is helpful to examine their universality relations or lack thereof.

\begin{figure}[tb]
  \centering
  \includegraphics[width=0.49\linewidth]{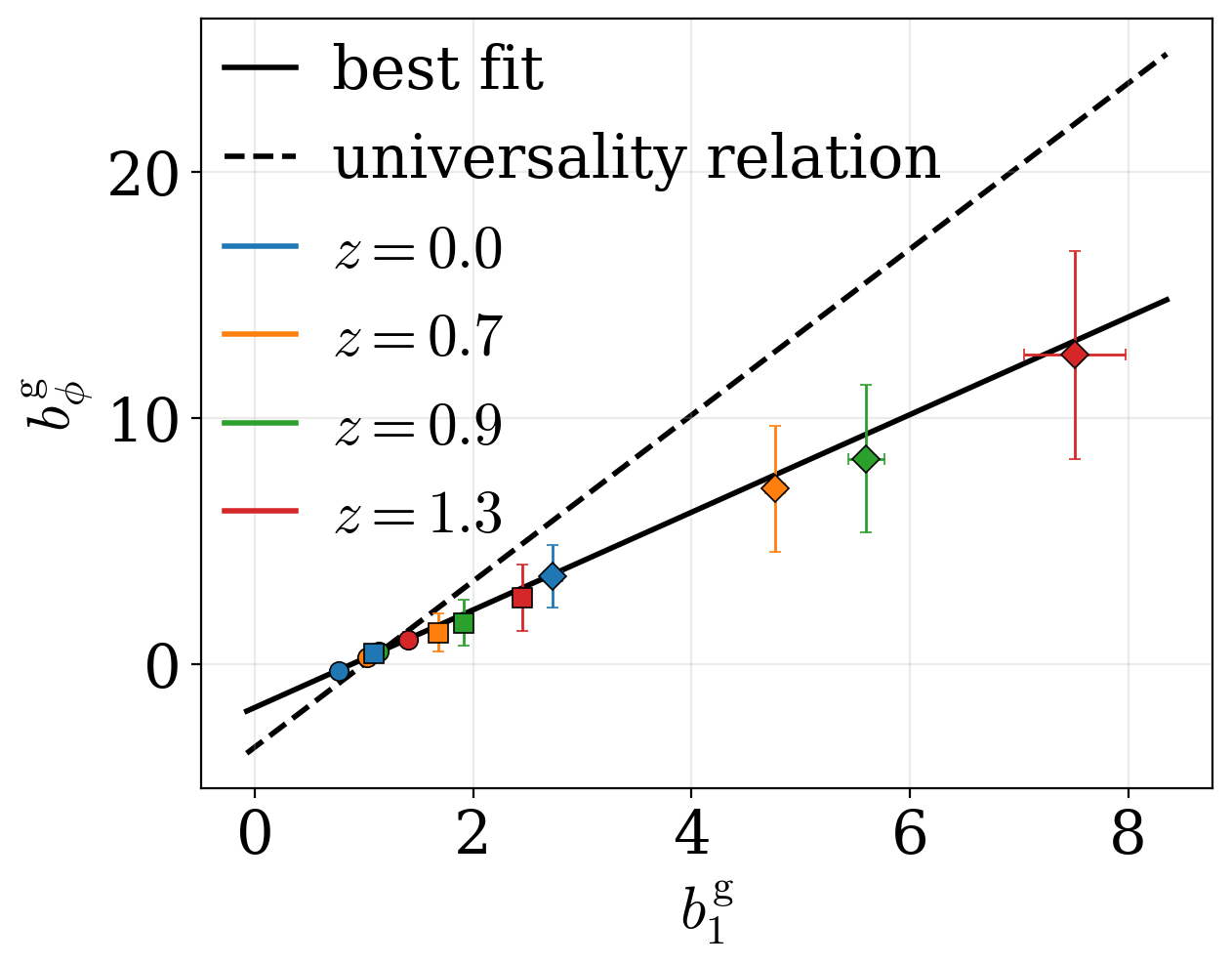}
  \includegraphics[width=0.49\linewidth]{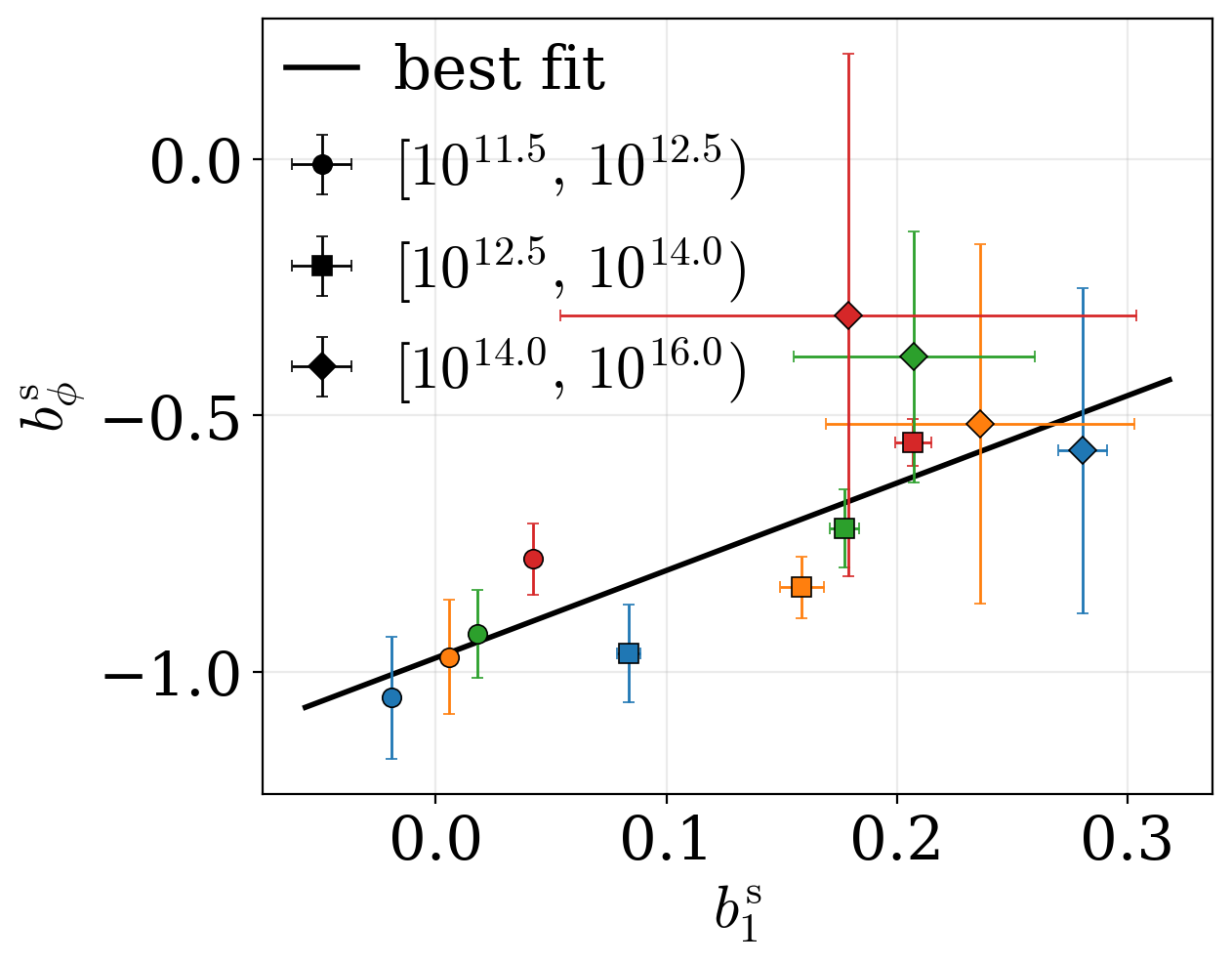}
  \caption{Left panel: Relation between Gaussian bias $b^{\rm g}_1$ and non-Gaussian bias $b^{\rm g}_\phi$ for galaxy numbers. Right panel: The analogous relation for galaxy sizes. In both panels, colors indicate the redshift snapshots and markers indicate the halo mass bins; the solid lines indicate the best-fit phenomenological relation in \cref{eq:ext_universality}. The dashed line in the left panel indicates the universality relation in \cref{eq:universality}.}
  \label{fig:b1-bphi_relations}
\end{figure}

In this appendix, we compare the $b^{\rm g,s}_1$--$b^{\rm g,s}_\phi$ relations of galaxy numbers and sizes in \cref{fig:b1-bphi_relations}. While galaxy numbers approximately follow the universality-inspired trend, galaxy sizes exhibit a qualitatively different pattern.
To quantify the departures from universality, we fit \cref{eq:ext_universality} to the measurements of the Gaussian and non-Gaussian bias coefficients $b^{\rm g,s}_1$ and $b^{\rm g,s}_\phi$ reported in \cref{sec:measurements}. We emphasize that, for sizes, \cref{eq:ext_universality} is used purely as an empirical summary of the measurements rather than as a consequence of strict universality. By fitting \cref{eq:ext_universality} to our Gaussian and non-Gaussian bias measurements for galaxy numbers or sizes, we obtain $c^{\rm g}=0.78$ and $p^{\rm g}=0.90$ for the number count bias, or $c^{\rm s}=0.55$ and $p^{\rm s}=0.57$ for the size bias.

For galaxy numbers, an approximate agreement with the standard universality relation is expected: the abundance is largely controlled by the peak height $\nu$, so both the Gaussian and non-Gaussian responses are approximately derivatives of the same one-parameter function. In contrast, for sizes, the existence of an approximately linear $b^{\rm s}_\phi$--$b^{\rm s}_1$ relation suggests only a weaker form of universality. In particular, this suggests that the peak height captures the dominant trend of the mean size, while additional internal variables provide subleading corrections.

\begin{figure}[tb]
  \centering
  \includegraphics[width=0.49\linewidth]{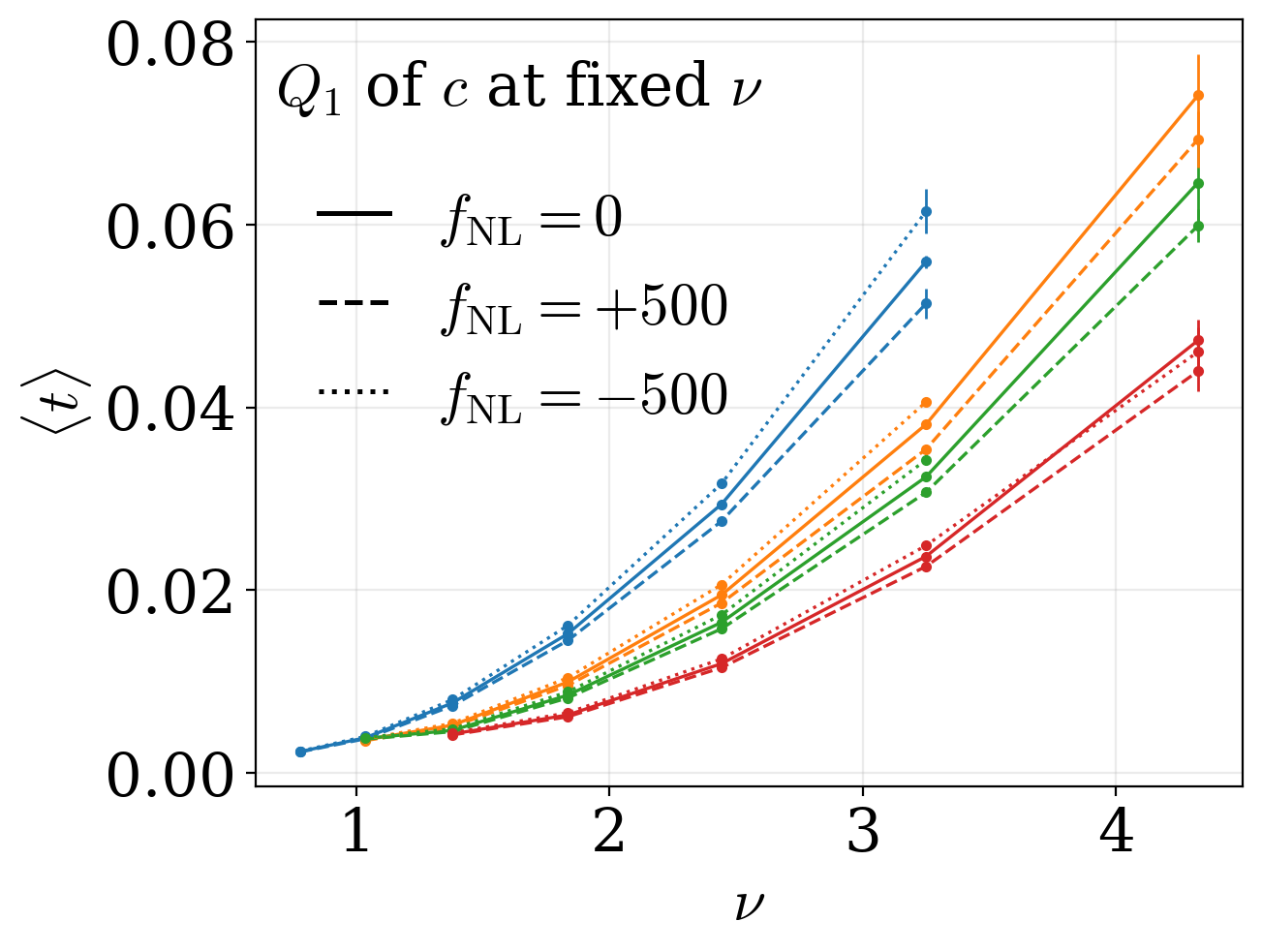}\hfill
  \includegraphics[width=0.49\linewidth]{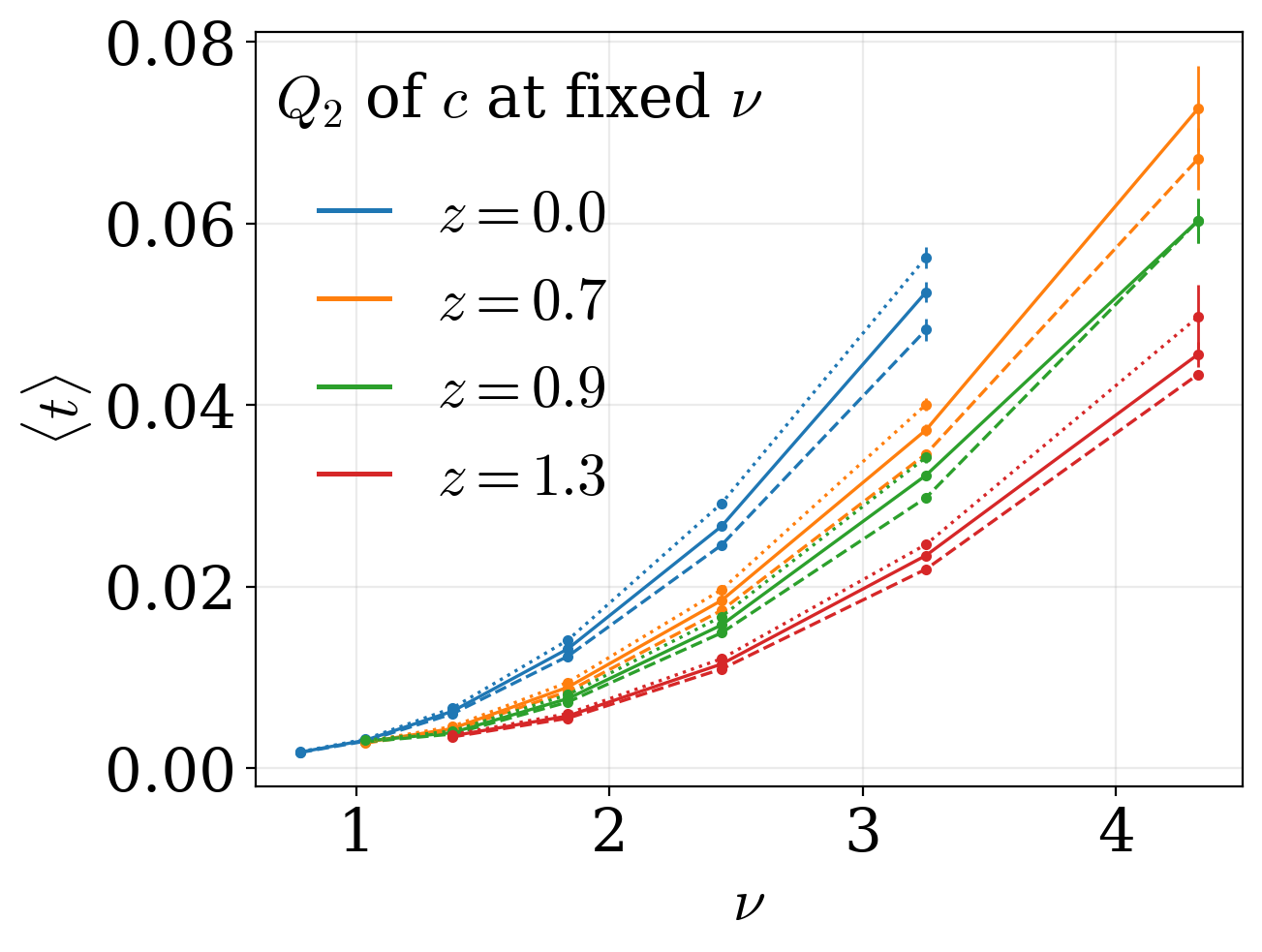}
  \vspace{0.3em}
  \includegraphics[width=0.49\linewidth]{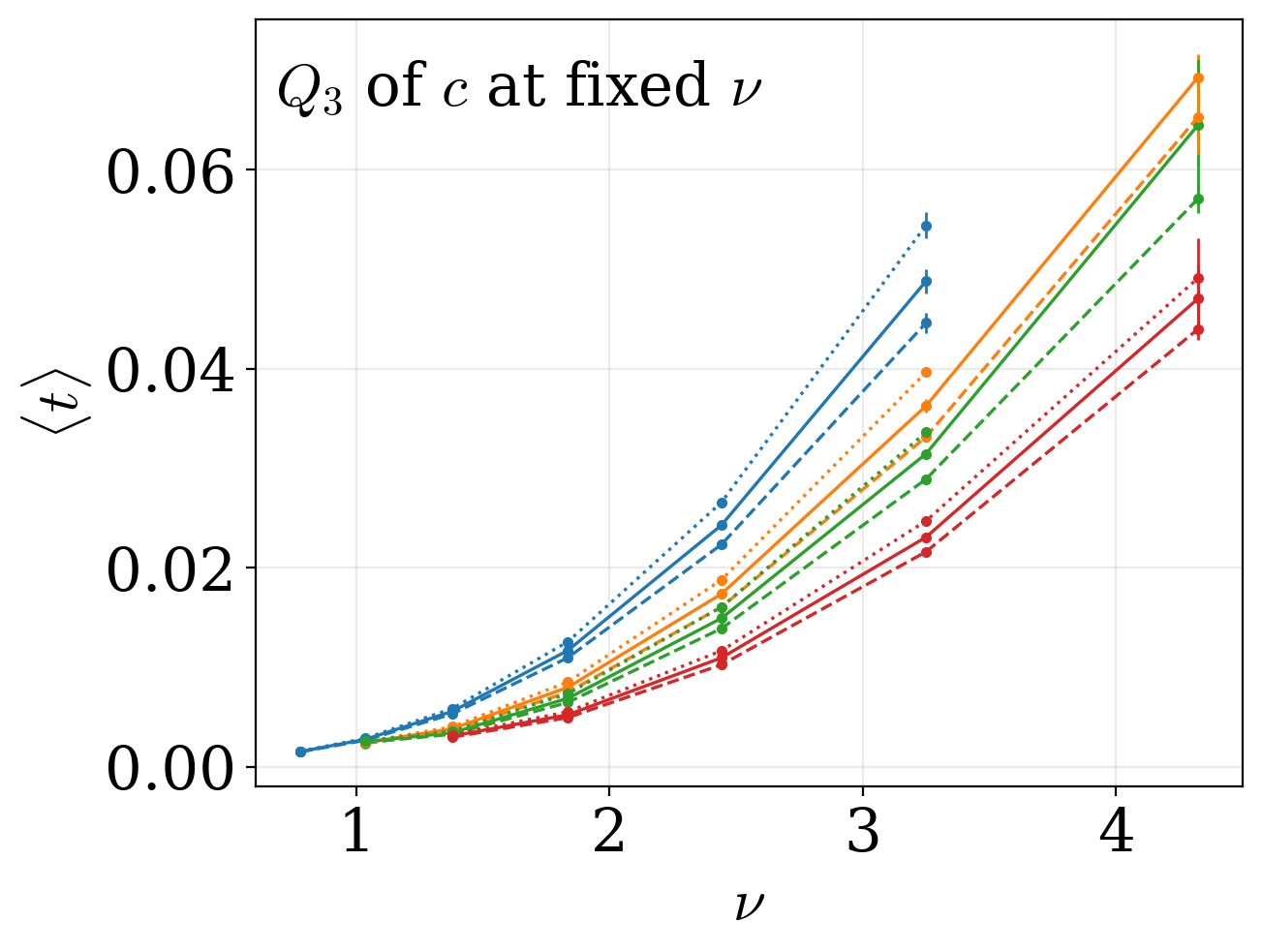}\hfill
  \includegraphics[width=0.49\linewidth]{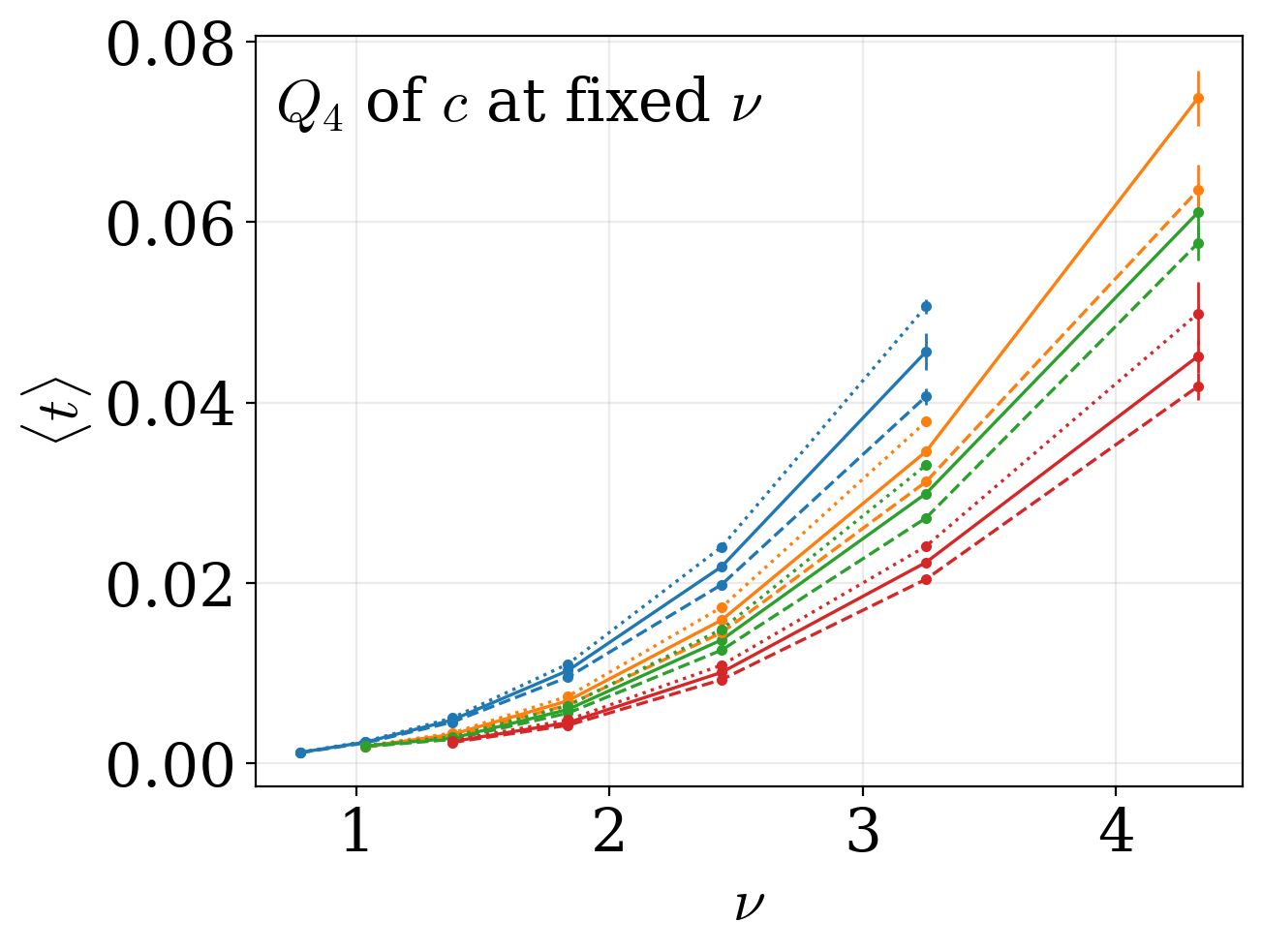}
  \caption{The mean halo size proxy $\<t\>$ as a function of the peak height $\nu$, after splitting halos into four quantiles of concentration $c$ at fixed values of $\nu$. In clockwise direction, from the top left panel: the first ($Q_1$) to the fourth ($Q_4$) quartile in increasing $c$. Line styles indicate cosmologies with different $\fnl$ and colors indicate redshift snapshots.}
  \label{fig:t_nu_cquant}
\end{figure}

To illustrate this point, in \cref{fig:t_nu_cquant}, we show the mean halo size proxy $\langle t\rangle$ as a function of the peak height $\nu$ after further splitting halos into concentration quartiles at a fixed $\nu$.
\cref{fig:t_nu_cquant} shows that $\nu$ provides the dominant ordering variable for the size proxy: within each concentration quartile, $\langle t\rangle$ varies smoothly and monotonically with $\nu$, and the overall shape of the trends remains similar across quartiles. At the same time, the four panels are not identical. Splitting by concentration shifts the normalization of the curves at a fixed $\nu$, indicating that the concentration retains a secondary but visible influence on the mean size. This is precisely the behavior expected from the chain-rule expansion discussed in \cref{sec:formalism:zerobias_SU}: the leading dependence of size on peak height gives rise to an approximately one-parameter relation between $b^{\rm s}_1$ and $b^{\rm s}_\phi$, while additional internal variables such as concentration generate departures from exact universality and modify the effective coefficients of the linear fit.

These results also connect directly to the discussion in \cref{sec:formalism:zerobias_SU} of why the Gaussian size bias is small, whereas the non-Gaussian one is not. The weak peak-height ordering observed in \cref{fig:t_nu_cquant} explains why $b^{\rm s}_1$ and $b^{\rm s}_\phi$ remain approximately linearly related. However, the residual concentration dependence at a fixed $\nu$ shows that additional internal variables contribute non negligibly to the chain-rule expansion of the size responses. As demonstrated explicitly in \cref{app:size_concentration}, these extra contributions are small and can partially cancel for the Gaussian response, whereas for the PNG response they tend to add coherently. Thus, the approximate $b^{\rm s}_\phi$--$b^{\rm s}_1$ linearity and the hierarchy $|b^{\rm s}_1|\ll |b^{\rm s}_\phi|$ are not in tension: the former reflects the dominant role of $\nu$, while the latter reflects how secondary internal variables modify the two responses differently.

\cref{fig:b1-bphi_relations,fig:t_nu_cquant} together support the following interpretation. For galaxy numbers, the approximate universality relation reflects the near one-parameter dependence of the abundance on $\nu$. For galaxy sizes, the analogous linear relation should instead be viewed as an effective approximate or weak universality relation, where the peak height captures the leading behavior, but additional halo properties remain relevant. In this sense, the empirical relation \cref{eq:ext_universality} provides a useful summary of the measured size responses without implying that the size field obeys the same strict universality arguments as the number counts.
This motivates the next two appendices: in \cref{app:size_proxy}, we clarify how our fiducial size proxy relates to other standard halo size measures; in \cref{app:size_concentration}, we isolate concentration as one explicit secondary variable in the chain-rule expansion of the size responses.

\section{Halo size proxies}
\label{app:size_proxy}

In the main study, we have adopted the trace of the halo inertia tensor as our definition of halo ``size''. In this appendix, we quantify the Pearson correlations between the trace of the halo inertia tensor and other halo properties that can be considered as alternative size proxies.

We consider the following additional halo properties:
\begin{itemize}

\item Virial radius $R_{\rm vir}$.
The virial radius is defined as the radius enclosing a virial overdensity,
\begin{equation}
M_{\rm vir} \equiv \frac{4\pi}{3}\,\Delta_{\rm vir}(z)\,\rho_{\rm crit}(z)\,R_{\rm vir}^3,
\end{equation}
where $\rho_{\rm crit}$ is the critical density and the virial threshold $\Delta_{\rm vir}$ is defined according to Ref.~\cite{Bryan:1997dn}. We use the \texttt{Rvir} values in the \code{MPI-Rockstar} catalogs.

\item Radius of maximum circular velocity $r_{v_{\max}}$.
Given the halo circular velocity profile
\begin{equation}
v_c(r) \equiv \sqrt{\frac{G\,M(<r)}{r}},
\end{equation}
we define $r_{v_{\max}}$ as the radius where $v_c(r)$ attains its maximum, i.e.
\begin{equation}
r_{v_{\max}} \equiv \arg\max_r\, v_c(r),
\end{equation}
and we use the \texttt{rvmax} values in the \code{MPI-Rockstar} catalogs.

\item Scale radius $r_s$ and concentration $c$.
For the Navarro--Frenk--White (NFW) profile of the form
\begin{equation}
\rho_{\rm NFW}(r)=\frac{\rho_s}{(r/r_s)\,(1+r/r_s)^2},
\end{equation}
where $\rho_s$ and $r_s$ are the characteristic density and scale radius of the halo, respectively.
The halo concentration is defined as $c=R_{\rm vir}/r_s$.
\code{Rockstar} and \code{MPI-Rockstar} follow the method in Ref.~\cite{Klypin:2010qw} to determine the concentration and scale radius from the halo virial mass $M_{\rm vir}$ and peak circular velocity $v_{\rm max}$. We adopt the \texttt{rs\_klypin} values in the \code{MPI-Rockstar} catalogs for the scale radius, from which we derive the halo concentration as
\begin{equation}
c_{\rm klypin}=R_{\rm vir}/r_{s,\rm klypin}.
\end{equation}

\item Half-mass radius $R_{1/2}$.
The half-mass radius is defined as the radius that encloses half of the halo mass. Here we define the halo mass by its virial mass, hence
\begin{equation}
M(<R_{1/2}) = \frac{1}{2}\,M_{\rm vir},
\end{equation}
and we use the \texttt{Halfmass\_Radius} values in the \code{MPI-Rockstar} catalogs.

\item Major-axis length $A$.
As \code{Rockstar} and \code{MPI-Rockstar} compute the halo second-moment tensor, they also output the principal axis vectors of the best-fit ellipsoid. We denote by $A$ the length of the major axis vector $\bm{A}$, i.e. the semimajor axis of the best-fit ellipsoid.
\end{itemize}

\begin{figure}[tb]
  \centering
  \includegraphics[width=0.49\linewidth]{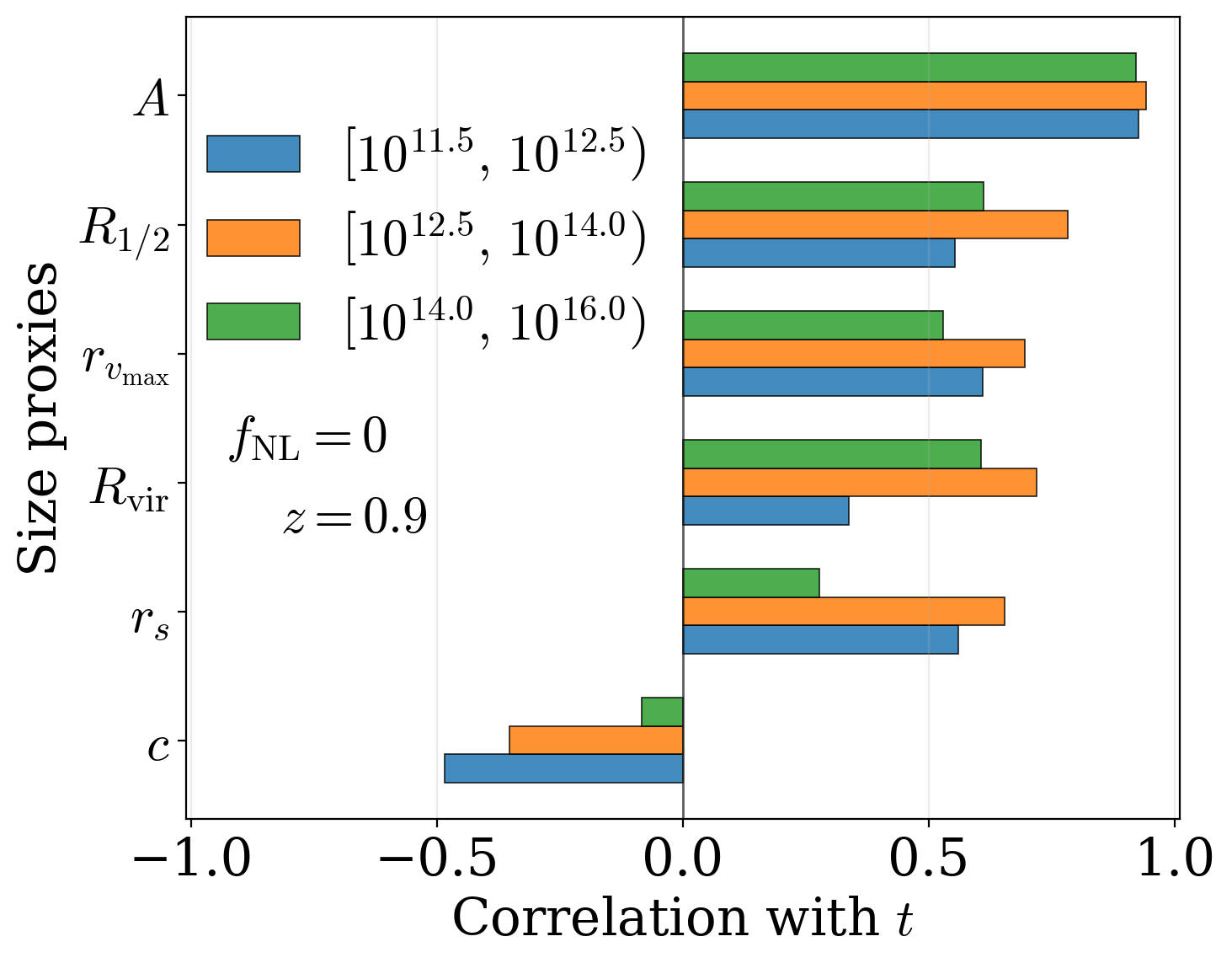}
  \includegraphics[width=0.49\linewidth]{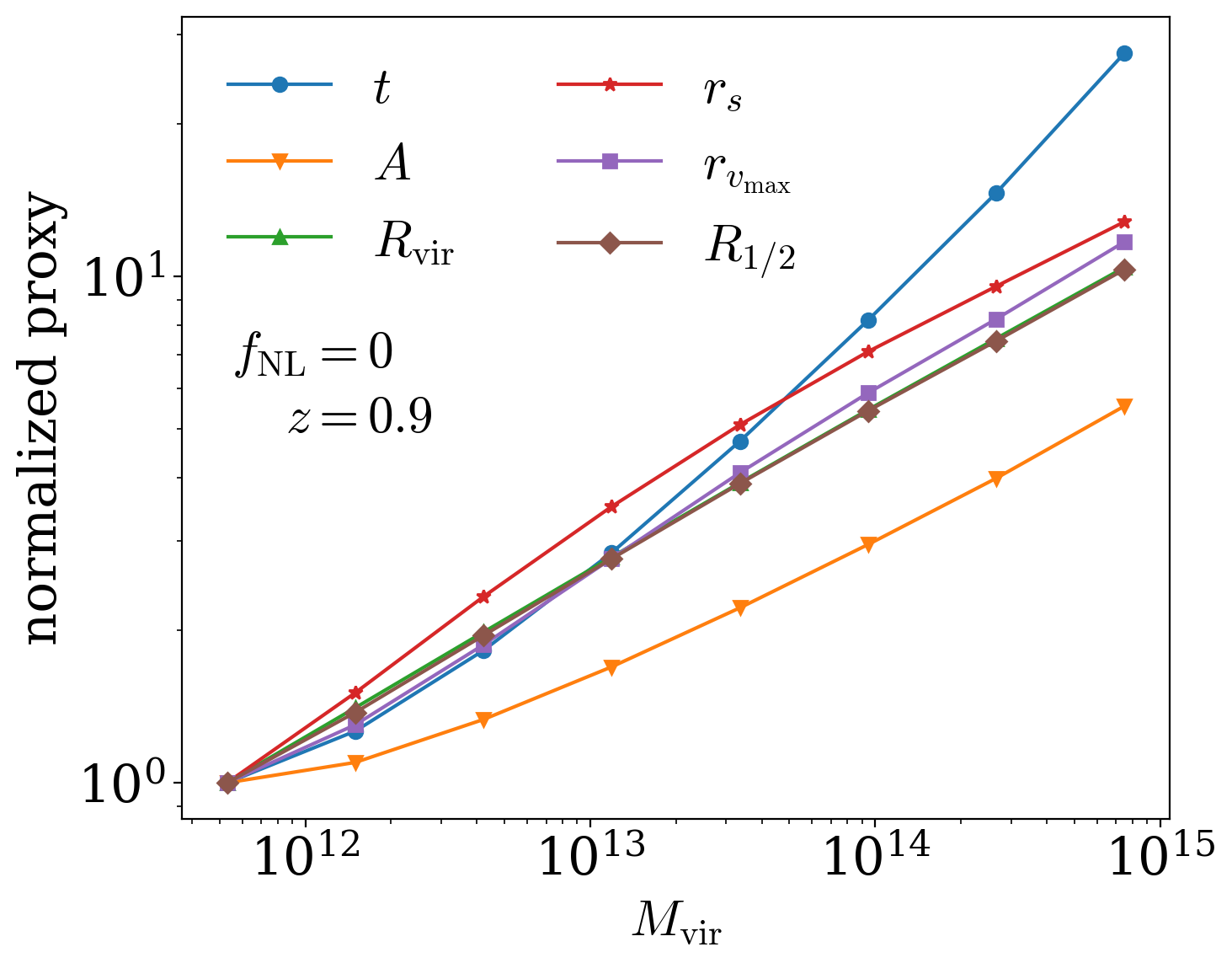}
  \caption{Left panel: Pearson correlation coefficients between different halo size proxies and the trace of the halo inertia tensor---our fiducial size proxy. We combine halos from all five realizations of the $\fnl=0$ simulations and bin them in mass. Right panel: Mean trends of the same halo size proxies as functions of halo mass for the $\fnl=0$ simulations at $z=0.9$. In both panels, the fiducial proxy is the 3D inertia-tensor trace $t$. The correlations are quantitatively different but qualitatively similar across other cosmologies ($\fnl=\pm500$) and redshifts ($z=0.0,0.7,1.3$).}
  \label{fig:size_proxy_correlation}
\end{figure}

\Cref{fig:size_proxy_correlation} shows that the trace of the halo inertia tensor is positively correlated with all radius-like size proxies and anticorrelated with halo concentration.
The strongest correlation is with the major axis length $A$, indicating that our fiducial proxy is closely related to the overall spatial extent of the halo mass distribution. The correlations with $R_{1/2}$, $r_{v_{\max}}$, and $r_s$ are also substantial, while the correlation with $R_{\rm vir}$ is weaker, particularly in the low-mass bin. This indicates that the inertia tensor trace is not simply a proxy for the virial radius. At the same time, the clear anticorrelation with concentration shows that the trace is also sensitive to the internal mass distribution.

The right panel further clarifies this. All proxies exhibit a systematic mass dependence, as expected for quantities that characterize the halo size or internal structure. However, the trends are not identical. In particular, $R_{\rm vir}$ shows the most direct geometric virial scaling with mass, whereas $r_s$, $r_{v_{\max}}$, $R_{1/2}$, and especially the inertia-tensor trace retain additional sensitivity to the internal structure of the halo. The concentration behaves opposite to the radius-like proxies. Taken together, these trends show that the inertia tensor trace cannot be reduced to a trivial rescaling of any single conventional size proxy.
We have verified that both the correlations and trends are qualitatively similar across different cosmologies and redshift snapshots considered in this work.

We thus conclude that the inertia tensor trace captures a broader notion of halo size, combining information about the halo extent, shape, and internal mass distribution. Therefore, we adopted the inertia tensor as our fiducial size proxy throughout this work, as it is strongly correlated with standard geometric size measures, yet remains sensitive to internal structure beyond what is captured by a simple virial radius description. The anticorrelation with concentration further motivates the decomposition presented in \cref{app:size_concentration}, where we explicitly test how much of the halo size response can be understood as mediation by halo concentration.

\section{Halo size and concentration}
\label{app:size_concentration}

As shown in \cref{app:size_proxy}, our fiducial size proxy is anticorrelated with halo concentration. Moreover, Ref.~\cite{Lazeyras:2022koc} \citep[see also, e.g.,][]{Reid:2010vc} showed that the halo non-Gaussian bias at fixed mass is correlated with halo concentration. Halo concentration therefore provides a natural first example of an additional variable, beyond the peak height, in the chain-rule expansion of the size responses in \cref{eq:bs_1_chain_X,eq:bs_phi_chain_X}. Since concentration itself responds to the long-wavelength perturbations $\d_L$ and $\phi_L$, it can mediate part of the halo size response. In this appendix, we quantify the extent to which the Gaussian and non-Gaussian size biases are inherited through concentration, and how much remains as a residual response at fixed concentration. This provides an empirical complement to the argument in \cref{sec:formalism:zerobias_SU}, and in particular helps explain why $|b^{\rm s}_1|\ll |b^{\rm s}_\phi|$.

To this end, we assume the NFW halo profile \citep{Navarro:1996gj} and define the halo concentration as
$c\equiv R_{\rm vir}/r_s$, where $R_{\rm vir}$ and $r_s$ are the halo virial and
scale radii\footnote{As for halo radius, there is no unique definition for neither the halo scale radius nor concentration. Here we adopt the Klypin et al.\ estimator \citep{Klypin:2010qw}, which yields a smaller scatter in our fits. We have verified that the qualitative conclusions below are unchanged for alternative concentration definitions.}.
In analogy with our size-density estimator, we define a concentration density perturbation field as
\begin{equation}
\d_c(\vx)=\frac{C(\vx)}{\overline{C}}-\frac{n(\vx)}{\nbar},
\qquad
C(\vx)\equiv \sum_i c_i\,\ddelta(\vx-\vx_i),
\qquad
\overline{C}=\ngbar\<c\>,
\label{eq:deltac_def}
\end{equation}
i.e.\ by replacing the per-object size proxy $t_i$ in $\ds$ by the per-object concentration $c_i$. With this definition, the corresponding Gaussian and non-Gaussian concentration biases can be interpreted as responses of the mean concentration at fixed mass,
\begin{equation}
b^{\rm c}_1(M,z)=\frac{\partial\ln\<c\>(M,z)}{\partial \d_L},\qquad
b^{\rm c}_\phi(M,z)=\frac{\partial\ln\<c\>(M,z)}{\partial\fnl\phi_L}.
\label{eq:bc_responses}
\end{equation}
In practice, we obtain $b^{\rm c}_1$ by fitting the large-scale ratio
$P_{mc}/P_{mm}$ in the $\fnl=0$ simulations, and $b^{\rm c}_\phi$ from the phase-matched $\fnl=\pm 500$ pair, in complete analogy with our measurements of the size bias coefficients.

Treating $\<t\>$ at fixed mass and redshift as a function of concentration, the chain rule implies
\begin{align}
b^{\rm s}_1(M,z)
\equiv \frac{\partial\ln\<t\>(M,z)}{\partial\d_L}\Big|_M
&=\left.\frac{\partial\ln\<t\>}{\partial\d_L}\right|_c
+\frac{\partial\ln\<t\>}{\partial\ln c}\Big|_{M,z}\,
\frac{\partial\ln c}{\partial\d_L}\Big|_{M,z}
\nonumber\\
&=\left.\frac{\partial\ln\<t\>}{\partial\d_L}\right|_c
+\R_{t|c}(M,z)\,b^{\rm c}_1(M,z),
\label{eq:b1_chain_tc}\\
b^{\rm s}_\phi(M,z)
\equiv \frac{\partial\ln\<t\>(M,z)}{\partial\fnl\phi_L}\Big|_M
&=\left.\frac{\partial\ln\<t\>}{\partial\fnl\phi_L}\right|_c
+\frac{\partial\ln\<t\>}{\partial\ln c}\Big|_{M,z}\,
\frac{\partial\ln c}{\partial\fnl\phi_L}\Big|_{M,z}
\nonumber\\
&=\left.\frac{\partial\ln\<t\>}{\partial\fnl\phi_L}\right|_c
+\R_{t|c}(M,z)\,b^{\rm c}_\phi(M,z),
\label{eq:bphi_chain_tc}
\end{align}
where $\R_{t|c}\equiv \partial\ln\<t\>/\partial\ln c$. Thus, even for a strong size--concentration anticorrelation, the concentration-mediated contribution to $b^{\rm s}_1$ is suppressed when $|b^{\rm c}_1|$ is small, whereas the mediated contribution to $b^{\rm s}_\phi$ can remain sizable when $|b^{\rm c}_\phi|$ is large. This provides an empirical complement to the separate-universe intuition discussed in \cref{sec:formalism:zerobias_SU}.

\begin{figure}[tb]
  \centering
  \includegraphics[width=0.95\linewidth]{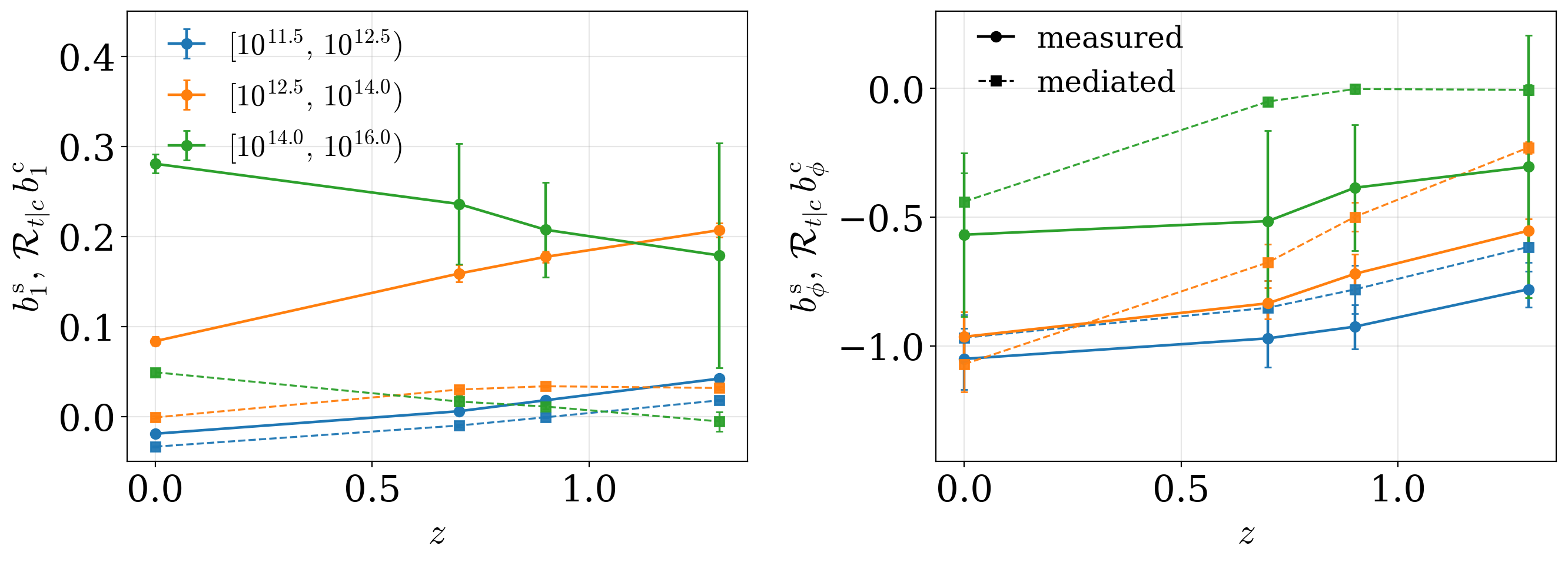}
  \caption{Comparison of the measured size responses (circle, solid) and the
  concentration-mediated contributions (square, dashed) implied by the chain-rule
  relations in \cref{eq:b1_chain_tc,eq:bphi_chain_tc}, as a function of redshift
  for the fiducial halo mass bins in the main text. Left panel: Gaussian size
  response. Right panel: non-Gaussian size response. The concentration-mediated
  contribution is generally subdominant for $b^{\rm s}_1$, but can account for a
  substantial fraction of $b^{\rm s}_\phi$ in some mass bins and redshifts.}
  \label{fig:bias_decomp_tc}
\end{figure}

In \cref{fig:bias_decomp_tc}, we compare the measured size responses $(b^{\rm s}_1,b^{\rm s}_\phi)$ to the concentration-mediated contributions $\R_{t|c}b^{\rm c}_1$ and $\R_{t|c}b^{\rm c}_\phi$ implied by \cref{eq:b1_chain_tc,eq:bphi_chain_tc}.
We obtain $\R_{t|c}$ by binning the halo inertia-tensor trace $t$ at fixed halo mass $M$ and redshift $z$ in halo concentration $c$ on logarithmic scales, and fitting the slope of $\ln\<t\>$ as a function of $\ln c$ via weighted least squares, assuming a power-law relation $\<t\>\propto c^{\R_{t|c}}$.
The weights are taken as the inverse variance of $\ln\<t\>$ in each bin, estimated as $(\sigma_i/\<t\>_i)^2/N_i$, where $\sigma_i$ and $N_i$ are the standard deviation and halo count in bin $i$; bins with $N_i < 50$ are excluded from the fit.
The comparison suggests two qualitative conclusions.
First, the concentration-mediated contribution is generally small for $b^{\rm s}_1$, but can be sizable for $b^{\rm s}_\phi$, consistent with the picture that internal halo properties respond much more weakly to $\d_L$ than to $\fnl\phi_L$.
Second, the residual fixed-$c$ contribution, inferred as the
difference between the measured total response and the mediated term, is bin dependent: for $b^{\rm s}_1$ the mediated term is typically subdominant, whereas for $b^{\rm s}_\phi$ it can account for a substantial fraction of the measured response in some mass bins and redshifts. Therefore, while the halo size
proxy defined by the trace of \cref{eq:Iij_g} with $w_p=r_p^{-2}$ exhibits some anticorrelation with halo concentration, concentration is not the sole secondary bias variable: the nonzero, mass-bin-dependent residual at fixed $c$ points toward a higher-dimensional assembly bias description.

\section{Halo trace definitions}
\label{app:trace_definition}

Throughout, our fiducial choice for evaluating halo traces and sizes is to use their reduced inertia tensors, i.e. to evaluate \cref{eq:Iij_g} with $w_p\propto r_p^{-2}$.
Here, for completeness, we report the Gaussian and non-Gaussian bias coefficients for the halo sizes defined by the simple inertia tensor, corresponding to \cref{eq:Iij_g} with $w_p=1$, in \cref{tab:bias_values_simple_tensor}.

The bias coefficients are expected to vary according to the size definition and estimator. More importantly, as defined by the simple inertia tensor, halo sizes are more correlated to the physical extensions of the halos, e.g., the halo virial radius $R_{\mathrm{vir}}$, and are also more noisy.
\begin{table}[t]
\centering
\caption{Similar to \cref{tab:bias_values} but with halo sizes defined by the trace of the simple (unreduced) inertia tensor (see text).}
\label{tab:bias_values_simple_tensor}
\small
\setlength{\tabcolsep}{1pt}
\begin{tabular}{lcc|cc|cc|cc}
\toprule
Mass [$\Msunh$] & \multicolumn{2}{c}{$z=0.0$} & \multicolumn{2}{c}{$z=0.7$} & \multicolumn{2}{c}{$z=0.9$} & \multicolumn{2}{c}{$z=1.3$} \\
\cmidrule(lr){2-3}\cmidrule(lr){4-5}\cmidrule(lr){6-7}\cmidrule(lr){8-9}
$[M_{\rm min},M_{\rm max})$ & $b^{\rm s}_1$ & $b^{\rm s}_\phi$ & $b^{\rm s}_1$ & $b^{\rm s}_\phi$ & $b^{\rm s}_1$ & $b^{\rm s}_\phi$ & $b^{\rm s}_1$ & $b^{\rm s}_\phi$ \\
\midrule
$[10^{11.5}\!,\!10^{12.5})$ & $0.02$ & $-0.33$ & $0.05$ & $-0.20$ & $0.07$ & $-0.17$ & $0.09$ & $-0.10$ \\
$[10^{12.5}\!,\!10^{14.0})$ & $0.17$ & $0.14$ & $0.28$ & $0.41$ & $0.30$ & $0.49$ & $0.34$ & $0.63$ \\
$[10^{14.0}\!,\!10^{16.0})$ & $0.35$ & $0.87$ & $0.31$ & $0.82$ & $0.29$ & $0.86$ & $0.22$ & $0.72$ \\
\bottomrule
\end{tabular}
\end{table}

\section{Shape-tidal alignment bias}
\label{app:shape-tidal_bias_measurement}

To measure the tidal alignment bias coefficient $b^{\rm tf}_1$ in our simulations, we project the shape 3D trace-free tensor field in \cref{eq:dtf_def} onto the $\ell=2$ spherical-tensor basis and fit the 3D cross-power spectrum between the matter density field and its $m=0$ component.

Specifically, following Ref.~\cite{Vlah:2019byq,Vlah:2020ovg}, we decompose the 3D trace-free shape tensor $\dtf_{ij}$ into a spherical tensor basis of $Y_{\ell=2,ij}^m$ as
\begin{equation}
    \dtf_{ij}(\vk)=\sum_{m=-2}^{2}\dtfm_{\ell=2}(\vk)Y_{\ell=2,ij}^m(\vk)\label{eq:dtfij_spherical_decomp}
\end{equation}
where the multipole order $\ell=2$ follows the fact that $\dtf_{ij}$ is a spin-2 field and each of the $\dtfm_{\ell=2}(\vk)$ is a scalar field defined by the contraction
\begin{equation}
    \dtfm_{\ell=2}(\vk)=\sum_{i,j}Y_{\ell=2,ij}^m(\vk)\dtf_{ij}(\vk)\label{eq:dtflm_spherical_decomp}.
\end{equation}
In particular, the $m=0$ component is given by
\begin{equation}
    \gamma^0_2(\vk)\equiv\left(\dtf\right)^{m=0}_{\ell=2}(\vk)=\sum_{i,j}Y_{\ell=2,ij}^{m=0}\dtf_{ij}(\vk)\label{eq:dtfm0_spherical_decomp}
\end{equation}
where $Y_{\ell=2,ij}^{m=0}=\sqrt{3/2}\mathcal{D}_{ij}$.
Using \cref{eq:gaussian_linear_bias_spin2,eq:Kij}, we rewrite \cref{eq:dtfm0_spherical_decomp} as
\begin{align*}
    \gamma^0_2(\vk)&=\sum_{i,j}\sqrt{\frac{3}{2}}\mathcal{D}_{ij}\left[b^{\rm tf}_1\mathcal{D}_{ij}\d(\vk)+\eps^{\rm tf}_{ij}(\vk)\right]\\
    &=\sqrt{\frac{3}{2}}\left[b^{\rm tf}_1\sum_{i,j}\mathcal{D}_{ij}\mathcal{D}_{ij}+\sum_{i,j}\mathcal{D}_{ij}\eps^{\rm tf}_{ij}\right]=\sqrt{\frac{2}{3}}b^{\rm tf}_1\d(\vk)+\eps_{\gamma^0_2}(\vk)\numberthis\label{eq:dflm0_exp},
\end{align*}
which implies the leading-order relation $P_{\gamma^0_2m}=\sqrt{2/3}b^{\rm tf}_1P_{\rm mm}$.

Similar to the cases of $b^{\rm g,s}_1$, we quantify the systematic uncertainties of $b^{\rm tf}_1$ by the shifts of their best-fit values when including the higher-derivative bias coefficient $b^{\rm tf}_{\lapl\d}$ and counterterm: 
\begin{equation}
  P_{\gamma^0_2m}(k)=\sqrt{2/3}\left(b^{\rm tf}_1+b^{\rm tf}_{\lapl\d}k^2\right)P_{\rm mm}(k)\label{eq:tree_Pgm}.
\end{equation}

\section{Galaxy number--shape power spectrum multipoles}
\label{app:shape_multipoles}

As discussed in \cref{sec:formalism:observations:projection}, line-of-sight (LOS) projection generically mixes the information carried by galaxy shapes and sizes in the projected observables. This suggests some relation between the galaxy number--shape multipoles and the galaxy number--size multipoles.

In this appendix, we derive the galaxy number--shape E-mode power spectrum multipoles and their relation with the galaxy number--size quadrupole at leading order. We then discuss how the projected trace perturbation $\d^{\rm tr2D}$, projected size perturbation $\delta^{\rm s2D}$, and projected spin-2 field $\gamma_E$ differ at the estimator level, and how this difference affects the measured multipoles.

We decompose the projected trace-free tensor field on the sky plane as
\begin{equation}
    \delta^{\rm tf2D}_{ab}(\vk)=\gamma^E(\vk)\,e^E_{ab}(\hat{\vk}^{\rm 2D})+\gamma^B(\vk)\,e^B_{ab}(\hat{\vk}^{\rm 2D}),
    \label{eq:EB-mode_decomp}
\end{equation}
where \(a,b\in\{x,y\}\), \(\hat{\vk}^{\rm 2D}\equiv \vk_\perp/|\vk_\perp|\), and
\begin{equation}
    e^E_{ab}(\hat{\vk}^{\rm 2D})\equiv \hat{k}^{\rm 2D}_a\hat{k}^{\rm 2D}_b-\frac{1}{2}\delta_{ab}.
    \label{eq:E-mode_basis}
\end{equation}
Here
\begin{equation}
    \hat{\vk}^{\rm 2D}=\frac{\vk_\perp}{|\vk_\perp|}=\frac{(\hat{k}_x,\hat{k}_y)}{\sqrt{1-\mu^2}},
\end{equation}
with \(\mu\equiv \hat{\vk}\cdot\hat{\mathbf z}\) and \(|\vk_\perp|=k\sqrt{1-\mu^2}\). We choose the normalization \(e^E_{ab}e^E_{ab}=1/2\), such that
\begin{equation}
    \gamma^E(\vk)=2\,\delta^{\rm tf2D}_{ab}(\vk)\,e^E_{ab}(\hat{\vk}^{\rm 2D}).
\end{equation}
For completeness, the parity-odd B-mode basis tensor may be chosen as
\begin{equation}
    e^B_{ab}(\hat{\vk}^{\rm 2D}) \equiv \frac{1}{2}\left(\epsilon_{ac}\hat{k}^{\rm 2D}_c\hat{k}^{\rm 2D}_b+\epsilon_{bc}\hat{k}^{\rm 2D}_c\hat{k}^{\rm 2D}_a\right),
\end{equation}
where \(\epsilon_{ab}\) is the 2D Levi-Civita tensor in the sky plane. In the Gaussian linear alignment model considered here, \(\gamma^B\) vanishes at the leading order.

Comparing \cref{eq:EB-mode_decomp,eq:E-mode_basis} with \cref{eq:gaussian_linear_bias_2Dprojected_spin2}, we obtain
\begin{equation}
    \gamma^E(\vk)=\frac{3}{2}\,b^{\rm tf}_1\left(1-\mu^2\right)\delta(\vk).
    \label{eq:gamma_E}
\end{equation}
The galaxy number--shape cross spectrum is therefore
\begin{equation}
    P_{\rm g\gamma^E}(k,\mu)=\frac{3}{2}\,b^{\rm g}_1 b^{\rm tf}_1 (1-\mu^2)\,P_{\rm mm}(k),
    \label{eq:PgE_mu}
\end{equation}
which yields the nonvanishing real-space multipoles
\begin{align}
P^{\ell=0}_{\rm g\gamma^E}(k)&=b^{\rm g}_1 b^{\rm tf}_1 P_{\rm mm}(k), \label{eq:PgE_ell0}\\
P^{\ell=2}_{\rm g\gamma^E}(k)&=-b^{\rm g}_1 b^{\rm tf}_1 P_{\rm mm}(k), \label{eq:PgE_ell2}
\end{align}
and
\begin{equation}
    P^{\ell=4}_{\rm g\gamma^E}(k)=0.
    \label{eq:PgE_ell4}
\end{equation}
Comparing \cref{eq:Pgs_ell2} with \cref{eq:PgE_ell0,eq:PgE_ell2}, we find
\begin{equation}
    P^{\ell=2}_{\rm gs2D}(k)=P^{\ell=2}_{\rm g\gamma^E}(k)=-P^{\ell=0}_{\rm g\gamma^E}(k)
    \label{eq:PgE_Pgs_relation}
\end{equation}
in the galaxy rest frame, i.e. in the absence of RSD.

For completeness, in redshift space, the monopole becomes
\begin{equation}
P^{\ell=0}_{\rm g\gamma^E}(k)=
\left(
b^{\rm g}_1 b^{\rm tf}_1+\frac{1}{5}b^{\rm tf}_1 f
\right)P_{\rm mm}(k).
\label{eq:PSgE_ell0}
\end{equation}
The higher redshift-space multipoles follow analogously from the \(\mu\)-dependence of the redshift-space kernel.

\begin{figure}[tb]
  \centering
  \includegraphics[width=0.49\linewidth]{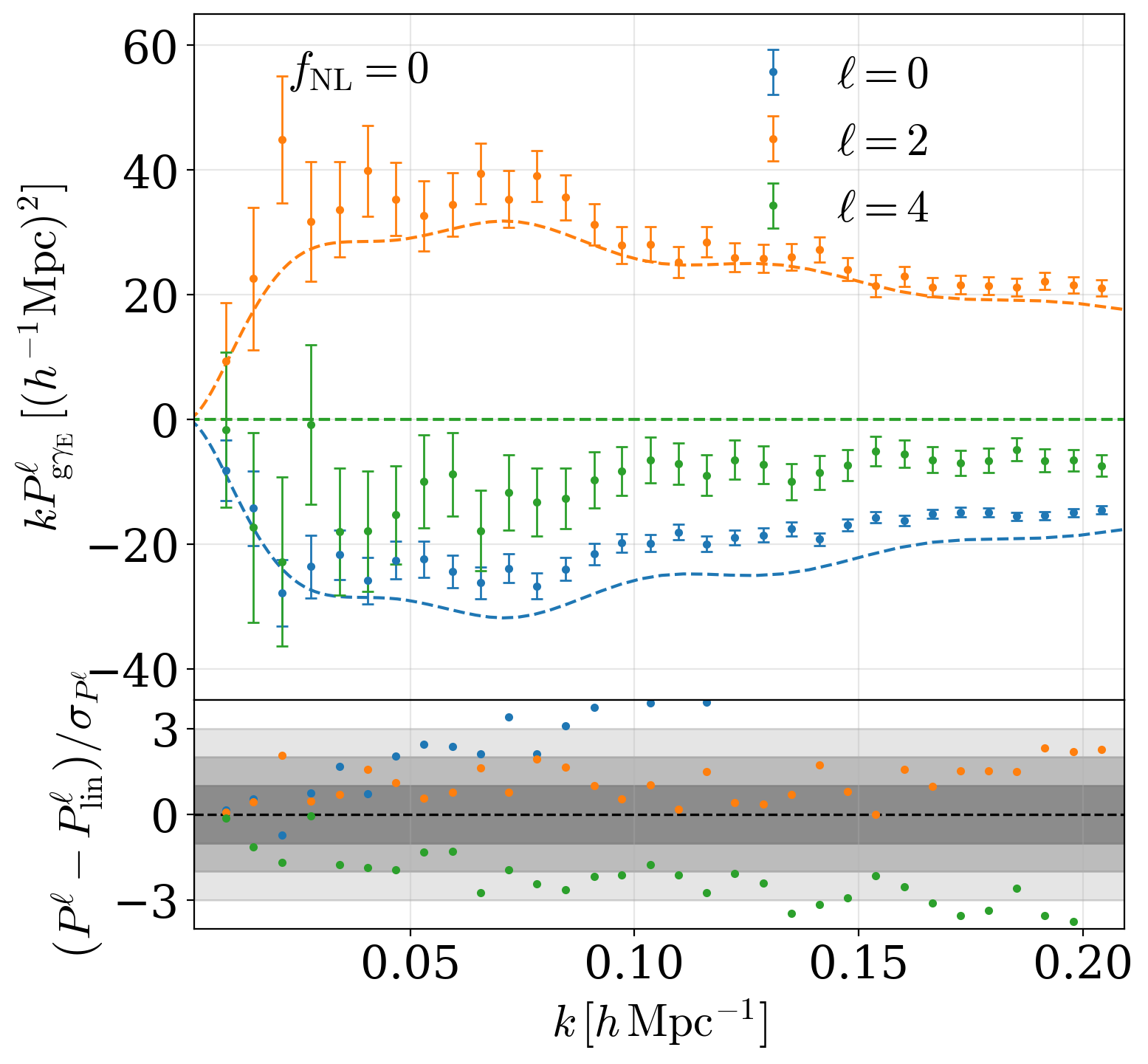}
  \includegraphics[width=0.49\linewidth]{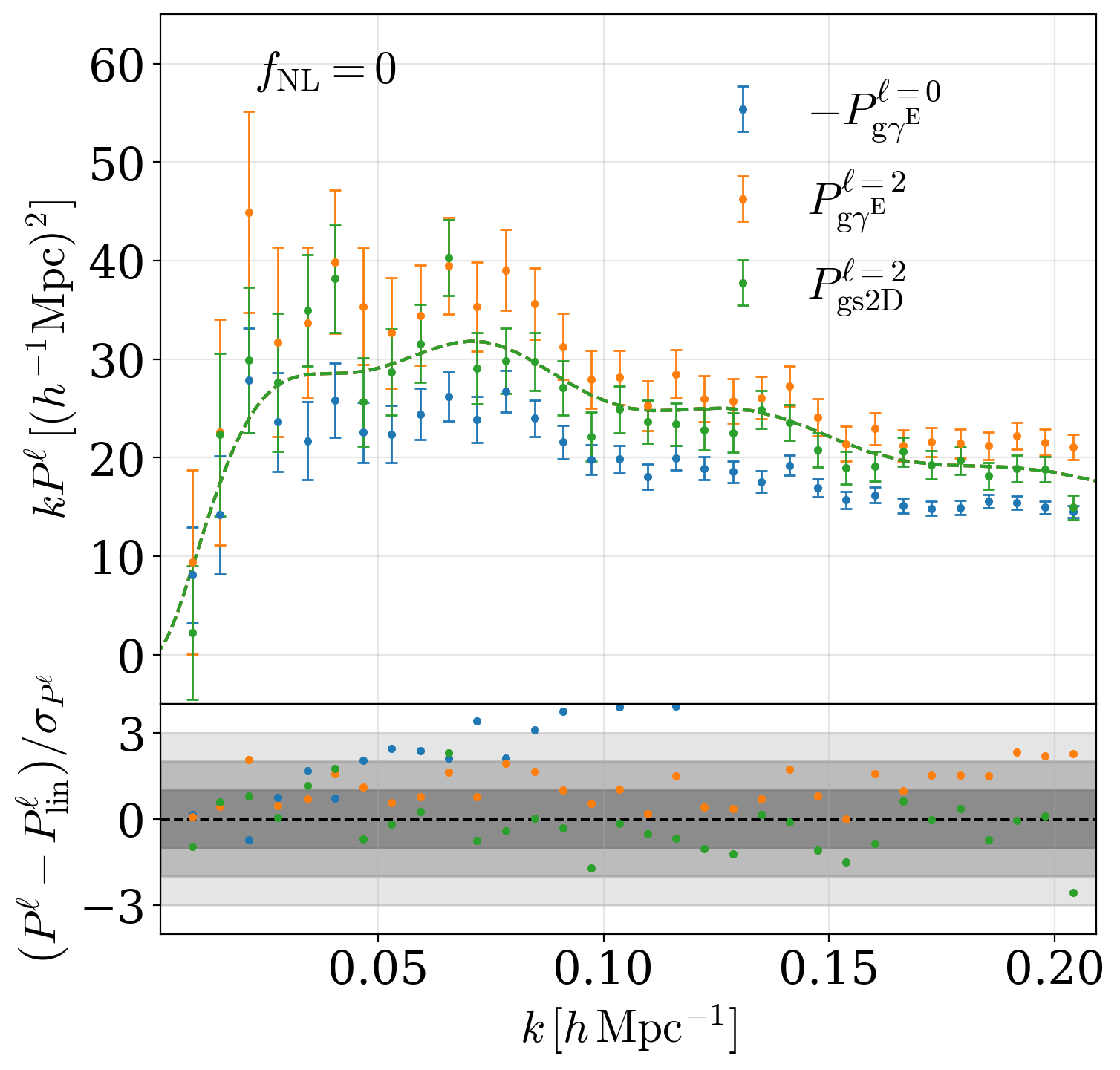}
  \caption{Left panel: The galaxy number--shape E-mode cross power spectrum multipoles for $M_h\in[10^{11.5},\,10^{12.5})\,\Msunh$ at $z=0.9$. Dashed curves indicate the best-fit theoretical predictions. Points indicate measurements from a simulation with $\fnl=0$. Error bars indicate the $1\sigma$ estimated by the Gaussian covariance. The subpanel shows the data-theory misfits relative to the Gaussian $1\sigma$. Bands indicate the $1\sigma$-$2\sigma$-$3\sigma$ ranges. Right panel: Comparison of the galaxy number--shape E-mode cross power spectrum monopole ($-P^{\ell=0}_{\rm g\gamma^E}$, blue), quadrupole ($P^{\ell=2}_{\rm g\gamma^E}$, orange) and galaxy--size cross power spectrum quadrupole ($P^{\ell=2}_{\rm gs2D}$, green). All theoretical prediction curves lie on top of each other.}
  \label{fig:PkgE_ell02_Pkgs_ell0_fnl0}
\end{figure}
In the left panel of \cref{fig:PkgE_ell02_Pkgs_ell0_fnl0}, we plot the galaxy number--shape E-mode cross power spectrum multipoles $P^{\ell}_{\rm g\gamma_E}$ and the corresponding leading-order predictions. The measurements deviate from the predictions in two ways that are consistent with the loop corrections identified by Ref.~\cite{Bakx:2023mld}: a mismatch between the monopole and quadrupole absolute amplitudes (cf.~\cref{eq:PgE_ell0,eq:PgE_ell2}), and a nonvanishing hexadecapole (cf.~\cref{eq:PgE_ell4}). Both features share a common origin: LOS-dependent loop corrections arising from the normalization of the projected shape field. The projection along the LOS picks out a preferred direction along which trace information is removed. As the projected trace enters the normalization of the shape field, it introduces additional angular dependencies that couple to the underlying large-scale tidal field. We refer readers to Sec.~2 and App.~B of Ref.~\cite{Bakx:2023mld} for a detailed discussion.

Ref.~\cite{Bakx:2023mld} suggested that these contributions are specific to object-by-object normalization, and that a sample-wise normalization using the sample-mean projected trace would avoid such complications. We find, however, that our sample-wise normalization (cf.~\cref{eq:dtf2D_est}) does not eliminate them. A sample-wise normalization replaces the per-object nonlinearity by a scalar modulation,
\begin{equation}
\gamma^{\langle t^{2D}\rangle}_{ab}=(1+\d^{\mathrm{tr2D}})\,\gamma^{t^{2D}}_{ab},
\end{equation}
where $\d^{\mathrm{tr2D}}$ acts as a modulation field for $\gamma_E$, generating a distinct but analogous set of mode-coupling corrections.

With this caveat in mind, we test the relation in \cref{eq:PgE_Pgs_relation} in the right panel of \cref{fig:PkgE_ell02_Pkgs_ell0_fnl0}. Since the galaxy number--shape E-mode monopole $P^{\ell=0}_{\rm g\gamma_E}$ is more strongly affected by the normalization-dependent corrections, we focus on the comparison between $P^{\ell=2}_{\rm g\gamma_E}$ and $P^{\ell=2}_{\rm gs2D}$. The two quadrupoles are in good statistical agreement over the $k$ range shown. We note, however, that this level of agreement may not generalize across estimators and normalization choices, as the galaxy number--shape E-mode multipoles are sensitive to these specifics.

Taken together, the three measurements therefore support the consistency relations in \cref{eq:PgE_Pgs_relation} only approximately, as these relations are sensitive to the details of the projected-shape estimator and normalization. We leave a more detailed investigation of these effects for future work.

\bibliographystyle{apsrev4-2}
\bibliography{references}

@article{Planck:2018vyg,
    author = "Aghanim, N. and others",
    collaboration = "Planck",
    title = "{Planck 2018 results. VI. Cosmological parameters}",
    eprint = "1807.06209",
    archivePrefix = "arXiv",
    primaryClass = "astro-ph.CO",
    doi = "10.1051/0004-6361/201833910",
    journal = "Astron. Astrophys.",
    volume = "641",
    pages = "A6",
    year = "2020",
    note = "[Erratum: Astron.Astrophys. 652, C4 (2021)]"
}

@article{Press:1973iz,
    author = "Press, William H. and Schechter, Paul",
    title = "{Formation of galaxies and clusters of galaxies by selfsimilar gravitational condensation}",
    doi = "10.1086/152650",
    journal = "Astrophys. J.",
    volume = "187",
    pages = "425--438",
    year = "1974"
}

@article{Bond:1990iw,
    author = "Bond, J. R. and Cole, S. and Efstathiou, G. and Kaiser, Nick",
    title = "{Excursion set mass functions for hierarchical Gaussian fluctuations}",
    reportNumber = "CFPA-TH-90-015",
    doi = "10.1086/170520",
    journal = "Astrophys. J.",
    volume = "379",
    pages = "440",
    year = "1991"
}

@article{Zentner:2006vw,
    author = "Zentner, Andrew R.",
    title = "{The Excursion Set Theory of Halo Mass Functions, Halo Clustering, and Halo Growth}",
    eprint = "astro-ph/0611454",
    archivePrefix = "arXiv",
    doi = "10.1142/S0218271807010511",
    journal = "Int. J. Mod. Phys. D",
    volume = "16",
    pages = "763--816",
    year = "2007"
}

@article{Desjacques:2016bnm,
    author = "Desjacques, Vincent and Jeong, Donghui and Schmidt, Fabian",
    title = "{Large-Scale Galaxy Bias}",
    eprint = "1611.09787",
    archivePrefix = "arXiv",
    primaryClass = "astro-ph.CO",
    doi = "10.1016/j.physrep.2017.12.002",
    journal = "Phys. Rept.",
    volume = "733",
    pages = "1--193",
    year = "2018"
}

@article{McDonald:2006mx,
    author = "McDonald, Patrick",
    title = "{Clustering of dark matter tracers: Renormalizing the bias parameters}",
    eprint = "astro-ph/0609413",
    archivePrefix = "arXiv",
    doi = "10.1103/PhysRevD.74.129901",
    journal = "Phys. Rev. D",
    volume = "74",
    pages = "103512",
    year = "2006",
    note = "[Erratum: Phys.Rev.D 74, 129901 (2006)]"
}

@article{McDonald:2009dh,
    author = "McDonald, Patrick and Roy, Arabindo",
    title = "{Clustering of dark matter tracers: generalizing bias for the coming era of precision LSS}",
    eprint = "0902.0991",
    archivePrefix = "arXiv",
    primaryClass = "astro-ph.CO",
    doi = "10.1088/1475-7516/2009/08/020",
    journal = "JCAP",
    volume = "08",
    pages = "020",
    year = "2009"
}

@article{Lazeyras:2015lgp,
    author = "Lazeyras, Titouan and Wagner, Christian and Baldauf, Tobias and Schmidt, Fabian",
    title = "{Precision measurement of the local bias of dark matter halos}",
    eprint = "1511.01096",
    archivePrefix = "arXiv",
    primaryClass = "astro-ph.CO",
    doi = "10.1088/1475-7516/2016/02/018",
    journal = "JCAP",
    volume = "02",
    pages = "018",
    year = "2016"
}

@article{Schmittfull:2018yuk,
    author = "Schmittfull, Marcel and Simonovi{\'c}, Marko and Assassi, Valentin and Zaldarriaga, Matias",
    title = "{Modeling Biased Tracers at the Field Level}",
    eprint = "1811.10640",
    archivePrefix = "arXiv",
    primaryClass = "astro-ph.CO",
    doi = "10.1103/PhysRevD.100.043514",
    journal = "Phys. Rev. D",
    volume = "100",
    number = "4",
    pages = "043514",
    year = "2019"
}

@article{Cabass:2023nyo,
    author = "Cabass, Giovanni and Simonovi{\'c}, Marko and Zaldarriaga, Matias",
    title = "{Cosmological information in perturbative forward modeling}",
    eprint = "2307.04706",
    archivePrefix = "arXiv",
    primaryClass = "astro-ph.CO",
    reportNumber = "CERN-TH-2023-130",
    doi = "10.1103/PhysRevD.109.043526",
    journal = "Phys. Rev. D",
    volume = "109",
    number = "4",
    pages = "043526",
    year = "2024"
}

@article{Foreman:2024kzw,
    author = "Foreman, Simon and Obuljen, Andrej and Simonovi{\'c}, Marko",
    title = "{Improving cosmological analyses of HI clustering by reducing stochastic noise}",
    eprint = "2405.18559",
    archivePrefix = "arXiv",
    primaryClass = "astro-ph.CO",
    doi = "10.1103/PhysRevD.110.063555",
    journal = "Phys. Rev. D",
    volume = "110",
    number = "6",
    pages = "063555",
    year = "2024"
}

@article{Dalal:2007cu,
    author = "Dalal, Neal and Dore, Olivier and Huterer, Dragan and Shirokov, Alexander",
    title = "{The imprints of primordial non-gaussianities on large-scale structure: scale dependent bias and abundance of virialized objects}",
    eprint = "0710.4560",
    archivePrefix = "arXiv",
    primaryClass = "astro-ph",
    doi = "10.1103/PhysRevD.77.123514",
    journal = "Phys. Rev. D",
    volume = "77",
    pages = "123514",
    year = "2008"
}

@article{Slosar:2008hx,
    author = "Slosar, Anze and Hirata, Christopher and Seljak, Uros and Ho, Shirley and Padmanabhan, Nikhil",
    title = "{Constraints on local primordial non-Gaussianity from large scale structure}",
    eprint = "0805.3580",
    archivePrefix = "arXiv",
    primaryClass = "astro-ph",
    doi = "10.1088/1475-7516/2008/08/031",
    journal = "JCAP",
    volume = "08",
    pages = "031",
    year = "2008"
}

@article{Dalal:2008zd,
    author = "Dalal, Neal and White, Martin and Bond, J. Richard and Shirokov, Alexander",
    title = "{Halo Assembly Bias in Hierarchical Structure Formation}",
    eprint = "0803.3453",
    archivePrefix = "arXiv",
    primaryClass = "astro-ph",
    doi = "10.1086/591512",
    journal = "Astrophys. J.",
    volume = "687",
    pages = "12--21",
    year = "2008"
}

@article{Reid:2010vc,
    author = "Reid, Beth A. and Verde, Licia and Dolag, Klaus and Matarrese, Sabino and Moscardini, Lauro",
    title = "{Non-Gaussian halo assembly bias}",
    eprint = "1004.1637",
    archivePrefix = "arXiv",
    primaryClass = "astro-ph.CO",
    doi = "10.1088/1475-7516/2010/07/013",
    journal = "JCAP",
    volume = "07",
    pages = "013",
    year = "2010"
}

@article{Barreira:2020kvh,
    author = "Barreira, Alexandre and Cabass, Giovanni and Schmidt, Fabian and Pillepich, Annalisa and Nelson, Dylan",
    title = "{Galaxy bias and primordial non-Gaussianity: insights from galaxy formation simulations with IllustrisTNG}",
    eprint = "2006.09368",
    archivePrefix = "arXiv",
    primaryClass = "astro-ph.CO",
    doi = "10.1088/1475-7516/2020/12/013",
    journal = "JCAP",
    volume = "12",
    pages = "013",
    year = "2020"
}

@article{Barreira:2022sey,
    author = "Barreira, Alexandre",
    title = "{Can we actually constrain f$_{NL}$ using the scale-dependent bias effect? An illustration of the impact of galaxy bias uncertainties using the BOSS DR12 galaxy power spectrum}",
    eprint = "2205.05673",
    archivePrefix = "arXiv",
    primaryClass = "astro-ph.CO",
    doi = "10.1088/1475-7516/2022/11/013",
    journal = "JCAP",
    volume = "11",
    pages = "013",
    year = "2022"
}

@article{Lazeyras:2021dar,
    author = "Lazeyras, Titouan and Barreira, Alexandre and Schmidt, Fabian",
    title = "{Assembly bias in quadratic bias parameters of dark matter halos from forward modeling}",
    eprint = "2106.14713",
    archivePrefix = "arXiv",
    primaryClass = "astro-ph.CO",
    doi = "10.1088/1475-7516/2021/10/063",
    journal = "JCAP",
    volume = "10",
    pages = "063",
    year = "2021"
}

@article{Lazeyras:2022koc,
    author = "Lazeyras, Titouan and Barreira, Alexandre and Schmidt, Fabian and Desjacques, Vincent",
    title = "{Assembly bias in the local PNG halo bias and its implication for f $_{NL}$ constraints}",
    eprint = "2209.07251",
    archivePrefix = "arXiv",
    primaryClass = "astro-ph.CO",
    doi = "10.1088/1475-7516/2023/01/023",
    journal = "JCAP",
    volume = "01",
    pages = "023",
    year = "2023"
}

@article{Barreira:2023rxn,
    author = "Barreira, Alexandre and Krause, Elisabeth",
    title = "{Towards optimal and robust f{\_}nl constraints with multi-tracer analyses}",
    eprint = "2302.09066",
    archivePrefix = "arXiv",
    primaryClass = "astro-ph.CO",
    doi = "10.1088/1475-7516/2023/10/044",
    journal = "JCAP",
    volume = "10",
    pages = "044",
    year = "2023"
}

@article{Hadzhiyska:2025rez,
    author = "Hadzhiyska, Boryana and Ferraro, Simone",
    title = "{Refining localtype primordial non-Gaussianity: Sharpened b{\ensuremath{\phi}} constraints through bias expansion}",
    eprint = "2501.14873",
    archivePrefix = "arXiv",
    primaryClass = "astro-ph.CO",
    doi = "10.1103/PhysRevD.111.103521",
    journal = "Phys. Rev. D",
    volume = "111",
    number = "10",
    pages = "103521",
    year = "2025"
}

@article{Johnston:2022nbv,
    author = "Johnston, Harry and Westbeek, Dana Sophia and Weide, Sjoerd and Chisari, Nora Elisa and Dubois, Yohan and Devriendt, Julien and Pichon, Christophe",
    title = "{Intrinsic correlations of galaxy sizes in a hydrodynamical cosmological simulation}",
    eprint = "2209.11063",
    archivePrefix = "arXiv",
    primaryClass = "astro-ph.CO",
    doi = "10.1093/mnras/stad201",
    journal = "Mon. Not. Roy. Astron. Soc.",
    volume = "520",
    number = "1",
    pages = "1541--1566",
    year = "2023"
}

@article{Ghosh:2024mlz,
    author = "Ghosh, Aritra and Urry, C. Megan and Powell, Meredith C. and Shimakawa, Rhythm and van den Bosch, Frank C. and Nagai, Daisuke and Mitra, Kaustav and Connolly, Andrew J.",
    title = "{Denser Environments Cultivate Larger Galaxies: A Comprehensive Study beyond the Local Universe with 3 Million Hyper Suprime-Cam Galaxies}",
    eprint = "2408.07128",
    archivePrefix = "arXiv",
    primaryClass = "astro-ph.GA",
    doi = "10.3847/1538-4357/ad596f",
    journal = "Astrophys. J.",
    volume = "971",
    number = "2",
    pages = "142",
    year = "2024"
}

@article{Shiveshwarkar:2025nac,
    author = "Shiveshwarkar, Charuhas and Loverde, Marilena and Hirata, Christopher M. and Jamieson, Drew",
    title = "{Where does non-Universality in Assembly Bias come from?}",
    eprint = "2508.11798",
    archivePrefix = "arXiv",
    primaryClass = "astro-ph.CO",
    month = "8",
    year = "2025"
}

@article{Vlah:2019byq,
    author = "Vlah, Zvonimir and Chisari, Nora Elisa and Schmidt, Fabian",
    title = "{An EFT description of galaxy intrinsic alignments}",
    eprint = "1910.08085",
    archivePrefix = "arXiv",
    primaryClass = "astro-ph.CO",
    doi = "10.1088/1475-7516/2020/01/025",
    journal = "JCAP",
    volume = "01",
    pages = "025",
    year = "2020"
}

@article{Ghosh:2020zfa,
    author = "Ghosh, Basundhara and Durrer, Ruth and Schaefer, Bjoern Malte",
    title = "{Intrinsic and extrinsic correlations of galaxy shapes and sizes in weak lensing data}",
    eprint = "2005.04604",
    archivePrefix = "arXiv",
    primaryClass = "astro-ph.CO",
    doi = "10.1093/mnras/stab1435",
    journal = "Mon. Not. Roy. Astron. Soc.",
    volume = "505",
    number = "2",
    pages = "2594--2609",
    year = "2021"
}

@article{Vlah:2020ovg,
    author = "Vlah, Zvonimir and Chisari, Nora Elisa and Schmidt, Fabian",
    title = "{Galaxy shape statistics in the effective field theory}",
    eprint = "2012.04114",
    archivePrefix = "arXiv",
    primaryClass = "astro-ph.CO",
    doi = "10.1088/1475-7516/2021/05/061",
    journal = "JCAP",
    volume = "05",
    pages = "061",
    year = "2021"
}

@article{Kurita:2025hmp,
    author = "Kurita, Toshiki and Jamieson, Drew and Komatsu, Eiichiro and Schmidt, Fabian",
    title = "{Parity Violation in Galaxy Shapes: Primordial Non-Gaussianity}",
    eprint = "2509.08787",
    archivePrefix = "arXiv",
    primaryClass = "astro-ph.CO",
    month = "9",
    year = "2025"
}

@article{Hirata:2003cv,
    author = "Hirata, Christopher M. and Seljak, Uros",
    title = "{Shear calibration biases in weak lensing surveys}",
    eprint = "astro-ph/0301054",
    archivePrefix = "arXiv",
    doi = "10.1046/j.1365-8711.2003.06683.x",
    journal = "Mon. Not. Roy. Astron. Soc.",
    volume = "343",
    pages = "459--480",
    year = "2003"
}

@article{Massey:2006ha,
    author = "Massey, Richard and others",
    title = "{The Shear TEsting Programme 2: Factors affecting high precision weak lensing analyses}",
    eprint = "astro-ph/0608643",
    archivePrefix = "arXiv",
    doi = "10.1111/j.1365-2966.2006.11315.x",
    journal = "Mon. Not. Roy. Astron. Soc.",
    volume = "376",
    pages = "13--38",
    year = "2007"
}

@article{Meert:2014pta,
    author = "Meert, Alan and Vikram, Vinu and Bernardi, Mariangela",
    title = "{A catalogue of 2D photometric decompositions in the SDSS-DR7 spectroscopic main galaxy sample: preferred models and systematics}",
    eprint = "1406.4179",
    archivePrefix = "arXiv",
    primaryClass = "astro-ph.GA",
    doi = "10.1093/mnras/stu2333",
    journal = "Mon. Not. Roy. Astron. Soc.",
    volume = "446",
    number = "4",
    pages = "3943--3974",
    year = "2015"
}

@article{Singh:2015sva,
    author = "Singh, Sukhdeep and Mandelbaum, Rachel",
    title = "{Intrinsic alignments of BOSS LOWZ galaxies {\textendash} II. Impact of shape measurement methods}",
    eprint = "1510.06752",
    archivePrefix = "arXiv",
    primaryClass = "astro-ph.CO",
    doi = "10.1093/mnras/stw144",
    journal = "Mon. Not. Roy. Astron. Soc.",
    volume = "457",
    number = "3",
    pages = "2301--2317",
    year = "2016"
}

@article{Schmidt:2015xka,
    author = "Schmidt, Fabian and Chisari, Nora Elisa and Dvorkin, Cora",
    title = "{Imprint of inflation on galaxy shape correlations}",
    eprint = "1506.02671",
    archivePrefix = "arXiv",
    primaryClass = "astro-ph.CO",
    doi = "10.1088/1475-7516/2015/10/032",
    journal = "JCAP",
    volume = "10",
    pages = "032",
    year = "2015"
}

@article{Kurita:2020hap,
    author = "Kurita, Toshiki and Takada, Masahiro and Nishimichi, Takahiro and Takahashi, Ryuichi and Osato, Ken and Kobayashi, Yosuke",
    title = "{Power spectrum of halo intrinsic alignments in simulations}",
    eprint = "2004.12579",
    archivePrefix = "arXiv",
    primaryClass = "astro-ph.CO",
    reportNumber = "IPMU20-0044, YITP-20-48",
    doi = "10.1093/mnras/staa3625",
    journal = "Mon. Not. Roy. Astron. Soc.",
    volume = "501",
    number = "1",
    pages = "833--852",
    year = "2021"
}

@article{Akitsu:2020jvx,
    author = "Akitsu, Kazuyuki and Kurita, Toshiki and Nishimichi, Takahiro and Takada, Masahiro and Tanaka, Satoshi",
    title = "{Imprint of anisotropic primordial non-Gaussianity on halo intrinsic alignments in simulations}",
    eprint = "2007.03670",
    archivePrefix = "arXiv",
    primaryClass = "astro-ph.CO",
    reportNumber = "IPMU20-0075, YITP-20-85",
    doi = "10.1103/PhysRevD.103.083508",
    journal = "Phys. Rev. D",
    volume = "103",
    number = "8",
    pages = "083508",
    year = "2021"
}

@article{Akitsu:2020fpg,
    author = "Akitsu, Kazuyuki and Li, Yin and Okumura, Teppei",
    title = "{Cosmological simulation in tides: power spectra, halo shape responses, and shape assembly bias}",
    eprint = "2011.06584",
    archivePrefix = "arXiv",
    primaryClass = "astro-ph.CO",
    doi = "10.1088/1475-7516/2021/04/041",
    journal = "JCAP",
    volume = "04",
    pages = "041",
    year = "2021"
}

@article{Akitsu:2023eqa,
    author = "Akitsu, Kazuyuki and Li, Yin and Okumura, Teppei",
    title = "{Quadratic shape biases in three-dimensional halo intrinsic alignments}",
    eprint = "2306.00969",
    archivePrefix = "arXiv",
    primaryClass = "astro-ph.CO",
    doi = "10.1088/1475-7516/2023/08/068",
    journal = "JCAP",
    volume = "08",
    pages = "068",
    year = "2023"
}

@article{Kurita:2023qku,
    author = "Kurita, Toshiki and Takada, Masahiro",
    title = "{Constraints on anisotropic primordial non-Gaussianity from intrinsic alignments of SDSS-III BOSS galaxies}",
    eprint = "2302.02925",
    archivePrefix = "arXiv",
    primaryClass = "astro-ph.CO",
    doi = "10.1103/PhysRevD.108.083533",
    journal = "Phys. Rev. D",
    volume = "108",
    number = "8",
    pages = "083533",
    year = "2023"
}

@article{Bakx:2023mld,
    author = "Bakx, Thomas and Kurita, Toshiki and Chisari, Nora Elisa and Vlah, Zvonimir and Schmidt, Fabian",
    title = "{Effective field theory of intrinsic alignments at one loop order: a comparison to dark matter simulations}",
    eprint = "2303.15565",
    archivePrefix = "arXiv",
    primaryClass = "astro-ph.CO",
    doi = "10.1088/1475-7516/2023/10/005",
    journal = "JCAP",
    volume = "10",
    pages = "005",
    year = "2023"
}

@article{Chua:2018sbi,
    author = "Chua, Kunting Eddie and Pillepich, Annalisa and Vogelsberger, Mark and Hernquist, Lars",
    title = "{Shape of Dark Matter Haloes in the Illustris Simulation: Effects of Baryons}",
    eprint = "1809.07255",
    archivePrefix = "arXiv",
    primaryClass = "astro-ph.GA",
    doi = "10.1093/mnras/sty3531",
    journal = "Mon. Not. Roy. Astron. Soc.",
    volume = "484",
    number = "1",
    pages = "476--493",
    year = "2019"
}

@article{Karmakar:2023dtt,
    author = "Karmakar, Tathagata and Genel, Shy and Somerville, Rachel S.",
    title = "{The relationship between galaxy and halo sizes in the Illustris and IllustrisTNG simulations}",
    eprint = "2301.10409",
    archivePrefix = "arXiv",
    primaryClass = "astro-ph.GA",
    doi = "10.1093/mnras/stad178",
    journal = "Mon. Not. Roy. Astron. Soc.",
    volume = "520",
    number = "2",
    pages = "1630--1641",
    year = "2023"
}

@article{Shi:2020cyz,
    author = "Shi, Jingjing and Kurita, Toshiki and Takada, Masahiro and Osato, Ken and Kobayashi, Yosuke and Nishimichi, Takahiro",
    title = "{Power Spectrum of Intrinsic Alignments of Galaxies in IllustrisTNG}",
    eprint = "2009.00276",
    archivePrefix = "arXiv",
    primaryClass = "astro-ph.GA",
    reportNumber = "YITP-20-113",
    doi = "10.1088/1475-7516/2021/03/030",
    journal = "JCAP",
    volume = "03",
    pages = "030",
    year = "2021"
}

@article{vanHeukelum:2025njq,
    author = "van Heukelum, Marloes and Neumann, Dennis and Escobar, Marta Garcia and Chisari, Nora Elisa and Hoekstra, Henk",
    title = "{Intrinsic alignments of galaxies in multiple projections}",
    eprint = "2509.07734",
    archivePrefix = "arXiv",
    primaryClass = "astro-ph.CO",
    month = "9",
    year = "2025"
}

@article{Behroozi:2021ovy,
    author = "Behroozi, Peter and Hearin, Andrew and Moster, Benjamin P.",
    title = "{Observational measures of halo properties beyond mass}",
    eprint = "2101.05280",
    archivePrefix = "arXiv",
    primaryClass = "astro-ph.GA",
    doi = "10.1093/mnras/stab3193",
    journal = "Mon. Not. Roy. Astron. Soc.",
    volume = "509",
    number = "2",
    pages = "2800--2824",
    year = "2021"
}

@article{Gangui:1993tt,
    author = "Gangui, Alejandro and Lucchin, Francesco and Matarrese, Sabino and Mollerach, Silvia",
    title = "{The Three point correlation function of the cosmic microwave background in inflationary models}",
    eprint = "astro-ph/9312033",
    archivePrefix = "arXiv",
    reportNumber = "SISSA-193-93-A, DFPD-93-A-80",
    doi = "10.1086/174421",
    journal = "Astrophys. J.",
    volume = "430",
    pages = "447--457",
    year = "1994"
}

@article{Komatsu:2001rj,
    author = "Komatsu, Eiichiro and Spergel, David N.",
    title = "{Acoustic signatures in the primary microwave background bispectrum}",
    eprint = "astro-ph/0005036",
    archivePrefix = "arXiv",
    doi = "10.1103/PhysRevD.63.063002",
    journal = "Phys. Rev. D",
    volume = "63",
    pages = "063002",
    year = "2001"
}

@article{Seljak:2008xr,
    author = "Seljak, Uros",
    title = "{Extracting primordial non-gaussianity without cosmic variance}",
    eprint = "0807.1770",
    archivePrefix = "arXiv",
    primaryClass = "astro-ph",
    doi = "10.1103/PhysRevLett.102.021302",
    journal = "Phys. Rev. Lett.",
    volume = "102",
    pages = "021302",
    year = "2009"
}

@article{Castorina:2018zfk,
    author = "Castorina, Emanuele and Feng, Yu and Seljak, Uros and Villaescusa-Navarro, Francisco",
    title = "{Primordial non-Gaussianities and zero bias tracers of the Large Scale Structure}",
    eprint = "1803.11539",
    archivePrefix = "arXiv",
    primaryClass = "astro-ph.CO",
    doi = "10.1103/PhysRevLett.121.101301",
    journal = "Phys. Rev. Lett.",
    volume = "121",
    number = "10",
    pages = "101301",
    year = "2018"
}

@article{Kokron:2025yma,
    author = "Kokron, Nickolas",
    title = "{Local primordial non-Gaussianity from 'zero-bias' 21cm radiation during reionization}",
    eprint = "2504.20025",
    archivePrefix = "arXiv",
    primaryClass = "astro-ph.CO",
    month = "4",
    year = "2025"
}

@article{Lewis:camb,
      author         = "Lewis, Antony and Challinor, Anthony and Lasenby,
                        Anthony",
      title          = "{Efficient computation of CMB anisotropies in closed FRW
                        models}",
      journal        = "\apj",
      volume         = "538",
      year           = "2000",
      pages          = "473-476",
      doi            = "10.1086/309179",
      eprint         = "astro-ph/9911177",
      archivePrefix  = "arXiv",
      primaryClass   = "astro-ph",
      SLACcitation   = "%%CITATION = ASTRO-PH/9911177;%%"
}

@article{Howlett:camb,
      author         = "Howlett, Cullan and Lewis, Antony and Hall, Alex and
                        Challinor, Anthony",
      title          = "{CMB power spectrum parameter degeneracies in the era of
                        precision cosmology}",
      journal        = "\jcap",
      volume         = "1204",
      year           = "2012",
      pages          = "027",
      doi            = "10.1088/1475-7516/2012/04/027",
      eprint         = "1201.3654",
      archivePrefix  = "arXiv",
      primaryClass   = "astro-ph.CO",
      SLACcitation   = "%%CITATION = ARXIV:1201.3654;%%"
}

@article{Adame:2025zhg,
    author = "Adame, Adrian G. and Avila, Santiago and Gonzalez-Perez, Violeta and Hahn, Oliver and Yepes, Gustavo and Manera, Marc",
    title = "{Accurate $N$-body simulations with local Primordial non-Gaussianities: initial conditions and aliasing}",
    eprint = "2506.06200",
    archivePrefix = "arXiv",
    primaryClass = "astro-ph.CO",
    month = "6",
    year = "2025"
}

@software{2020ascl.soft08024H,
       author = {{Hahn}, Oliver and {Michaux}, Micha{\"e}l and {Rampf}, Cornelius and {Uhlemann}, Cora and {Angulo}, Raul E.},
        title = "{MUSIC2-monofonIC: 3LPT initial condition generator}",
 howpublished = {Astrophysics Source Code Library, record ascl:2008.024},
         year = 2020,
        month = aug,
          eid = {ascl:2008.024},
archivePrefix = {ascl},
       eprint = {2008.024},
       adsurl = {https://ui.adsabs.harvard.edu/abs/2020ascl.soft08024H}
}

@article{Springel:2005mi,
    author = "Springel, Volker",
    title = "{The Cosmological simulation code GADGET-2}",
    eprint = "astro-ph/0505010",
    archivePrefix = "arXiv",
    doi = "10.1111/j.1365-2966.2005.09655.x",
    journal = "Mon. Not. Roy. Astron. Soc.",
    volume = "364",
    pages = "1105--1134",
    year = "2005"
}

@article{Navarro:1996gj,
    author = "Navarro, Julio F. and Frenk, Carlos S. and White, Simon D. M.",
    title = "{A Universal density profile from hierarchical clustering}",
    eprint = "astro-ph/9611107",
    archivePrefix = "arXiv",
    doi = "10.1086/304888",
    journal = "Astrophys. J.",
    volume = "490",
    pages = "493--508",
    year = "1997"
}

@article{Bryan:1997dn,
    author = "Bryan, G. L. and Norman, M. L.",
    title = "{Statistical properties of x-ray clusters: Analytic and numerical comparisons}",
    eprint = "astro-ph/9710107",
    archivePrefix = "arXiv",
    doi = "10.1086/305262",
    journal = "Astrophys. J.",
    volume = "495",
    pages = "80",
    year = "1998"
}

@article{Klypin:2010qw,
    author = "Klypin, A. and Trujillo-Gomez, S. and Primack, J.",
    title = "{Halos and galaxies in the standard cosmological model: results from the Bolshoi simulation}",
    eprint = "1002.3660",
    archivePrefix = "arXiv",
    primaryClass = "astro-ph.CO",
    doi = "10.1088/0004-637X/740/2/102",
    journal = "Astrophys. J.",
    volume = "740",
    pages = "102",
    year = "2011"
}

@article{Behroozi:2011ju,
    author = "Behroozi, Peter S. and Wechsler, Risa H. and Wu, Hao-Yi",
    title = "{The Rockstar Phase-Space Temporal Halo Finder and the Velocity Offsets of Cluster Cores}",
    eprint = "1110.4372",
    archivePrefix = "arXiv",
    primaryClass = "astro-ph.CO",
    reportNumber = "SLAC-PUB-15331",
    doi = "10.1088/0004-637X/762/2/109",
    journal = "Astrophys. J.",
    volume = "762",
    pages = "109",
    year = "2013"
}

@article{Tokuue:2024ney,
    author = "Tokuue, Tomoyuki and Ishiyama, Tomoaki and Osato, Ken and Tanaka, Satoshi and Behroozi, Peter",
    title = "{MPI-Rockstar: a Hybrid MPI and OpenMP Parallel Implementation of the Rockstar Halo finder}",
    eprint = "2412.18629",
    archivePrefix = "arXiv",
    primaryClass = "astro-ph.IM",
    month = "12",
    year = "2024"
}

@article{Knollmann:2009pb,
    author = "Knollmann, Steffen R. and Knebe, Alexander",
    title = "{Ahf: Amiga's Halo Finder}",
    eprint = "0904.3662",
    archivePrefix = "arXiv",
    primaryClass = "astro-ph.CO",
    doi = "10.1088/0067-0049/182/2/608",
    journal = "Astrophys. J. Suppl.",
    volume = "182",
    pages = "608--624",
    year = "2009"
}

@article{Sirko:2005uz,
    author = "Sirko, Edwin",
    title = "{Initial conditions to cosmological N-body simulations, or how to run an ensemble of simulations}",
    eprint = "astro-ph/0503106",
    archivePrefix = "arXiv",
    doi = "10.1086/497090",
    journal = "Astrophys. J.",
    volume = "634",
    pages = "728--743",
    year = "2005"
}

@article{Wagner:2014aka,
    author = "Wagner, Christian and Schmidt, Fabian and Chiang, Chi-Ting and Komatsu, Eiichiro",
    title = "{Separate Universe Simulations}",
    eprint = "1409.6294",
    archivePrefix = "arXiv",
    primaryClass = "astro-ph.CO",
    doi = "10.1093/mnrasl/slu187",
    journal = "Mon. Not. Roy. Astron. Soc.",
    volume = "448",
    number = "1",
    pages = "L11--L15",
    year = "2015"
}

@article{Kaiser:1987qv,
    author = "Kaiser, N.",
    title = "{Clustering in real space and in redshift space}",
    doi = "10.1093/mnras/227.1.1",
    journal = "Mon. Not. Roy. Astron. Soc.",
    volume = "227",
    pages = "1--27",
    year = "1987"
}

@article{DESI:2016fyo,
    author = "Aghamousa, Amir and others",
    collaboration = "DESI",
    title = "{The DESI Experiment Part I: Science,Targeting, and Survey Design}",
    eprint = "1611.00036",
    archivePrefix = "arXiv",
    primaryClass = "astro-ph.IM",
    reportNumber = "FERMILAB-PUB-16-517-AE",
    month = "10",
    year = "2016"
}

@article{Chaussidon:2024qni,
    author = "Chaussidon, E. and others",
    title = "{Constraining primordial non-Gaussianity with DESI 2024 LRG and QSO samples}",
    eprint = "2411.17623",
    archivePrefix = "arXiv",
    primaryClass = "astro-ph.CO",
    reportNumber = "FERMILAB-PUB-24-0884-PPD",
    doi = "10.1088/1475-7516/2025/06/029",
    journal = "JCAP",
    volume = "06",
    pages = "029",
    year = "2025"
}

@article{Euclid:2024yrr,
    author = "Mellier, Y. and others",
    collaboration = "Euclid",
    title = "{Euclid. I. Overview of the Euclid mission}",
    eprint = "2405.13491",
    archivePrefix = "arXiv",
    primaryClass = "astro-ph.CO",
    doi = "10.1051/0004-6361/202450810",
    journal = "Astron. Astrophys.",
    volume = "697",
    pages = "A1",
    year = "2025"
}

@article{Dore:2014,
  author = "Dor{\'e}, Olivier and Bock, Jamie and Ashby, Matthew and Capak, Peter and Cooray, Asantha and de Putter, Roland and Eifler, Tim and Flagey, Nicolas and Gong, Yan and Habib, Salman and Heitmann, Katrin and Hirata, Chris and Jeong, Woong-Seob and Katti, Raj and Korngut, Phil and Krause, Elisabeth and Lee, Dae-Hee and Masters, Daniel and Mauskopf, Phil and Melnick, Gary and Mennesson, Bertrand and Nguyen, Hien and {\"O}berg, Karin and Pullen, Anthony and Raccanelli, Alvise and Smith, Roger and Song, Yong-Seon and Tolls, Volker and Unwin, Steve and Venumadhav, Tejaswi and Viero, Marco and Werner, Mike and Zemcov, Mike",
  title = "{Cosmology with the SPHEREX All-Sky Spectral Survey}",
  eprint = "1412.4872",
  archivePrefix = "arXiv",
  primaryClass = "astro-ph.CO",
  year = "2014"
}

@article{Wang:2021oec,
    author = "Wang, Yun and others",
    title = "{The High Latitude Spectroscopic Survey on the Nancy Grace Roman Space Telescope}",
    eprint = "2110.01829",
    archivePrefix = "arXiv",
    primaryClass = "astro-ph.CO",
    doi = "10.3847/1538-4357/ac4973",
    journal = "Astrophys. J.",
    volume = "928",
    number = "1",
    pages = "1",
    year = "2022"
}

@article{Nguyen:2024yth,
    author = "Nguyen, Nhat-Minh and Schmidt, Fabian and Tucci, Beatriz and Reinecke, Martin and Kosti{\'c}, Andrija",
    title = "{How Much Information Can Be Extracted from Galaxy Clustering at the Field Level?}",
    eprint = "2403.03220",
    archivePrefix = "arXiv",
    primaryClass = "astro-ph.CO",
    reportNumber = "LCTP-24-05",
    doi = "10.1103/PhysRevLett.133.221006",
    journal = "Phys. Rev. Lett.",
    volume = "133",
    number = "22",
    pages = "221006",
    year = "2024"
}

@article{Akitsu:2025boy,
    author = "Akitsu, Kazuyuki and Simonovi{\'c}, Marko and Chen, Shi-Fan and Cabass, Giovanni and Zaldarriaga, Matias",
    title = "{Cosmology inference with perturbative forward modeling at the field level: a comparison with joint power spectrum and bispectrum analyses}",
    eprint = "2509.09673",
    archivePrefix = "arXiv",
    primaryClass = "astro-ph.CO",
    reportNumber = "KEK-Cosmo-0386, RBI-ThPhys-2025-34",
    month = "9",
    year = "2025"
}

@MISC{Pylians,
    author = {{Villaescusa-Navarro}, Francisco},
    title = "{Pylians: Python libraries for the analysis of numerical simulations}",
    howpublished = {Astrophysics Source Code Library, record ascl:1811.008},
    year = 2018,
    month = nov,
    eid = {ascl:1811.008},
    pages = {ascl:1811.008},
    archivePrefix = {ascl},
    eprint = {1811.008},
    adsurl = {https://ui.adsabs.harvard.edu/abs/2018ascl.soft11008V}
}

@software{andrew_collette_2022_6575970,
  author       = {Andrew Collette and
                  Thomas Kluyver and
                  Thomas A Caswell and
                  James Tocknell and
                  Jerome Kieffer and
                  Aleksandar Jelenak and
                  Anthony Scopatz and
                  Darren Dale and
                  Chen and
                  Thomas VINCENT and
                  Matt Einhorn and
                  payno and
                  juliagarriga and
                  Pierlauro Sciarelli and
                  Valentin Valls and
                  Satrajit Ghosh and
                  Ulrik Kofoed Pedersen and
                  jakirkham and
                  Martin Raspaud and
                  Cyril Danilevski and
                  Hameer Abbasi and
                  John Readey and
                  Kai Mühlbauer and
                  Andrey Paramonov and
                  Lawrence Chan and
                  V. Armando Solé and
                  jialin and
                  Daniel Hay Guest and
                  Yu Feng and
                  Mark Kittisopikul},
  title        = {h5py/h5py: 3.7.0},
  month        = may,
  year         = 2022,
  publisher    = {Zenodo},
  version      = {3.7.0},
  doi          = {10.5281/zenodo.6575970},
  url          = {https://doi.org/10.5281/zenodo.6575970},
}

@article{Li:2015jsz,
    author = "Li, Yin and Hu, Wayne and Takada, Masahiro",
    title = "{Separate Universe Consistency Relation and Calibration of Halo Bias}",
    eprint = "1511.01454",
    archivePrefix = "arXiv",
    primaryClass = "astro-ph.CO",
    doi = "10.1103/PhysRevD.93.063507",
    journal = "Phys. Rev. D",
    volume = "93",
    number = "6",
    pages = "063507",
    year = "2016"
}

@article{Baldauf:2015vio,
    author = "Baldauf, Tobias and Seljak, Uro{\v{s}} and Senatore, Leonardo and Zaldarriaga, Matias",
    title = "{Linear response to long wavelength fluctuations using curvature simulations}",
    eprint = "1511.01465",
    archivePrefix = "arXiv",
    primaryClass = "astro-ph.CO",
    doi = "10.1088/1475-7516/2016/09/007",
    journal = "JCAP",
    volume = "09",
    pages = "007",
    year = "2016"
}

@article{Smith:2006ne,
    author = "Smith, Robert E. and Scoccimarro, Roman and Sheth, Ravi K.",
    title = "{The Scale Dependence of Halo and Galaxy Bias: Effects in Real Space}",
    eprint = "astro-ph/0609547",
    archivePrefix = "arXiv",
    doi = "10.1103/PhysRevD.75.063512",
    journal = "Phys. Rev. D",
    volume = "75",
    pages = "063512",
    year = "2007"
}

@article{Hamaus:2010im,
    author = "Hamaus, Nico and Seljak, Uros and Desjacques, Vincent 
              and Smith, Robert E. and Baldauf, Tobias",
    title = "{Minimizing the stochasticity of halos in large-scale 
              structure surveys}",
    eprint = "1004.5377",
    archivePrefix = "arXiv",
    primaryClass = "astro-ph.CO",
    doi = "10.1103/PhysRevD.82.043515",
    journal = "Phys. Rev. D",
    volume = "82",
    pages = "043515",
    year = "2010"
}

@article{Baldauf:2013hka,
    author = "Baldauf, Tobias and Seljak, Uro\v{s} and Smith, Robert E. and Hamaus, Nico and Desjacques, Vincent",
    title = "{Halo stochasticity from exclusion and nonlinear clustering}",
    eprint = "1305.2917",
    archivePrefix = "arXiv",
    primaryClass = "astro-ph.CO",
    doi = "10.1103/PhysRevD.88.083507",
    journal = "Phys. Rev. D",
    volume = "88",
    number = "8",
    pages = "083507",
    year = "2013"
}

\end{document}